\documentclass[12pt]{article}
\pdfoutput=1
\usepackage{jheppub}

\usepackage{amsmath, amssymb, amsfonts, amsthm}
\usepackage{mathtools}
\usepackage{bm}
\usepackage{dsfont}
\usepackage{enumerate}
\usepackage{bbold}
\usepackage{physics}
\usepackage[table]{xcolor}
\usepackage{pgf}
\usepackage{simplewick}
\usepackage{tikz}
\usetikzlibrary{tikzmark}
\usetikzlibrary{calc}
\usepackage{array}
\usepackage{url}
\usepackage{graphicx}
\usepackage{subcaption}
\usepackage{comment}

\numberwithin{equation}{section}

\newcommand{\be}{\begin{equation}}
\newcommand{\ee}{\end{equation}}
\newcommand{\bea}{\begin{aligned}}
\newcommand{\eea}{\end{aligned}}
\newcommand{\mfrac}[2]{%
  \mathchoice
    {\scalebox{0.85}{$\displaystyle\frac{#1}{#2}$}}
    {\scalebox{0.85}{$\textstyle\frac{#1}{#2}$}}
    {\scalebox{0.85}{$\scriptstyle\frac{#1}{#2}$}}
    {\scalebox{0.85}{$\scriptscriptstyle\frac{#1}{#2}$}}%
}

\newcommand{\sker}[3]{\mathbb{S}_{#1 #2}[#3]}

\usetikzlibrary{shapes}
\tikzset{
    partial ellipse/.style args={#1:#2:#3}{
        insert path={+ (#1:#3) arc (#1:#2:#3)}
    }
}

\definecolor{myforestgreen}{RGB}{24, 150, 144}

\usepackage[normalem]{ulem}
\definecolor{ForestGreen}{RGB}{34,139,34}

\def\bbi{\mathbb{1}}

\def\tilde{\widetilde}

\def\e{\mathrm{e}}
\def\Re{\mathrm{Re}}
\def\Im{\mathrm{Im}} 
\def\hat{\widehat}

\newcommand{\Li}{\text{Li}}

\newcommand{\tetra}{
  \mathord{
    \begin{tikzpicture}[scale=0.5,baseline=-0.5ex]
      \draw[line width=0.75] (0,0) circle (0.5);
      \draw[line width=0.75] (0,0)--({0.5*cos(30)},{-0.5*sin(30)});
      \draw[line width=0.75] (0,0)--({-0.5*cos(30)},{-0.5*sin(30)});
      \draw[line width=0.75] (0,0)--(0,0.5);
    \end{tikzpicture}
  }
}

\begin{document}
\tikzset{every picture/.style={line width=0.9pt}} 

\title{3D de Sitter from complex Virasoro TQFT: Observers and knots}

\author[1]{Jeevan Chandra,}
\affiliation[1]{Leinweber Institute for Theoretical Physics and Department of Physics,
University of California, Berkeley, USA}
\emailAdd{jcn1998@berkeley.edu}

\author[2,3]{Diego Liska,}
\affiliation[2]{Department of Theoretical Physics, University of Geneva, 1211 Geneva, Switzerland}
\affiliation[3]{Laboratory for Theoretical Fundamental Physics, Institute of Physics, Ecole Polytechnique
F\'ed\'erale de Lausanne (EPFL), 1015 Lausanne, Switzerland}
\emailAdd{diego.liska@epfl.ch}

\author[4]{Boris Post}
\affiliation[4]{Mathematical Institute, University of Oxford, OX2 6GG Oxford, United
Kingdom}
\emailAdd{boris.post@maths.ox.ac.uk}

\abstract{

The Euclidean saddles of 3D Einstein gravity with positive cosmological constant are spherical three-manifolds. When one adds a heavy observer, the backreacted saddles are spherical cone manifolds, with a conical singularity along the observer worldline. We propose exact gravitational partition functions at fixed topology on a class of spherical cone manifolds that undergo hyperbolic-to-spherical transitions as the cone angles increase. Our construction relies on a careful analytic continuation of the product of two Virasoro TQFT partition functions. We test the proposal by showing that the exact partition functions exponentiate in the semiclassical limit to the spherical volume, in agreement with the gravitational on-shell action. This leads to a spherical volume conjecture for the product of continued Virasoro TQFT amplitudes. As further evidence, we relate the theta and tetrahedron graph amplitudes to the two-point function and out-of-time-ordered four-point function of operators on a massive observer's worldline. In appropriate limits, we recover the static patch scalar two-point function with finite observer backreaction and the semiclassical effects of observer recoil. In addition, we analyze examples with knotted worldlines and reproduce their spherical volumes. We also apply our prescription to Seifert manifolds, which are off-shell in AdS$_3$ gravity but admit dS$_3$ saddles, and find agreement with their spherical volumes semiclassically. Finally, using Dehn surgery on torus knots and links, we obtain partition functions on smooth spherical three-manifolds. 
}

\maketitle

\newpage

\section{Introduction}\label{sec:Intro}

Gravitational partition functions are among the most concrete quantities that one can hope to compute in quantum gravity. Even at the semiclassical level, they encode thermodynamic quantities, most notably horizon entropies, that appear to admit a microscopic interpretation \cite{Gibbons:1976ue,Gibbons:1977mu,Bekenstein:1973ur,Hawking:1975vcx,Strominger:1996sh}. De Sitter space is particularly interesting in this regard because the microscopic origin of its horizon entropy remains poorly understood.

3D gravity provides an especially tractable setting in which to explore this question. In three dimensions, pure Einstein gravity has no local degrees of freedom \cite{Witten:1988hc}; instead, the theory is governed by global degrees of freedom and topology. This structure has led to substantial progress in understanding the quantum theory with a negative cosmological constant. In particular, Virasoro TQFT provides a reformulation of AdS$_3$ gravity in terms of the quantization of Teichmüller space \cite{Collier:2023fwi}. Within this framework, Euclidean gravitational partition functions can be computed exactly in $G_N$ for a large class of 3-manifolds, see e.g. \cite{Collier:2024mgv,Chandra:2025fef,Belin:2026pko}.

More recently, several developments have indicated that the structures underlying Virasoro TQFT may also admit a de Sitter analogue. Analytically continued Virasoro TQFT amplitudes have been used to construct wavefunctions for inflating cosmologies \cite{Collier:2025lux}, and closely related structures have appeared in proposals for the quantization and microscopic description of dS$_3$ gravity \cite{Blommaert:2025eps,Blommaert:2026ofx}  (see also, \cite{Cotler:2019nbi,Collier:2024kmo,Verlinde:2024znh,Godet:2024ich,Chakravarty:2025sbg}). These results suggest that suitably continued Virasoro TQFT amplitudes may provide a natural prescription for defining 3D de Sitter partition functions to all orders in Newton’s constant $G_N$. They also raise the following question: \emph{for which 3-manifolds can such a continuation be given a gravitational interpretation?}

This paper focuses on a class of 3-manifolds that admit a continuation between hyperbolic and spherical geometries. The geometries we consider contain one or more line defects inside $S^3$, along which the metric has a conical singularity. Each such defect is characterized by a cone angle $\alpha$, and can be interpreted as the worldline of a massive spin-zero point particle whose action is 
\begin{equation}
  I_{\text{WL}} = m \int d\tau \sqrt{g_{\mu\nu}\frac{dx^\mu}{d\tau}\frac{dx^\nu}{d\tau}}\quad\text{where}\quad m =\frac{ 2\pi - \alpha}{8\pi G_N }.
\end{equation}
Here $x(\tau)$ parametrizes the particle's worldline. The topology of this worldline can be nontrivial, and in several of our examples the defects form knotted closed curves, such as the figure-eight knot:
\begin{equation}
   M_\alpha \;=\; \vcenter{\hbox{ \begin{tikzpicture} 
   \begin{scope}[scale=0.70,xscale=0.75]      
        \draw[very thick, out = 90, in = 180, black] (-5/2,-3/2) to (-1,0);
        \draw[very thick, out = 0, in = 180, black] (-1,0) to (0,0);
        \draw[fill=white,draw=white] (-1,0) circle (1/10);
        \draw[very thick, out = 180, in = 90, black] (0,1) to (-1,0);
\draw[very thick, out = 270, in = 150, black] (-1,0) to (0,-1);
        \draw[very thick, out = 330, in = 90, black] (0,-1) to (1,-2);
        \draw[fill=white,draw=white] (0,-1) circle (1/10);
\draw[very thick, out = 90, in = 270, black] (-1,-2) to (1,0);
        \draw[very thick, out = 90, in = 0, black] (1,0) to (0,1);
        \draw[fill=white,draw=white] (1,0) circle (1/10);
        \draw[very thick, out = 0, in = 180, black] (0,0) to (1,0);
        \draw[very thick, out = 0, in = 90, black] (1,0) to (5/2,-3/2);
        \draw[very thick, out = 270, in = 330, black] (5/2,-3/2) to (0,-3);
\draw[very thick, out = 150, in = 270, black] (0,-3) to (-1,-2);
        \draw[fill=white,draw=white] (0,-3) circle (1/5);
\draw[very thick, out = 270, in= 0, black] (1,-2) to (0,-3);
        \draw[very thick, out = 180, in = 270, black] (0,-3) to (-5/2,-3/2);
        \end{scope}
        \draw[very thick, black] (0,-0.7) node[below right=1.4cm] {$S^3$} circle (2);
        \node at (0.75,0.5){$\alpha$};
\end{tikzpicture}}}.
\end{equation}
As the cone angle is varied, the manifold admits a constant-curvature metric for different signs of the cosmological constant: hyperbolic, flat or spherical. In this way, the cone angle $\alpha$ provides a parameter that interpolates between AdS$_3$ and dS$_3$ geometries (in Euclidean signature). The figure-eight knot provides a well-studied example: the corresponding cone manifold is hyperbolic if $\alpha\in [0,2\pi/3)$, Euclidean if $\alpha = 2\pi/3$, and 
spherical if $\alpha \in (2\pi/3, 4\pi/3)$ \cite{Mednykh2003,MednykhRasskazov:2006}. 

More generally, the singular locus $\Gamma_{\vec{\alpha}}$ of these manifolds does not need to be a collection of disjoint closed worldlines. Instead, defect lines may meet at junctions, so that $\Gamma_{\vec{\alpha}}$ forms an embedded graph in $S^3$. Two examples that will be particularly important for us are those where $\Gamma_{\vec{\alpha}}$ is a theta graph or forms the skeleton of a tetrahedron:
\begin{equation}\label{eq:intropictures}
    \Gamma_{\vec{\alpha}}\; =\; 
    \vcenter{\hbox{
    \begin{tikzpicture}
        \draw[very thick, black] (0,0) node[right=1.5cm]{$\alpha_3$}  node[left=0.8cm]{$\alpha_1$} ellipse (1.5 and 1);
        \draw[very thick, black] (0,-1) -- node[right]{$\alpha_2$} (0,1);
    \end{tikzpicture}
    }}\quad \text{or} \qquad 
    \vcenter{
    \hbox{
\begin{tikzpicture}[scale=1.7,xscale=0.95]
    \draw[very thick, black] (-1,0) -- (1,0);
    \draw[line width=5pt, white] (0,-0.6) -- (0,1);
    \draw[very thick, black] (0,-0.6) -- (0,1);
    \draw[very thick,black] (-1,0) --node[below left]{$\alpha_6$} (0,-0.6) --node[below right]{$\alpha_4$} (1,0) --node[above right]{$\alpha_3$} (0,1) --node[above left]{$\alpha_1$} cycle;
    \node at(-0.35,0.15) {$\alpha_5$};
    \node at(0.2,0.35) {$\alpha_2$};
\end{tikzpicture}
    }
    }.
\end{equation}
We will denote $\vec{\alpha} = (\alpha_1,\dots,\alpha_n)$ to keep track of the angles in a cone manifold $M_{\vec{\alpha}}$.

An appealing feature of our setup is that the singular locus itself can be interpreted as the worldline of a heavy observer. This is especially interesting in de Sitter quantum gravity, where observers are an essential part of the description. 
Since de Sitter space has no asymptotic boundary, the observer worldline  provides a natural reference for defining gravitational observables \cite{Anninos:2011af,Anninos:2017hhn,Chandrasekaran:2022cip,Witten:2023xze}. 
Cone manifolds with junctions like the theta graph and the tetrahedron contribute to the observer 2-point and 4-point function, respectively. These correlators have been the focus of considerable recent research \cite{Kolchmeyer:2024fly,Narovlansky:2025tpb,Cui:2026bcd,Harlow:2026pwe,Chen:2026boh,Milekhin:2026tbi}.
Recent work has also shown that including an observer can play an important role in understanding de Sitter entropy and the associated Hilbert space \cite{Maldacena:2024spf,Harlow:2025pvj,Blommaert:2025bgd,Tietto:2025oxn,Abdalla:2025gzn,deBoer:2025tmh}.

\paragraph{dS$_3$ partition functions.} The main proposal of this paper is that the gravitational path integral on a manifold $M_{\vec{\alpha}}$ with a hyperbolic-to-spherical transition can be obtained, to all orders in $G_N$, from a careful analytic continuation of two copies of a Virasoro TQFT amplitude. These two copies are naturally associated with the left- and right-moving sectors of the gravitational theory:
\begin{equation}
\label{eq:zgrav}
    Z_{\text{grav}}^{\Lambda>0} [M_{\vec{\alpha}}] =
    Z_{\text{Vir}}^{(b_+)}[M_{\vec{\alpha}}]\,
    Z_{\text{Vir}}^{(b_-)}[M_{\vec{\alpha}}],
\end{equation}
where each Virasoro TQFT amplitude $Z_{\text{Vir}}^{(b)}$ depends on a quantization parameter $b$, related to the central charge of the Virasoro algebra by $c=1+6(b+b^{-1})^2$. It also depends on the Liouville momenta $\vec{P}$, which will parametrize the cone angles $\vec{\alpha}$. 

In the hyperbolic regime, $Z_{\rm Vir}^{(b)}$ is well-defined and analytic in $b$. We can thus analytically continue the partition function to the spherical regime. In \eqref{eq:zgrav}, the two factors are analytic continuations of the same underlying Virasoro TQFT amplitude to two complex conjugate values of the quantization parameter $b$, together with complex conjugate identifications between the cone angles and the Virasoro TQFT momenta. To state the prescription, we introduce a positive real $\kappa$:
\begin{equation}
    \kappa = \frac{i}{b_+^2} = -\frac{i}{b_-^2} \in \mathbb{R}_{>0}.
\end{equation}
The left- and right-moving amplitudes are then defined by the following analytic continuations, with further details of the prescription discussed in section \ref{sec:Observer}:
\begin{equation}\label{eq:prescriptionintro}
Z_{\text{Vir}}^{(b_+)}[M_{\vec{\alpha}}] = \lim_{b\rightarrow b_+}
Z_{\text{Vir}}^{(b)}\left[P = \frac{i\alpha}{4\pi b}\right] \quad \text{and}\quad 
Z_{\text{Vir}}^{(b_-)}[M_{\vec{\alpha}}] = \lim_{b\rightarrow b_-}
Z_{\text{Vir}}^{(b)}\left[P =-\frac{i\alpha}{4\pi b}\right].
\end{equation}
Here, the limits are taken continuously along the trajectories
\begin{equation}
b_{\pm}
=
\lim_{\varphi\to\frac{\pi}{4}}
\frac{\e^{\pm i\varphi}}{\sqrt{\kappa}}\,.
\end{equation}
The continuous deformation is important in the semiclassical limit $\kappa\rightarrow\infty$. In this limit, Virasoro TQFT amplitudes may develop branch cuts, and the continuation has to be performed along a specified path in parameter space in order to reach the correct sheet. Finally, the relationship between the quantization parameters $b_{\pm}$ and Newton's constant $G_N$ is given by the Brown-Henneaux formula \cite{Brown:1986nw,Strominger:2001pn}
\begin{equation}
    \kappa  \sim \frac{\ell_{\text{dS}}}{4 G_N} \quad \text{as}\quad \kappa \rightarrow\infty.
\end{equation}

This analytic continuation gives complex conjugate Virasoro
central charges $c_\pm=13\mp6i(\kappa-\kappa^{-1})$, with
$c_++c_-=26$. The same central charge-relation appears in the
complex Liouville string (CLS), whose worldsheet theory consists
of two Liouville CFTs together with the reparametrization ghosts
\cite{Collier:2024kmo,Collier:2024kwt}. The conformal weights
associated with our continued momenta also satisfy
$h_++h_-=1$, which is the on-shell condition for physical CLS
vertex operators. CLS string amplitudes are obtained by
integrating worldsheet correlators over moduli and have a
proposed description in terms of a double-scaled two-matrix
integral \cite{Collier:2024lys}. In \cite{Collier:2025lux}, these amplitudes were related to
squared norms of a proposed Liouville cosmological wavefunction,
with the inner product defined by integration over the moduli
of future infinity. This state is motivated by expanding
cosmologies with compact hyperbolic spatial slices and a
Milne-type big bang singularity. The same work also discussed
a conjectural $S^3$ partition function based on analytically
continued Virasoro TQFT. Here we study fixed-topology Euclidean
partition functions on a broader class of spherical
cone manifolds, including nontrivial worldline graphs and knots.
Understanding the relation between these amplitudes and CLS
observables remains an interesting question.

\paragraph{Relation to observer worldline correlators.}
The theta graph and tetrahedron in \eqref{eq:intropictures} also connect our proposal to correlators of operators inserted on a massive observer's worldline. For example, the theta graph amplitude in Virasoro TQFT is inversely proportional to the DOZZ Liouville three-point function. Analytically continuing it following the prescription \eqref{eq:prescriptionintro}, we propose an exact expression for the observer two-point function.
We show that its heavy-heavy-light limit reproduces the static patch scalar correlator while retaining finite observer backreaction. In the weak-backreaction limit, this agrees with the two-point function of \cite{Chen:2026boh}.

Similarly, the tetrahedron amplitude is given in Virasoro TQFT by the Virasoro $6j$ symbol. Its analytic continuation through \eqref{eq:prescriptionintro} gives a proposal for the observer out-of-time-ordered four-point function (OTOC).
In a semiclassical heavy-light limit, its Lorentzian continuation reproduces the sign of the growing saddle contribution associated with the de Sitter time advance \cite{Aalsma:2020aib,Aalsma:2021kle,Kolchmeyer:2024fly,Narovlansky:2025tpb,Cui:2026bcd,Harlow:2026pwe,Chen:2026boh,Milekhin:2026tbi}. In the pure recoil limit, where gravitational backreaction vanishes but observer recoil is retained, the action reduces to the tetrahedral worldline action studied in \cite{Chen:2026boh}, connecting our calculation to earlier work on observer recoil. 

\paragraph{The spherical volume conjecture.} For 3D gravity with $\Lambda>0$ coupled to a heavy observer worldline, the Euclidean on-shell action is proportional to the volume of the spherical cone manifold, computed in the spherical metric. Thus it is natural to expect that our exact proposal \eqref{eq:zgrav}, as a product of complex Virasoro TQFT partition functions, should agree with the gravitational on-shell action in a saddle-point approximation, in the semiclassical limit $\kappa \rightarrow \infty$. This leads to the central conjecture of this paper, which we call the \emph{Spherical Volume Conjecture}:
\begin{equation}
\label{eq:volconj}
    \lim_{\kappa\to\infty} \frac{\pi \ell^3_{\text{dS}}}{\kappa}  \log\left| Z_{\text{Vir}}^{(b_+)}[M_{\vec{\alpha}}]\,
    Z_{\text{Vir}}^{(b_-)}[M_{\vec{\alpha}}]
    \right|\;=\; \text{Vol}_{\text{sph}}(M_{\vec{\alpha}}).
\end{equation}
In sections \ref{sec:Observer}-\ref{sec:Seifert}, we verify this conjecture in many examples.

This conjecture differs from the hyperbolic volume conjecture
for Virasoro TQFT \cite{Collier:2023fwi,Collier:2024mgv},
which concerns the semiclassical limit $b\to0$ with real $b$, and which for asymptotically AdS boundaries involves the renormalized volume.
(For earlier hyperbolic volume conjectures
for finite-volume hyperbolic knot complements, see
\cite{kashaev1997hyperbolic,murakami2010introduction}; for a figure-eight knot analysis in Virasoro TQFT, see \cite{Collier:2024mgv}.) One important difference is how the volume is recovered
in the hyperbolic and spherical cases. For a geometric hyperbolic saddle, the real part of the semiclassical exponent encodes the volume, while its imaginary part is related to the Chern--Simons invariant. In the spherical case, we instead find that the volume is determined by the combined semiclassical exponent of the two continued amplitudes. In the knot examples of section~\ref{sec:hyptosph}, this involves
the difference of the relevant saddle-point actions in the
two sectors.

The conjecture is particularly interesting for the knotted
cone manifolds studied in section~\ref{sec:hyptosph}, whose spherical volumes are nontrivial functions of the cone angles. The analytic continuation is performed on the exact Virasoro TQFT amplitudes, before taking the semiclassical limit. As in Witten's analytic continuation of Chern--Simons theory \cite{Witten:2010cx}, the integration contours must be transported along with the parameters. For the figure-eight and three-twist knots, we determine the contributing saddles through a Lefschetz-thimble analysis, keeping track of Stokes phenomena and the branches of the semiclassical action. Together with the saddle-point analysis of the Whitehead link, these calculations provide nontrivial checks of the conjecture.

An earlier spherical volume conjecture for colored Alexander
invariants was proposed and checked in examples using formal
saddle-point limits of their phases
\cite{Murakami:2008Alexander}. A spherical tetrahedron volume also appears
in the oscillatory asymptotics of $U_q(\mathfrak{sl}_2)$ $6j$ symbols
\cite{TaylorWoodward:2006} and in a formal semiclassical
analysis of the Turaev--Viro invariant
\cite{TaylorWoodward:2004Spherical}. Our spherical volume conjecture instead applies to the exponential growth of a product of two analytically
continued Virasoro TQFT amplitudes.

\paragraph{Seifert-fibered manifolds.} 

In the last part of this paper, we consider cone manifolds
whose singular loci have Seifert-fibered complements. We focus
on two basic examples: the Hopf link and the trefoil knot.
These complements do not admit complete finite-volume
hyperbolic metrics, but the corresponding cone manifolds admit
spherical structures for suitable cone angles. So for $\Lambda>0$, they are on-shell saddles. For the trefoil,
these structures and their volumes were studied in
\cite{KolpakovMednykhSpherical}. In both examples, we show that the product
of analytically continued Virasoro TQFT amplitudes in (\ref{eq:zgrav})
reproduces the spherical volume in the leading semiclassical
exponent.

We also study spherical Dehn fillings of these Seifert cone manifold.
For the Hopf link, the spherical fillings are lens spaces
\cite{Bojkovic:2023Hopf}. For generic lens spaces, we also
recover the known one-loop coefficient up to an overall
normalization \cite{Castro:2011xb}. For the trefoil, we study four integer fillings which give smooth spherical
manifolds with binary icosahedral, octahedral, tetrahedral and
dihedral fundamental groups
\cite{MoserTorusKnotSurgery,HikamiSphericalSeifert}. For all of
these examples, the leading semiclassical exponent of the
proposed partition function reproduces the spherical volume.

We use these amplitudes to revisit
the de Sitter Farey-tail sum of \cite{Castro:2011xb}.
For the sum over Lens spaces, we recover the residual divergence found there:
the sum remains divergent under the same zeta-function
regularization, even after including higher-loop corrections.
Our results for spherical Dehn fillings of the trefoil also
provide explicit amplitudes for non-cyclic spherical manifolds.
It would be interesting to incorporate their contributions into the sum over spherical topologies.

\subsection*{Outline}

This paper is organized as follows.\\[-10pt]

In section \ref{sec:Sphericalcone}, we review the classical geometry of cone manifolds and their hyperbolic-to-spherical transitions. We also discuss the Schl\"afli formula, an important result that relates the volume of a cone manifold to the geodesic length of its singular locus.\\[-10pt]

In section \ref{sec:Observer}, we present our proposal for the dS$_3$ partition function on these manifolds. Using this prescription, we study the observer partition function, the observer two-point function, and the observer four-point function in various limits. \\[-10pt]

Section \ref{sec:hyptosph} deals with knot and link complements in $S^3$ that undergo a hyperbolic-to-spherical transition. We study the examples of the figure-eight knot, the three-twist knot, and the Whitehead link. An interesting finding in this section is that the contour of integration which defines their exact partition functions must be treated carefully. A Lefschetz-thimble analysis shows that, as the contour is varied, different saddles contribute to the left- and right-moving partition functions. Interestingly, neither chiral factor satisfies the volume conjecture on its own; only their product does.\\[-10pt]

In section \ref{sec:Seifert}, we consider a broader class of spherical manifolds, including Seifert manifolds and Seifert cone manifolds. Although these manifolds do not undergo a hyperbolic-to-spherical transition, we find that they also satisfy our volume conjecture.\\[-10pt]

We conclude with a discussion and open questions in section \ref{sec:discussion}. In appendix \ref{app:specialfunctions}, we collect useful facts about the double sine functions and the quantum dilogarithm. In appendix \ref{app:saddlepoint6j}, we provide details of the saddle-point equation for the $6j$ symbol. Appendix \ref{sec:tetra-recoil} examines the pure-recoil limit of the spherical tetrahedron amplitude. Finally, appendix \ref{app:Trefoil} contains technical details about the evaluation of the trefoil Virasoro TQFT partition function.

\newpage
\section{Spherical cone manifolds} \label{sec:Sphericalcone}

A three-dimensional spherical cone manifold is a manifold $M$ equipped with a metric of constant positive sectional curvature
$\ell_{\rm dS}^{-2}>0$
away from a codimension-two singular locus $\Gamma$. In the cases of interest for this paper, $\Gamma$ is an embedded knot, link, or trivalent graph. Along a smooth component $\gamma_i\subset\Gamma$, the metric is locally modeled on that of conical dS$_3$ in Euclidean signature,
\begin{equation}\label{eq:conicaldSmetric}
    ds_E^2 = \ell_{\rm dS}^2\left(d\theta^2 + \cos^2(\theta) d\psi^2 +\left(\frac{\alpha_i}{2\pi}\right)^2 \sin^2(\theta)d\phi^2\right),
\end{equation}
with local coordinates $\theta \in [0,\frac{\pi}{2}]$ and 
$\psi, \phi \in [0,2\pi)$.
Away from $\gamma_i$ it is a round $S^3$ metric, but it has a conical singularity along $\theta = 0$ with deficit angle
\begin{equation}
    \delta_i=2\pi-\alpha_i.
\end{equation}
Globally, the geometry of $M$ is encoded by a developing map into $S^3$ and a holonomy representation 
$\pi_1(M\setminus \Gamma)\to \text{Isom}^+(S^3)$, for which a meridian around $\gamma_i$ has elliptic holonomy parametrized by $\alpha_i$. We use `cone manifold' in this metric sense, without requiring the angles to satisfy $\alpha = \frac{2\pi}{n}$, $n\in \mathbb{Z}_{>0}$; the latter restriction would give an orbifold $S^3/\mathbb{Z}_n$. General references on three-dimensional cone manifolds are
\cite{CooperHodgsonKerckhoff2000,BoileauMaillotPorti2003}.

\subsection{Hyperbolic-to-spherical transitions of knots and links}

A useful class of spherical cone manifolds consists of knot  complements in the three-sphere $S^3$.\footnote{One can also consider knots embedded in any other smooth spherical manifold, such as a Lens space $L(p,q)$, but for ease of exposition we restrict attention to $S^3$.} We will denote them by $(S^3, K, \alpha)$, with singular locus $K\subset S^3$ along the knot, and cone angle $\alpha$.
For a hyperbolic knot, the complete hyperbolic metric on the complement $S^3\setminus K$ is recovered in the limit
    $\alpha\to 0.$ Increasing $\alpha$ gives a family of hyperbolic cone metrics (also called deformed hyperbolic structures \cite{thurston1982geometry}) until, at a critical angle $\alpha=\alpha_*$, the geometry degenerates. After an appropriate rescaling, the limiting structure is a Euclidean cone manifold. Continuing past $\alpha>\alpha_*$, in some cases the metric may regenerate as a family of spherical cone manifolds \cite{Porti1998,Porti2004}. Schematically,
\begin{equation}
\begin{array}{ccl}
    0<\alpha<\alpha_\ast
        &:& \mathbb H^3\text{-cone structure},\\[1mm]
    \alpha=\alpha_\ast
        &:& \mathbb E^3\text{-cone structure after rescaling},\\[1mm]
    \alpha > \alpha_\ast
        &:& S^3\text{-cone structure}.
\end{array}
\end{equation}
The spherical cone structure continues to exist until some $\alpha_{\rm max}$, at which point the knot self-intersects. 
Not all knot complements feature such a geometric transition to a spherical regime, but in this paper we focus on those $(S^3,K,\alpha)$ that do. Many familiar examples of hyperbolic knots, such as the figure-eight knot or the three-twist knot, have a hyperbolic-to-spherical transition, as shown in section \ref{sec:hyptosph}. In fact, for any non-toric two-bridge knot  there is a critical angle $\alpha_\ast\in[2\pi/3,\pi)$, with hyperbolic cone structures for  $\alpha<\alpha_\ast$ and spherical cone structures for $\alpha_\ast <\alpha <2\pi - \alpha_\ast$ \cite{Porti2004}. 
This generalizes to link complements, with cone angles $\alpha_i$ associated to the meridians of the different link components, and a critical locus $\alpha_i^*$ above which a spherical cone metric exists. 

There is a related but distinct transition from hyperbolic to spherical cone metrics that can occur when the meridian holonomies are continued from hyperbolic to elliptic monodromy $e^{\pm l/2} \to e^{\pm i\alpha/2}$. The elementary example is the unknot in $S^3$: its complement (after removing a tubular neighborhood) is topologically a solid torus,
\begin{equation}
    S^3\setminus\bigcirc \cong D^2\times S^1.
\end{equation}
 The meridian $\mu$ encircling the unknot is a non-contractible cycle in $\pi_1(S^3\setminus\bigcirc) \cong \mathbb{Z}$. If this cycle carries  hyperbolic holonomy $\mathrm{diag}(e^{l/2},e^{-l/2})$ then the geometry is that of the Euclidean BTZ black hole, which is hyperbolic (with infinite volume), and $l$ is related to the horizon radius by $l = 2\pi r_+$. If, on the other hand, $\mu$ carries an elliptic holonomy $\mathrm{diag}(e^{i\alpha/2},e^{-i\alpha/2})$, it describes a conical defect with conical angle $\alpha\in (0,2\pi)$, and the same topology now admits a finite-volume spherical cone metric:
 \begin{equation}
     ds^2_E = f(r)d t_E^2+\frac{d r^2}{f(r)}+r^2 d\phi^2,
 \qquad
 f(r)=\frac{\alpha^2}{4\pi^2}-\frac{r^2}{\ell_{\rm dS}^2},
\end{equation}
with $\phi \in [0,2\pi)$. This is just the Euclidean continuation $t\to -it_E$ of the Schwarzschild-dS$_3$ metric, which describes 
the worldline of a heavy non-spinning point-mass in the static patch of dS$_3$ \cite{DeserJackiw1984}.\footnote{One could also easily include angular momentum: if we give different left and right holonomies to $\mu$, the geometry becomes rotating BTZ for hyperbolic holonomies $l_L, l_R$ \cite{BanadosTeitelboimZanelli1992} and Kerr-dS$_3$ for elliptic holonomies $\alpha_L,\alpha_R$ \cite{DeserJackiw1984}.} The angle $\alpha$ is related to the mass $m$ by
\begin{equation}\label{eq:massformula}
    \alpha= 2\pi (1-4G_Nm).
\end{equation}
Sometimes the alternative parametrization $\alpha = 2\pi \sqrt{1-8G_NM}$ is used; but the two differ only at $\order{G_N}$. The cosmological horizon radius is $2\pi r_H = \ell_{\rm dS}\alpha$. Note that the substitution $r = \frac{\alpha}{2\pi} \ell_{\rm dS} \sin\theta$ and $t_E = \frac{2\pi \ell_{\rm dS}}{\alpha}\psi$ brings the metric into the standard form \eqref{eq:conicaldSmetric}. These coordinates also show that the defect locus is a great circle of length $2\pi \ell_{\rm dS}$.

 This example shows that the same topology can support a hyperbolic metric when the meridian holonomies $\alpha_i$ take values in one domain $U_{\text{hyp}}\subset \mathbb{C}$ and a spherical cone metric in another domain $ U_{\text{sph}}\subset \mathbb{C}$. In the next subsection, we will generalize this observation from knot and link complements to the complements of trivalent graphs in the three-sphere, which will naturally connect to Virasoro TQFT.

\subsection{Spherical theta graph and spherical tetrahedron}\label{sec:sphericaltheta}
Another class of three-manifolds with hyperbolic-to-spherical transitions is provided by graph complements $S^3\setminus \Gamma$, where the singular locus $\Gamma$ is a closed trivalent graph. We denote the resulting cone manifold by
$(S^3, \Gamma, \vec{\alpha})$, where $\vec{\alpha}$ are the conical angles around the edges of $\Gamma$ (i.e.~holonomy variables).

The first example we will analyze is the complement of the theta graph in $S^3$:
\begin{equation}
    S^3\setminus \Theta_{\vec{\alpha}},\quad \text{where} \quad 
    \Theta_{\vec{\alpha}} = 
    \vcenter{
    \hbox{
    \begin{tikzpicture}
    \draw[thick] (0,0) node[left=7pt] {$\alpha_1$}
    node[right=-3pt] {$\alpha_2$}
    node[right=27pt] {$\alpha_3$}
    ellipse (1 and 0.6);
    \draw[thick] (0,-0.6) -- (0,0.6);    
    \end{tikzpicture}
    }
    }.
\end{equation}
Topologically, this is $\Sigma_{0,3}\times I$ where $\Sigma_{0,3}$ is $S^2$ with three cone points. If the sum of the opening angles satisfies $\alpha_1+\alpha_2+\alpha_3 < 2\pi$, this topology admits a hyperbolic cone metric of the Euclidean wormhole form
\begin{equation}
    ds_E^2 = \ell_{\rm AdS}^2\left(d\rho^2 + \cosh^2(\rho) d\Sigma_{0,3}^2\right)
\end{equation}
where $\rho\in(-\infty,\infty)$. Here $d\Sigma_{0,3}^2$ is the hyperbolic metric on the sphere with three cone points, which is unique up to isometry once $\alpha_{1,2,3}$ are given. This type of wormhole geometry is well-studied in the AdS$_3$ gravity literature \cite{Maldacena:2004rf,Belin:2020hea,Chandra:2022bqq,Collier:2023fwi,Collier:2024mgv}.\footnote{For real cone angles, removing a small neighborhood of each vertex gives the 3-punctured sphere wormhole; whereas analytic continuation to length variables $\ell_a = i\alpha_a$ describes the genus-two handlebody in the sunset channel \cite{Belin:2026pko}.}
If, on the other hand, the angles add up to $\alpha_1+\alpha_2+\alpha_3>2\pi$, it can admit a spherical cone metric
\begin{equation}\label{eq:thetagraphmetric}
    ds_E^2 = \ell^2_{\rm dS} \left(d\tau^2 + \cos^2(\tau) d\widetilde{\Sigma}_{0,3}^2\right)
\end{equation}
where $\tau\in (-\frac{\pi}{2},\frac{\pi}{2})$ and $d\widetilde{\Sigma}_{0,3}^2$ is a metric of \emph{positive} constant sectional curvature one 
on the sphere with three cone points of angles $\alpha_{1,2,3}$. The condition for a spherical metric to exist on $\Sigma_{0,3}$ is the so-called Eremenko condition
\cite{eremenko2004metrics},
\begin{equation}\label{eq:Eremenko_condition}
\cos\left(\mfrac{\alpha_1+\alpha_2+\alpha_3}{4}\right)\cos\left(\mfrac{\alpha_1+\alpha_2-\alpha_3}{4}\right)\cos\left(\mfrac{\alpha_1-\alpha_2+\alpha_3}{4}\right)\cos\left(\mfrac{\alpha_1-\alpha_2-\alpha_3}{4}\right)<0.
\end{equation}
This is equivalent to the spherical triangle inequalities for conical defects, but the condition also applies for conical excesses when none of the angles are integer multiples of $2\pi$.

This example shows that there is a transition in $S^3\setminus \Theta_{\vec{\alpha}}$ from a hyperbolic to a spherical regime when $\sum_a \alpha_a$ becomes larger than $2\pi$ and the Eremenko condition is satisfied. The spherical geometry of the theta graph complement $S^3$, in turn, describes the worldlines of 3 heavy particles running between the poles. Alternatively, in Lorentzian signature, it describes a massive observer worldline in the static patch of dS$_3$ with two antipodal operator insertions $\mathcal{O},\mathcal{O}^\dagger$ separated by geodesic distance $\pi \ell_{\rm dS}$. This interpretation will be made precise in section \ref{sec:observer2point}.

The second graph complement that we study is the case that $\Gamma$ is homeomorphic to a tetrahedron graph:\vspace{-2mm}
\begin{equation}
\label{eq:tetragraph}
    S^3\setminus \tetra_{\vec{\alpha}}\quad \text{where}\quad    \tetra_{\vec{\alpha}} \;\coloneqq 
\vcenter{
    \hbox{
\begin{tikzpicture}[scale=0.8]
    \draw[thick] (0,0) circle (1.5);
    \draw[thick] (0,0)-- node[above]{$\alpha_4$} ({1.5*cos(30)},{-1.5*sin(30)});
    \draw[thick] (0,0)-- node[above]{\!$\alpha_6$} ({-1.5*cos(30)},{-1.5*sin(30)});
    \draw[thick] (0,0) --node[left]{$\alpha_2$}  (0,1.5);
    \node[above] at (0,-1.5) {$\alpha_5$};
    \node[left] at ({-1.5*cos(30)},{1.5*sin(30)}) {$\alpha_1$};
    \node[right] at ({1.5*cos(30)},{1.5*sin(30)}) {$\alpha_3$};
\end{tikzpicture}
    }
    }
\end{equation}
The six edges of $\Gamma$ are labeled by conical angles $\alpha_i$. The spherical cone geometry can be constructed by filling in the interior of a spherical tetrahedron, with interior dihedral angles $\theta_i = \frac{\alpha_i}{2} \in (0,\pi)$, and \emph{doubling} it along its four faces. This creates a closed cone manifold whose singular locus is the tetrahedron graph $\Gamma$. 

The condition for the existence of a spherical metric on a non-degenerate tetrahedron can be stated in terms of its Gram matrix \cite{murakamiyano}. Let $i,j \in \{ 1,\dots,4\}$ label the four faces, and let $\theta_{ij}$ be the dihedral angle between face $i$ and $j$ ($i<j$).  Then the Gram matrix $G$ is defined as
\begin{equation}\label{eq:grammatrix}
G_{ij} =\begin{cases}
    1 & \quad  i=j \\
    -\cos(\theta_{ij}) & \quad i\neq j.
\end{cases}
\end{equation}
A non-degenerate spherical tetrahedron $T_{\vec{\alpha}}$ with these dihedral angles exists if and only if $G$ is positive-definite. This in particular implies that $\det G >0$. 

The tetrahedron also has a hyperbolic regime, with $\det G <0$, whose geometry was recently connected to the semiclassical asymptotics of the Virasoro $6j$ symbol \cite{Hartman:2025cyj,Hartman:2025ula}. There, the Liouville momenta $P_k$ defining the Virasoro primary weights parametrize the six edge lengths of the tetrahedron by $P_k = \frac{\ell_k}{4\pi b}$. Interestingly, in section \ref{sec:observerOTOC} we find that the spherical geometry emerges from the analytically continued Virasoro $6j$ symbol, in a semiclassical limit with $P_k=\frac{i\alpha_k}{4\pi b}$, where now $\theta_k = \frac{\alpha_k}{2}$ is a dihedral angle of the spherical tetrahedron instead of a length variable.

The Lorentzian de Sitter interpretation of the tetrahedron graph complement is an observer with two pairs of operator insertions along its worldline. In section \ref{sec:observerOTOC}, we will make this precise and connect to the recent OTOC calculations in de Sitter \cite{Aalsma:2020aib,Kolchmeyer:2024fly,Narovlansky:2025tpb,Cui:2026bcd,Harlow:2026pwe,Chen:2026boh,Milekhin:2026tbi}.

\subsection{Spherical volumes and Schl\"afli formula} \label{sec:Schlafli}

To evaluate the semiclassical gravitational path integral on a spherical cone manifold $M_\alpha$, we need to compute its Euclidean on-shell action $-\log Z \approx I_E[M_\alpha]$. The on-shell action is proportional to the volume of $M_\alpha$ in the spherical metric, as explained below. So let us discuss their spherical volumes.
In simple cases, the volume may be computed directly, if an explicit metric is available. For example, the unknot in $S^3$ has 
\begin{equation}\label{eq:unknotvolume}
    \text{Vol}(S^3\setminus\bigcirc_\alpha) =  \ell_{\rm dS}^3 \pi\alpha\, , 
\end{equation}
where we used the metric \eqref{eq:conicaldSmetric}. 
At $\alpha = 2\pi$, we recover the volume of the three-sphere, $2\pi^2\ell_{\rm dS}^3$. 
The theta graph complement is also simple, if we use the metric \eqref{eq:thetagraphmetric}:
\begin{equation}\label{eq:thetavolume}
     \text{Vol}\left(
S^3\setminus \Theta_{\vec{\alpha}}
\right) = \frac{\pi}{2} \ell_{\rm dS}^3\,\text{Area}(\Tilde{\Sigma}_{0,3}).
\end{equation}
The area of the sphere with 3 cone points and sectional curvature one 
is found by Gauss-Bonnet to be
\begin{equation}
    \text{Area}(\Sigma_{0,3})=\alpha_1+\alpha_2+\alpha_3 - 2\pi.
\end{equation}
Again note that in the smooth limit $\alpha_{1,2,3} = 2\pi$ we recover the volume of $S^3$. In section \ref{sec:observer2point}, we show how the volume \eqref{eq:thetavolume} arises from the semiclassical limit of the analytically continued DOZZ formula.

The spherical tetrahedron volume is more complicated. First, note that since the graph complement arises by doubling a solid spherical tetrahedron $T_{\vec{\alpha}}$ across its faces, the volume obeys
\begin{equation}\label{eq:volumetetra}
    \text{Vol}\left(
        S^3\setminus \tetra_{\vec{\alpha}}  
\right) = 2 \text{Vol}(T_{\vec{\alpha}}).
\end{equation}
Next, we use Murakami-Yano \cite{murakamiyano,murakami2012volume} to write down the volume of a spherical tetrahedron, which we will need in section \ref{sec:observerOTOC}. First, form the cofactor matrix $C_{ij}$ of the Gram matrix $G_{ij}$. This determines the edge lengths of $T_{\vec{\alpha}}$ through
\begin{equation}
    \cos\frac{L_{ij}}{\ell_{\rm dS}}
    =
    \frac{C_{ij}}{\sqrt{C_{ii}C_{jj}}}\,.
\end{equation}
Here $L_{ij}$ denotes the length of the edge joining the vertices $v_i,v_j$ that lie opposite to the faces $i$ and $j$, respectively. 
(For example, the edge associated with $\theta_{12}$ has length $L_{34}$). Then the spherical volume satisfies the so-called \emph{Schl\"afli formula} \cite{Milnor1994,rivin2000schlafli}
\begin{equation}\label{eq:tetravolume}
    d\text{Vol}(T_{\vec{\alpha}})
    =
    \frac{\ell_{\rm dS}^2}{2}
    \sum_{i<j}L_{kl}\,d\theta_{ij}, \qquad
\{k,l\}=\{1,2,3,4\}\setminus\{i,j\}..
\end{equation}
We integrate this differential to get the volume as a function of the dihedral angles $\theta_{ij}$. A closed form that we will use in section \ref{sec:Observer} was found by Murakami in \cite{murakami2012volume} (Theorem~1.1) in terms of the dilogarithm function $\mathrm{Li}_2(z)$:
\begin{equation}\label{eq:murakamivolume}
   \ell_{\rm dS}^{-3}\text{Vol}(T_{\vec{\alpha}}) = \mathrm{Re}\,U(z_0;e^{-i\theta_j})-\frac{1}{2} \sum_{j=1}^3 \theta_{j}\theta_{j+3}-\frac{\pi}{2}\sum_{j=1}^6\theta_j
 +\pi \text{Im}\log z_0
 -\frac{ \pi^2}{2}\,.
\end{equation}
Here we labeled the dihedral angles $\theta_i = \frac{\alpha_i}{2}$ on the six edges of $T_{\vec{\alpha}}$ using the convention in \eqref{eq:tetragraph}, and the function $U$ is given in terms of $x_j \equiv e^{-i\theta_j}$ by
\begin{multline}
 U(z;x_j)=\frac12\Big[
 \Li_2(z)
 +\Li_2(x_1x_2x_4x_5z)
 +\Li_2(x_1x_3x_4x_6z)
 +\Li_2(x_2x_3x_5x_6z)
\\
-\Li_2(-x_1x_2x_3z)
-\Li_2(-x_1x_5x_6z)
-\Li_2(-x_2x_4x_6z)
-\Li_2(-x_3x_4x_5z)
\Big].
\label{eq:U-action}
\end{multline}
In the above, $z_0$ is a solution of the equation
\begin{equation}
    \exp(2 z\frac{\partial}{\partial z} U(z,x_j)) = 1.
\end{equation}
This is a quadratic equation for $z$, and $z_0$ is the root given in \eqref{eq:root}. Notably, in our conventions $z_0$ is the root with $|z_0|>1$, whereas Murakami evaluates the volume function on the other root; this explains the relative minus sign between our expression \eqref{eq:murakamivolume} and that in \cite{murakami2012volume}.\footnote{To make the comparison between \eqref{eq:murakamivolume} and Thm 1.1 of \cite{murakami2012volume}, use the dictionary $x_i = e^{-i\theta_i} = a_i^{-1}$ and use the fact that their volume function is defined modulo $2\pi^2$, which reflects the choice of branch of the complex logarithm in $\pi \text{Im}(\log(z_0)) = \pi \text{arg}(z_0) + 2\pi^2 k$.}
We have checked explicitly that \eqref{eq:murakamivolume} indeed obeys the spherical tetrahedral Schl\"afli formula.

 Previous work found a relation between the spherical tetrahedron volume and the asymptotics of the 6$j$ symbol of $\mathcal{U}_q(\mathfrak{sl}_2)$ at $q$ a primitive root of unity \cite{murakamiyano, TaylorWoodward:2006}. In section \ref{sec:observerOTOC}, we will instead derive the volume formula \eqref{eq:murakamivolume} from the semiclassical limit of the analytically continued Virasoro $6j$ symbol. 

\subsubsection*{Schl\"afli formula for spherical cone manifolds}
For more complicated spherical cone manifolds, often the only efficient way of computing the volume is to use a Schl\"afli formula. For knot complements, for example, this amounts to computing the length of the singular geodesic along the knot and integrating the Schl\"afli formula to obtain the volume, see e.g.~\cite{Mednykh2021,Mednykh2003, tran2026exact}. 

We can motivate the Schl\"afli formula for spherical cone manifolds from 3D gravity with $\Lambda>0$, by a simple adaption of the Lewkowycz-Maldacena argument \cite{Lewkowycz:2013nqa}.
Consider the Euclidean gravitational action on a spherical cone manifold $M_\alpha$, with opening angle $\alpha$ around a codimension-two defect $\Gamma$. Away from the defect the geometry is locally spherical, hence
\begin{equation}
    R_{\rm reg}=6\Lambda=\frac{6}{\ell_{\rm dS}^2}\,.
\end{equation}
The regular part of the on-shell Einstein--Hilbert action is therefore
\begin{equation}
    I_{\rm EH}^{\rm reg}[M_\alpha]
=
-\frac{1}{16\pi G_N}
\int_{M_\alpha\setminus\Gamma} d^3x\,\sqrt g\,(R_{\rm reg}-2\Lambda)
=
-\frac{1}{4\pi G_N \ell_{\rm dS}^2}\text{Vol}(M_\alpha).
\end{equation}
Writing the deficit angle as $\delta=2\pi-\alpha$, the Ricci scalar also receives a distributional contribution of the defect:
\begin{equation}
    \int_{M_\alpha}d^3x\,\sqrt g\,R
=
\int_{M_\alpha\setminus\Gamma}d^3x\,\sqrt g\,R_{\rm reg}
+
2 L_\alpha \delta\,,
\end{equation}
where $L_\alpha$ is the length of the defect. Hence the conical contribution to the Einstein--Hilbert action is
\begin{equation}\label{eq:conicalaction}
    I_{\rm EH}^{\rm cone}
=
-\frac{2\pi-\alpha}{8\pi G_N}L_\alpha.
\end{equation}
Since the action given by the sum of the Einstein-Hilbert term and the matter worldline term is on-shell, its first variation in $\alpha$ vanishes,
\begin{equation}
  0 =  \frac{d}{d\alpha} I_{\rm EH}[M_\alpha]+m\frac{dL_\alpha}{d\alpha} = \frac{d}{d\alpha}\left(-\frac{1}{4\pi G_N \ell_{\rm dS}^2}\text{Vol}(M_\alpha) -\frac{2\pi-\alpha}{8\pi G_N}L_\alpha\right),
\end{equation}
which can be rearranged to give the Schl\"afli formula:
\begin{equation}
    \frac{d}{d\alpha} \text{Vol}(M_\alpha) = \frac{\ell_{\rm dS}^2}{2} L_\alpha.
\end{equation}
This argument is easily generalized to include multiple defects, in which case the Schl\"afli formula becomes
\begin{equation}
    d\text{Vol}(M_{\vec{\alpha}}) = \frac{\ell_{\rm dS}^2}{2}\sum_{i}L_{\alpha_i}d\alpha_i,
\end{equation}
where $L_{\alpha_i}$ is the length of the $i^{\rm th}$ singular geodesic with cone angle $\alpha_i$.

If we use the relation between the opening angle $\alpha$ and the mass
 \eqref{eq:massformula}, we see that the conical contribution $I^{\rm cone}_{\rm EH}$ is minus the standard Euclidean worldline action $mL_\alpha$ for a particle of mass $m$ sourcing the defect,
\begin{equation}
    I_{\rm WL}[\Gamma] =mL_\alpha
=
\frac{2\pi-\alpha}{8\pi G_N}L_\alpha.
\end{equation}
Thus the total on-shell Einstein-Hilbert action coupled to the worldline is simply
\begin{equation}
    I_{\rm EH+WL}[M_\alpha]
=
-\frac{1}{4\pi G_N\ell_{\rm dS}^2}\text{Vol}(M_\alpha).
\end{equation}
So the semiclassical Euclidean partition function on $M_\alpha$ is weighted by the exponential of \emph{plus} the spherical volume
\begin{equation}\label{eq:semiclassvolume}
     Z_{\rm grav}\sim e^{-I_{\rm EH + WL}} =\exp(\frac{1}{4\pi G_N\ell_{\rm dS}^2}\text{Vol}(M_\alpha)).
\end{equation}
This should be contrasted to the AdS$_3$ case, where the exponent is minus the hyperbolic volume (or renormalized volume for infinite-volume manifolds \cite{Schlenker:2022dyo}).
Equation \eqref{eq:semiclassvolume} is the on-shell action that we will match to in the next sections, where we analyze the semiclassical limit of complex Virasoro TQFT amplitudes evaluated on various spherical cone manifolds.

\section{Observer worldline correlators from complex Virasoro TQFT}\label{sec:Observer}

In this section we test our proposal \eqref{eq:zgrav} for the exact gravitational partition function on any spherical cone manifold with a hyperbolic-to-spherical transition. Recall from the introduction that our proposal is defined by the following analytic continuation of Virasoro TQFT partition functions:
\begin{equation}\label{eq:proposal}
    Z_{\rm grav}^{\Lambda>0}[M_{{\alpha}}] = Z_{\rm Vir}^{(b_+)}[M_{{\alpha}}]Z_{\rm Vir}^{(b_-)}[M_{{\alpha}}]
\end{equation}
Here $M_{{\alpha}}$ denotes the spherical cone manifold with cone angles $\alpha = (\alpha_1,\dots,\alpha_k)$, and
\begin{equation}\label{eq:zdef}
   Z_{\rm Vir}^{(b_+)}[M_{\vec{\alpha}}] \coloneqq \lim_{b\to b_{+}} Z^{(b)}_{\rm Vir}\!\left[ P =\tfrac{i\alpha}{4\pi b}\right],  \qquad Z_{\rm Vir}^{(b_-)}[M_{{\alpha}}] \coloneqq \lim_{b\to b_{-}} Z^{(b)}_{\rm Vir}\!\left[{P} = -\tfrac{i\alpha}{4\pi b}\right].
\end{equation}
The precise meaning of the notation $\lim_{b\to b_{\pm}}$ is the following analytic continuation:
\begin{equation}
    \lim_{b\to b_{\pm}} f(b) \coloneqq \lim_{\varphi\to \frac{\pi}{4}} f(|b|e^{\pm i\varphi})
\end{equation}
where $\varphi$ approaches $\frac{\pi}{4}$ from below. Here $b$ is the standard Liouville parameter of Virasoro TQFT, which is real $b\in(0,1)$ at the starting point $\varphi=0$. At the endpoints, $b_+^2= - b_-^2$ is purely imaginary. We introduce a real positive parameter $\kappa$ by
\begin{equation}
     \kappa = \frac{i}{b_+^2} = -\frac{i}{b_-^2} \in \mathbb{R}_{>0}.
\end{equation}
So $b_+^2$ lies on the positive imaginary axis and $b_-^2$ lies on the negative imaginary axis. The corresponding central charges are also complex conjugates $c_\pm = c_\mp^*$, with $c_\pm \in 13+i\mathbb{R}$. 

Semiclassically, 
\begin{equation}
        \kappa \sim \frac{\ell_{\rm dS}}{4 G_N}.
    \end{equation}
    This follows from the standard Brown-Henneaux relation $b^{-2} \sim \frac{\ell_{\rm AdS}}{4G_N}$ together with $\ell_{\rm AdS}\to i \ell_{\rm dS}$ \cite{Strominger:2001pn,Klemm:2001ea}. The one-loop renormalization of the central charge by $c= C+13$ is the same for dS$_3$ and AdS$_3$ \cite{Cotler:2019nbi}. 
    Although this is useful to keep in mind, we stress that our proposal is stated in terms of $\kappa$, which is \emph{exact} in the Liouville parameter $b$. The semiclassical relation to $G_N^{-1}$ is only valid asymptotically as $\kappa\gg 1$, which we will consistently indicate with $\sim$.

As a further comment, we note that our proposal eliminates the framing anomaly inherent to Virasoro TQFT. Virasoro TQFT has a framing anomaly \cite{Collier:2023fwi}, which is an ambiguity in braiding prefactors of the form $e^{2\pi in h}$, where $h=\frac{Q^2}{4}+P^2$ and $n\in \mathbb{Z}$. The full gravitational partition function should not depend on a choice of framing. The framing dependence must therefore cancel in the combination \eqref{eq:proposal}. This requires that\footnote{In the original relation of Virasoro TQFT to AdS$_3$ gravity \cite{Collier:2023fwi}, $e^{2\pi i h}$ is a pure phase, and the framing cancels if $h-\bar h\in \mathbb{Z}$. In dS$_3$, $h_\pm$ are complex conjugates and add up to 1: this is because the left- and right-moving partition functions are obtained by analytic continuation of the same Virasoro TQFT amplitude and, in particular, involve the same representations of the braiding kernels.}
\begin{equation}
    \exp\Big[2\pi i(h_+ + h_-)\Big] = 1,
    \qquad
    h_{\pm}
    =
    \frac{(b_{\pm}+b_{\pm}^{-1})^2}{4}
    +
    \left(
    \pm \frac{i\alpha}{4\pi b_{\pm}}
    \right)^2.
\end{equation}
Using the continuation \eqref{eq:zdef} above,  one finds explicitly that
\begin{equation}
    h_+ + h_- = 1,
\end{equation}
and hence the framing dependence of the two Virasoro TQFT amplitudes cancels between the left- and right-moving sectors in \eqref{eq:proposal}. 
Additionally, any projective ambiguity of the form $e^{\frac{i}{b^2}A}$, with $A$ independent of $P$, also drops out because $b_+^2 = - b_-^2$.

As stated in the introduction, our Spherical Volume Conjecture  is that the product of analytically continued Virasoro TQFT amplitudes \eqref{eq:proposal} exponentiates, in the  limit $\kappa\to\infty$, to the volume of $M_\alpha$ computed in the spherical cone metric:
\begin{equation}
 Z_{\rm Vir}^{(b_+)}[M_{{\alpha}}]Z_{\rm Vir}^{(b_-)}[M_{{\alpha}}]  \sim \exp(\frac{1}{4\pi G_N \ell_{\rm dS}^2}\text{Vol}(M_\alpha)).
\end{equation}
Since the volume is the on-shell action of $\Lambda>0$ gravity coupled to a worldline action, as explained in section \ref{sec:Schlafli}, this provides evidence that our proposal indeed describes dS$_3$ quantum gravity with observers. 

We stress that quantum gravity on a fixed topology (including all $G_N$ corrections) is not the same as a \emph{microscopic} proposal for a worldline quantum mechanics. This is analogous to the way Virasoro TQFT exactly quantizes AdS$_3$ gravity with defects topology-by-topology, but does not yet by itself define a microscopic CFT$_2$ dual \cite{Belin:2026pko}. Indeed, the microscopic interpretation of the gravitational worldline correlators that we compute in this section is still an open question, to which we will return in the discussion section.

In this section, we verify the spherical volume conjecture for the simplest class of Virasoro TQFT amplitudes: the unknot, the theta graph and the tetrahedron graph in $S^3$. For each of these computations, we give a gravitational interpretation as a worldline correlator for a heavy observer in the static patch, finding a match to known semiclassical results in dS$_3$ with observers.

\subsection{Observer partition function from the Cardy formula}\label{sec:observerpartitionfunction}

The simplest instance of our proposal is the unknot inside $S^3$, carrying Liouville momentum $P$. Its Virasoro TQFT partition function is given by the Cardy density:
\begin{equation}
   Z_{\rm Vir}^{(b)} \big[\bigcirc_P\big] = \rho_0^{(b)}(P)= 4\sqrt{2}\,
    \sinh(2\pi bP)\sinh(2\pi b^{-1}P).
\end{equation}
In AdS$_3$ quantum gravity, the right-hand side is the
exact density of primary states of the Euclidean BTZ black hole \cite{Maloney:2007ud}. It is also  the identity S-kernel in 2d CFT \cite{Collier:2018exn}, where it explains the Cardy growth of the density of states at high energies \cite{Collier:2019weq}.

In dS$_3$, the same topology can be understood as a heavy observer in the static patch, which backreacts to create a conical defect with opening angle $\alpha$. Following our prescription \eqref{eq:proposal},
we analytically continue the $\rho_0$ formula in $b$ and $P$, giving 
\begin{equation}  
\label{eq:observer-rho}
  Z^{\Lambda>0}_{\rm grav}[\bigcirc_\alpha] =
    \rho_0^{(b_+)}
   \! \left(\frac{i\alpha}{4\pi b_+}\right)
    \rho_0^{(b_-)}
    \!\left(-\frac{i\alpha}{4\pi b_-}\right)
    =
    32\sin^2\left(\frac{\alpha}{2}\right)\sinh^2\left(\frac{\kappa\alpha}{2}\right)\, .
\end{equation}
We can factorize the exact answer into a leading growing exponential, a one-loop piece, and a correction that is non-perturbatively small at large $\kappa$:
\begin{align}  \label{eq:observer-factorized}
   Z^{\Lambda>0}_{\rm grav}[\bigcirc_\alpha] 
    &=
    e^{\kappa\alpha}
    \left[8\sin^2(\tfrac{\alpha}{2})\right]
    \left(1-e^{-\kappa\alpha}\right)^2 \\[1em]
    &= \exp(\frac{\text{Vol}(\alpha)}{4\pi G_N \ell_{\rm dS}^2}) Z_{\rm 1-loop}(\alpha) \left[ 1 +\order{e^{-1/G_N}}\right].
\end{align}
In the second line we recognized the leading exponential as the spherical volume of the conical de Sitter geometry \eqref{eq:unknotvolume}, 
and used $\kappa\sim\ell_{\rm dS}/(4G_N)$. This confirms our volume conjecture in this simple example. 

Using the fact that the backreacted cosmological horizon area is $A =2\pi r_H = \ell_{\rm dS} \alpha$, we can also interpret the leading exponent as the horizon entropy in conical de Sitter,
\begin{equation}
   S_{\rm hor} = \kappa\hspace{0.2mm} \alpha = \frac{A}{4 G_N}.
\end{equation}
This agrees with earlier observations relating the Cardy formula to the de Sitter entropy in the dS/CFT literature \cite{Bousso:2001mw,Balasubramanian:2001nb}. However, here our interpretation is different: following recent ideas about observer holography \cite{Susskind:2021omt,Maldacena:2024spf,Harlow:2025pvj,Chen:2026boh, Blommaert:2026ofx}, we think of \eqref{eq:observer-factorized} as computing the state counting partition function for a static patch 
observer
$
    Z_{\rm obs}(m)
$. This is motivated by the fact that the Hamiltonian constraint in a closed universe relates the observer energy
to that of the remaining degrees of freedom. For a static observer, the proper mass is related to the conical angle $\alpha$ to leading order in $G_N$ by (recall \eqref{eq:massformula}):
\begin{equation}
    m = \frac{2\pi-\alpha}{8\pi G_N}, \qquad
    \nu\equiv m\ell_{\rm dS}
    =\kappa\left(1-\frac{\alpha}{2\pi}\right).
    \label{eq:observer-mass}
\end{equation}
 Indeed, substituting this relation into 
the on-shell action gives $
    S_{\rm hor} = S_{\rm dS} - 2\pi \nu,
$
where $S_{\rm dS} = \frac{\ell_{\rm dS}\pi}{2G_N}$ is the pure de Sitter entropy and $2\pi \nu$ is the worldline action $S_{\rm WL} =m L$, with $L = 2\pi \ell_{\rm dS}$ the geodesic length of the observer worldline on a great circle of $S^3$. With $\alpha\in (0,2\pi)$, the mass takes a finite range $\nu/\kappa \in (0,1)$.

We further note that our proposal \eqref{eq:observer-rho} for the observer partition function is manifestly real and positive, agreeing with the positive sign for the observer 1-loop determinant argued for in \cite{Maldacena:2024spf, Chen:2025jqm}. Moreover, in the heavy probe regime $1\ll \nu \ll \kappa$, with $\kappa\to \infty$ and $\nu$ fixed, our answer \eqref{eq:observer-factorized} reduces to
\begin{equation} \label{eq:observer-probe}
    Z_{\rm obs}(\nu)
    \sim
    \frac{8\pi^2}{\kappa^2}\,
    \nu^2 e^{S_{\rm dS}-2\pi\nu}\, .
\end{equation}
Up to an overall normalization, the $\nu^2e^{-2\pi\nu}$ dependence agrees
with the one-loop result for a heavy observer in three-dimensional de
Sitter space found in \cite{Maldacena:2024spf}.

Lastly, we remark that the exact proposal $Z_{\rm obs} =\rho_0^{(b_-)}\rho_0^{(b_+)}$ is closely related to the observer quantization of 3D Kerr-de Sitter of 
\cite{Blommaert:2026ofx}. That paper additionally includes spin, and incorporates a sum over $SL(2,\mathbb{Z})$ images. We have instead restricted to a single topology, but the same prescription works for any of the modular images computed using Virasoro TQFT. We regard our proposal as a generalization of \cite{Blommaert:2026ofx} to more general spherical cone manifolds, to which we turn next.

\subsection{Observer two-point function from the DOZZ formula}\label{sec:observer2point}

The second example that we study is the unknotted theta graph complement in $S^3$. Its Virasoro TQFT partition function is the inverse of the $C_0$ function \cite{Collier:2023fwi}:\footnote{Throughout, we use the normalization convention of \cite{Collier:2023fwi} for Virasoro TQFT partition functions.}
\begin{equation}
    Z_{\rm Vir}^{(b)}\left[
        \vcenter{
    \hbox{
    \begin{tikzpicture}
    \draw[thick] (0,0) node[left=7pt] {$P_1$}
    node[right=-3pt] {$P_2$}
    node[right=27pt] {$P_3$}
    ellipse (1 and 0.6);
    \draw[thick] (0,-0.6) -- (0,0.6);    
    \end{tikzpicture}
    }
    }
\!\!\right] = \frac{1}{C^{(b)}_0(P_1,P_2,P_3)}
\end{equation}
where $C_0$ is closely related to the Liouville 3-point DOZZ formula \cite{Dorn:1994xn,Zamolodchikov:1995aa,Teschner:1995yf,Zamolodchikov:2005fy,Harlow:2011ny}. An explicit expression in terms of Barnes double gamma functions $\Gamma_b$ is given for $b^2>0$ by
\begin{equation}\label{eq:C0function}
     C^{(b)}_0(P_1,P_2,P_3)=\frac{\Gamma_b(2Q)\Gamma_b\left(\frac{Q}{2}\pm iP_1 \pm iP_2 \pm iP_3\right)}{\sqrt{2}\Gamma_b(Q)^3\prod_{a=1}^3 \Gamma_b(Q\pm 2iP_a)}\,.
\end{equation}
Here the $\pm$ notation means to take the product over all choices of sign, and $Q = b+b^{-1}$.
The meromorphic function $\Gamma_b(z)$ has no zeroes and has isolated poles at
\begin{equation}
    z=-mb-nb^{-1}, \qquad m,n \in \mathbb{Z}_{\geq 0}\,.
\end{equation}
In AdS$_3$, this amplitude (times its conjugate) is the microcanonical partition function for a genus-two handlebody in the sunset channel, or, equivalently, a two-boundary wormhole with two three-punctured sphere boundaries \cite{Belin:2020hea,Chandra:2022bqq,Collier:2024mgv}. In dS$_3$, we propose that it computes the partition function on the spherical theta graph complement discussed in subsection \ref{sec:sphericaltheta}. More precisely, applying our general prescription \eqref{eq:proposal}, our 
 proposal for the exact $\Lambda>0$ gravitational partition function on the theta graph complement is
\begin{equation}\label{eq:C0proposal}
    Z_{\rm grav}^{\Lambda>0}\left[\Theta_{\vec{\alpha}}\right]=\frac{1}{C^{(b_+)}_0\!\left(\frac{i\alpha_1}{4\pi b_+},\frac{i\alpha_2}{4\pi b_+},\frac{i\alpha_3}{4\pi b_+}\right)C^{(b_-)}_0\!\left(-\frac{i\alpha_1}{4\pi b_-},-\frac{i\alpha_2}{4\pi b_-},-\frac{i\alpha_3}{4\pi b_-}\right)}\,.
\end{equation}
This proposal requires an analytic continuation of $C_0$ which we will describe below in detail. First note this proposal is real and positive, because the two 
$C_0$ factors are complex conjugates of each other (this follows from $\Gamma_{\bar b}(\bar z) = \overline{\Gamma_b(z)}$ and $b_+ = \overline{b_-}$):
\begin{equation}
    C^{(b_+)}_0\!\left(\frac{i\alpha_1}{4\pi b_+},\frac{i\alpha_2}{4\pi b_+},\frac{i\alpha_3}{4\pi b_+}\right)= \overline{C^{(b_-)}_0\!\left(-\frac{i\alpha_1}{4\pi b_-},-\frac{i\alpha_2}{4\pi b_-},-\frac{i\alpha_3}{4\pi b_-}\right)}\,.
\end{equation}
Second, note that the momenta $P_{k,\pm}$ give conformal weights in the principal continuous series representation of SL$(2,\mathbb{C})$, $h_\pm\in \frac{1}{2}+i\mathbb{R}$. Lastly, from the definition it is clear that $C_0$ is invariant under $P_k\to-P_k$, for each $k\in\{1,2,3\}$.

As before we define the analytic continuation as the endpoint of the following paths:
\begin{equation}
    b_\eta(\varphi)= |b| e^{i\eta \varphi}, \quad P_{k,\eta}(\varphi)=\eta \frac{i\alpha_k}{4\pi b_\eta(\varphi)}, \quad 0\leq \varphi \leq \frac{\pi}{4}\,,
\end{equation}
where $\eta=\pm 1$ labels the two different paths. The two paths do not encounter any poles of $C_0$ so the analytic continuation is well-defined. The reason to rotate $P_k$ and $b$ continuously in this way is that in the semiclassical limit, the $C_0$ asymptotics contain multi-valued functions (see below). The above prescription automatically selects the correct branch of these functions.

We will now analyze the semiclassical limit of $Z^{\Lambda>0}_{\rm grav}[\Theta_{\vec{\alpha}}]$ and verify the spherical volume conjecture in this example. Let us focus on $b_+^2 = \frac{i}{\kappa}$, with the semiclassical limit being $\kappa \to \infty$. 
We can show that 
\begin{align}\label{eq:C0_spherical_asymptotics}
&\lim_{\kappa\to \infty}\text{Re}\left[\frac{2\pi}{\kappa}\log C_0\right] =\nonumber\\ &\hspace{1.5cm} 2\pi \text{Im}\left[\sum_{i=1}^3\psi^{(-2)}
\left(1-\frac{\alpha_i}{2\pi}\right)-\sum_{\pm_1,\pm_2,\pm_3} \psi^{(-2)}\!\left(\frac{2\pi\pm_1 \alpha_1\pm_2 \alpha_2\pm_3 \alpha_3}{4\pi}\right)\right],
\end{align}
where the function $\psi^{(-2)}(y)$ is a second antiderivative of the digamma function:
\begin{equation}\label{eq:G_function}
\psi^{(-2)}(y)=\int_{\frac{1}{2}}^y dz \log \Gamma(z).
\end{equation}
This follows from the asymptotics of the double gamma function \cite{Collier:2018exn}:
\begin{equation}\label{eq:Barnes_double_Gamma_asymptotics}
b^2 \log \Gamma_b(b^{-1}y)=\tfrac{1}{2}(y-\tfrac{1}{2})\log (2\pi)+\tfrac{1}{2}(y-\tfrac{1}{2})^2 \log b-\psi^{(-2)}(y)+\order{1}
\end{equation}
as $|b|\to 0$.
For positive real argument $y>0$, $\psi^{(-2)}(y)$ is real. For the present purpose, we also need to evaluate $\psi^{(-2)}(y)$ when $y<0$. For this, we analytically continue $\Gamma(z)$ to $z<0$ via the upper half plane and observe that
\begin{equation}\label{eq:Gamma_negative_axis_phase}
\text{Im}\log \Gamma (z+i0)=-\pi (n+1), \qquad -(n+1)<z<-n,
\end{equation}
for a non-negative integer $n$. To derive this, we used
\begin{equation}\label{eq:Gamma_recursion_negative}
\Gamma(z)=\frac{\Gamma(z+n+1)}{z(z+1)\dots (z+n)}.
\end{equation}
In the range of interest, $\Gamma(z+n+1)>0$, so the phases come entirely from the terms in the denominator which are negative. This immediately gives an expression for 
\begin{equation}\label{eq:Im_G_negative}
\text{Im}\,\psi^{(-2)}(y+i0)=-\pi(\lfloor -y \rfloor +1)\left(y+\frac{\lfloor -y \rfloor}{2}\right), \qquad y<0,
\end{equation}
where $\lfloor . \rfloor$ is the floor function.

Using this result for $\Im\,\psi^{(-2)}$, we can show that whenever the Eremenko condition \eqref{eq:Eremenko_condition} for the existence of a spherical saddle is satisfied, for the $b_+$ sector, we have
\begin{equation}\label{eq:C0_volume_result}
- \lim_{\kappa\to \infty}\text{Re}\left[\frac{2\pi }{\kappa}\log C_0\right] = \frac{\pi}{2}(\alpha_1+\alpha_2+\alpha_3-2\pi).
\end{equation}
This is easiest to see for the  case of equal angles with $0<\alpha<2\pi$. Plugging in the result for $\text{Im}\,\psi^{(-2)}(y)$, we see that of all the terms, only the term $\psi^{(-2)}\left(\frac{1}{2}-\frac{3\alpha}{4\pi}\right)$ could have a negative argument, i.e. when $\alpha>\frac{2\pi}{3}$, and hence
\begin{equation}\label{eq:C0_equal_angles}
- \lim_{\kappa\to \infty}\text{Re}\left[\frac{2\pi }{\kappa}\log C_0\right]  =
\begin{cases}
\frac{\pi}{2}(3\alpha-2\pi), \qquad &\frac{2\pi}{3}<\alpha<2\pi,\\
0, \qquad &0<\alpha<\frac{2\pi}{3}.
\end{cases}
\end{equation}
Thus the classical action becomes non-zero precisely when the spherical metric exists.

Next we derive the result in the unequal angles case with $0<\alpha_i<2\pi$ using the Eremenko condition \eqref{eq:Eremenko_condition}. First, we establish some notation. Define the 4 combinations that show up in the Eremenko condition and also in the asymptotics of $C_0$,
\begin{equation}\label{eq:C0_s_variables}
\begin{split}
&s_1 \equiv \frac{\alpha_1+\alpha_2+\alpha_3}{2\pi}, \\
&s_2 \equiv \frac{\alpha_1+\alpha_2-\alpha_3}{2\pi}, \\
&s_3 \equiv \frac{\alpha_1-\alpha_2+\alpha_3}{2\pi}, \\
&s_4 \equiv \frac{\alpha_1-\alpha_2-\alpha_3}{2\pi}.
\end{split}
\end{equation}
Defining the integers $k_a=\lfloor \frac{s_a+1}{2} \rfloor$, we see that the Eremenko condition turns into a parity condition which is easier to work with,
\begin{equation}\label{eq:Eremenko_parity_condition}
\prod_{a=1}^4 \cos\left(\frac{\pi s_a}{2}\right)<0 \quad \text{implies that}\quad \sum_{a=1}^4 k_a \in 2\mathbb{Z}+1.
\end{equation}
In terms of the new parameters, the $C_0$ asymptotics can be expressed as
\begin{equation}\label{eq:C0_phase_k_variables}
- \lim_{\kappa \to \infty}\text{Re}\left[\frac{2\pi }{\kappa}\log C_0\right] = \pi^2 \sum_{a=1}^4 (k_a s_a-k_a^2).
\end{equation}
The Eremenko condition implies the two identities
\begin{equation}\label{eq:C0_k_identities}
2\pi\sum_{a=1}^4 k_a s_a = \alpha_1+\alpha_2+\alpha_3,
\qquad\text{and}\qquad \sum_{a=1}^4 k_a^2 =1 \,,
\end{equation}
which immediately give \eqref{eq:C0_volume_result}. This completes the derivation of the semiclassics of the spherical $C_0$ formula. Substituting these semiclassical $C_0$ asymptotics into \eqref{eq:C0proposal}, we have thus shown that
\begin{equation}
    Z_{\rm grav}^{\Lambda>0}\left[\Theta_{\vec{\alpha}}\right]\sim 
    \exp[
    \frac{\text{Vol}(S^3\setminus\Theta_{\vec{\alpha}})}{4\pi G_N\ell_{\rm dS}^2}
    ],\quad 
    \text{Vol}(S^3\setminus\Theta_{\vec{\alpha}}) = 
     \frac{\pi}{2}\ell_{\text{dS}}^3(\alpha_1+\alpha_2+\alpha_3-2\pi).
\end{equation}
Satisfactorily, the right-hand side precisely matches with $\big(4\pi G_N\ell_{\rm dS}^2\big)^{-1}$
      times the volume of the spherical theta graph complement, computed in \eqref{eq:thetavolume}, thus confirming the spherical volume conjecture in this example. This match provides further evidence for our conjecture that \eqref{eq:C0proposal} is a non-perturbative completion of $\Lambda>0$ quantum gravity on a fixed spherical topology with conical defects. 

      \subsubsection*{Observer two-point function and its heavy-light limit}
      As in the previous subsection, we can give the above computation an interpretation in terms of an observer in the static patch of dS$_3$.  
If we identify $\alpha_3\equiv\alpha_W$ as parametrising the dimension $\Delta_W \in 1+i\mathbb{R}$ of an operator $W$
inserted on the observer worldline, we must distinguish the
theta graph partition function from the observer two-point
function, which also includes observer spectral weights and the
normalization of the physical insertions. We propose that the correct normalization is 
\begin{equation}\label{eq:C0_observer_trace}
Z_{\rm obs}(\alpha_1,\alpha_2,\alpha_W)
=\mathcal{N}_{12;W}\,
Z_{\rm obs}(\alpha_1)Z_{\rm obs}(\alpha_2)
\left|C_0(\alpha_1,\alpha_2,\alpha_W)\right|^2 
\end{equation}
Here $Z_{\rm obs}(\alpha)$ is the observer
partition function defined in \eqref{eq:observer-rho}, while
$\mathcal{N}_{12;W}$ is a positive normalization symmetric under
$\alpha_1\leftrightarrow\alpha_2$. The proposal therefore retains
positivity and the observer-label symmetry of $C_0$. There is no
factor of $Z_{\rm obs}(\alpha_W)$, since $W$ is a light operator insertion rather
than an intermediate heavy state.

To check how this proposal relates to the known semiclassical observer two-point function in dS$_3$ \cite{Chen:2026boh}, consider a heavy-heavy-light limit in which
the observer remains fully backreacting while $W$ creates a probe
excitation of finite energy. More precisely, take
$\kappa=\ell_{\rm dS}/(4G_N)\to\infty$ while keeping
$\alpha\in(0,2\pi)$, $\omega$, and $\nu$ fixed, with
\begin{equation}
\alpha_{1,2}=\alpha\mp\frac{\pi\omega}{\kappa},
\qquad
\alpha_W=2\pi\left(1-\frac{\nu}{\kappa}\right).
\end{equation}
Using the relation between $\alpha$ and mass \eqref{eq:observer-mass},  $\omega=(m_1-m_2)\ell_{\rm dS}$ is the energy transferred by
$W$, while $\nu$ is the principal-series parameter of the probe
scalar, whose physical mass obeys
$M_W^2\ell_{\rm dS}^2=1+\nu^2$.
In de Sitter units, matching to the canonically normalized scalar
on the conical background gives the normalization prescription
\begin{equation}
\mathcal{N}_{12;W} =  \frac{2\pi^2}{\kappa\, (\alpha_1\alpha_2)^{3/2}}
e^{-\pi\nu}\left|\Gamma(1+i\nu)\right|^2 + \order{\kappa^{-2}},\quad \text{as}\quad \kappa\rightarrow\infty.
\end{equation}
The factor $(\alpha_1\alpha_2)^{-3/2}$ normalizes the observer legs;
the remaining factor fixes the scalar normalization. This matching
fixes the leading probe-limit behavior.

After dividing by the observer partition function, the
heavy-heavy-light limit of the continued $C_0$ formula \eqref{eq:C0function} gives
the microcanonical two-point function
\begin{equation}\label{eq:C0_microcanonical_kernel}
\begin{aligned}
K_{\alpha,\nu}(\omega)
&\equiv \lim_{\kappa\to\infty}
\frac{Z_{\rm obs}(\alpha_1,\alpha_2,\alpha_W)}{Z_{\rm obs}(\alpha)} \\
&=\frac{1}{8\pi^2\alpha}
\prod_{\epsilon_1,\epsilon_2=\pm1}
\Gamma\!\left(
\frac12+\frac{i}{2}(\epsilon_1\nu+\epsilon_2\omega)
\right) \\
&=\frac{1}{4\alpha}\,
\frac{1}{\cosh(\pi\omega)+\cosh(\pi\nu)}.
\end{aligned}
\end{equation}
In the weak-backreaction, heavy-observer regime, this reproduces
the dS$_3$ two-point kernel of \cite{Chen:2026boh}.
At fixed conical backreaction, the dependence on the energy
transfer is unchanged, while the canonical normalization supplies
the overall factor $1/\alpha$.

The Lorentzian correlator is obtained by Fourier transforming with respect to the energy difference \(\omega\). Fixing the average observer energy while Fourier transforming the difference is precisely the microcanonical-to-time representation considered for static-patch observer correlators in \cite{Chen:2026boh}; geometrically it amounts to analytically continuing the antipodal Euclidean separation according to \(\theta=\pi+i t/\ell_{\rm dS}\). Here, $t$ is the observer's proper time in terms of which the Lorentzian static patch metric is given by
\begin{equation}\label{eq:metric2}
    ds_L^2=-\left(1-\frac{R^2}{\ell^2_{\rm dS}}\right) dt^2 + \frac{dR^2}{1-\frac{R^2}{\ell^2_{\rm dS}}}+\frac{\alpha^2}{4\pi^2} R^2 d\phi^2\,.
\end{equation}
After taking the Fourier transform of \eqref{eq:C0_microcanonical_kernel}, the resulting two-sided correlator is
\begin{equation}
G^{LR}_{\alpha,\nu}(t)
=
\frac{1}{2\alpha\ell_{\rm dS}}
\frac{\sin(\nu t/\ell_{\rm dS})}
{\sinh(\pi\nu)\sinh(t/\ell_{\rm dS})},
\end{equation}
which is \(2\pi/\alpha\) times the standard static-patch Green's function \(G_\nu(\pi+i t/\ell_{\rm dS})\) appearing in the observer-correlator literature \cite{Chen:2026boh, Anninos:2011af}.\footnote{The factor of $2\pi/\alpha$ arises from the normalization of the 2-point function of a probe scalar propagating in conical dS$_3$. An easy way to see it is to write the action of a probe scalar propagating in the metric \eqref{eq:metric2}; then the $\sqrt{-g}$ gives a factor of $\alpha$. Comparing with the 2-point function of a canonically normalized scalar field explains the factor of $2\pi/\alpha$.} The poles of the energy-space kernel (\ref{eq:C0_microcanonical_kernel}) in the lower half-plane occur at
\begin{equation}
\omega_{n,\pm}=\pm\nu-i(2n+1),
\qquad n=0,1,\ldots,
\end{equation}
and hence the late-time correlator exhibits exponentially damped oscillations,
\begin{equation}
G^{LR}_{\alpha,\nu}(t)
\sim
\frac{e^{-t/\ell_{\rm dS}}}
{\alpha\ell_{\rm dS}\sinh(\pi\nu)}
\sin\!\left(\frac{\nu t}{\ell_{\rm dS}}\right).
\end{equation}
Thus, when time is measured in the observer's proper time, the finite backreaction changes the residues but not the quasinormal frequencies or the decay scale. For a heavy probe, \(\nu\gg1\), the two oscillatory contributions admit the usual interpretation in terms of the pair of complex geodesics \(L_\pm=\pi\ell_{\rm dS}\pm it\). The exponential late-time decay is the familiar behavior of de Sitter worldline correlators, while here it follows directly from the HHL limit of the analytically continued Virasoro \(C_0\) amplitude.

\subsection{Observer four-point OTOC from the $6j$ symbol}\label{sec:observerOTOC}

As a third application of our general proposal \eqref{eq:proposal}, we study the partition function on the tetrahedron graph complement \eqref{eq:tetragraph}, continued to the spherical regime. As in the previous subsections, the starting point is the exact partition function in Virasoro TQFT, which is proportional to the Virasoro $6j$ symbol \cite{Collier:2023fwi, Collier:2024mgv}: 
\begin{equation}\label{eq:6jVir}
Z^{(b)}_{\rm Vir}\!\left(
\vcenter{
    \hbox{
\begin{tikzpicture}[scale=0.8]
    \draw[thick] (0,0) circle (1.5);
    \draw[thick] (0,0)-- node[above]{$P_4$} ({1.5*cos(30)},{-1.5*sin(30)});
    \draw[thick] (0,0)-- node[below]{$P_6$} ({-1.5*cos(30)},{-1.5*sin(30)});
    \draw[thick] (0,0) --node[left]{$P_2$}  (0,1.5);
    \node[above right] at (0,-1.5) {$P_5$};
    \node[right] at ({-1.5*cos(15)},{1.5*sin(15)}) {\!$P_1$};
    \node[left] at ({1.5*cos(30)},{1.5*sin(30)}) {$P_3$};
\end{tikzpicture}
    }
    }
\right) = \frac{1}{\sqrt{\mathsf{C}_{123}\mathsf{C}_{156}\mathsf{C}_{246}\mathsf{C}_{345}}}\,\left\{\begin{matrix}
        P_1 & P_2 & P_3 \\ P_4 & P_5 & P_6
    \end{matrix}\right\}_b\,.
\end{equation}
Here we used the shorthand $\mathsf{C}_{ijk} = C_0(P_i,P_j,P_k)$, with the $C_0$ function given in \eqref{eq:C0function} and the Liouville momenta $P_j$ parametrizing the conformal weights as $h_j = \frac{Q^2}{4}+P_j^2$ with $Q = b+b^{-1}$. In AdS$_3$ quantum gravity, this is the exact gravitational partition function on the tetrahedron graph complement if we multiply by the right-moving partition function $Z_{\overline{\text{Vir}}}(\bar{P}_j)$. 

In the hyperbolic regime, it was shown in \cite{Teschner:2012em, Hartman:2025ula} that the semiclassical limit $b\to 0$ of the Virasoro $6j$ symbol, with $b$ real, matches the on-shell action of pure gravity on the corresponding hyperbolic geometry. For pure imaginary momenta with $4\pi b P_j = i\alpha_j$ held fixed as $b\to 0$, the geometry is a four-boundary wormhole with three-punctured sphere boundaries \cite{Collier:2024mgv}. For  
real $P$ with $4\pi b P_j = l_j$ fixed, the geometry is the convex core of a genus-3 handlebody, and its volume is twice the volume of a hyperbolic truncated tetrahedron \cite{Hartman:2025ula}. In a mixed regime where four momenta are heavy and two light, the $6j$ symbol was related to the out-of-time-ordered correlator (OTOC) in \cite{Jackson:2014nla}, whose classical geometry is a shockwave in an AdS$_3$ black hole background. A further scaling limit to AdS$_2$ gravity and the Schwarzian OTOC was analyzed in \cite{Mertens:2017mtv}.

The spherical regime has not been analyzed so far. As in the previous subsections, we propose the analytic continuation to the spherical regime through \eqref{eq:proposal}. That is, we take two copies of \eqref{eq:6jVir} with $b$ continued to complex $b_+,b_-$ with
\begin{equation}\label{eq:6jcontinuation}
   b_\pm^2 = \pm \frac{i}{\kappa}, \qquad P_{j,\pm}= \pm\frac{i\alpha_j}{4\pi b_\pm}, \qquad \alpha_j\in (0,2\pi), \qquad \kappa \in \mathbb{R}_{>0}.
\end{equation}
This is well-defined because the $6j$ symbol (which is proportional to the Virasoro fusion kernel) is an analytic function of the central charge $c$ away from $(-\infty,1]$ \cite{Eberhardt:2023mrq}.\footnote{A related analytic continuation of the Virasoro $6j$ symbol to complex $b$ was studied in \cite{Meng:2026xlo}, but in that paper the angles $\alpha_j$ are in the hyperbolic regime.}\footnote{Closed form expressions for the fusion kernel at irrational and rational central charges with $c<1$ were given in \cite{Roussillon:2024wmr,Roussillon:2025tmv}.} 
Moreover, the $6j$ symbol at $(b_+,P_{j,+})$ is the complex conjugate of the $6j$ symbol at $(b_-,P_{j,-})$, so the total gravity partition function is the absolute value squared: 
\begin{equation}
    Z_{\rm grav}^{\Lambda>0}\big[S^3 \setminus
    \tetra_{\vec{\alpha}}
\big] = \left|\frac{1}{\sqrt{\mathsf{C}^{+}_{123}\mathsf{C}^{+}_{156}\mathsf{C}^{+}_{246}\mathsf{C}^{+}_{345}}}\,\left\{\begin{matrix}
        \frac{i\alpha_1}{4\pi  b_+} & \frac{i\alpha_2}{4\pi  b_+}& \frac{i\alpha_3}{4\pi  b_+} \\[0.5em] \frac{i\alpha_4}{4\pi  b_+} & \frac{i\alpha_5}{4\pi  b_+} & \frac{i\alpha_6}{4\pi  b_+}
    \end{matrix}\right\}_{b_+} \right |^2
\end{equation}
We will analyze this proposal in the semiclassical limit $\kappa\to \infty$, and verify the spherical volume conjecture. Namely, we  now show that the semiclassical limit agrees with the spherical volume of the tetrahedral graph complement \eqref{eq:volumetetra}.

\subsubsection*{Spherical tetrahedron volume from complex Virasoro $6j$ symbol}

The Virasoro $6j$ symbol admits the following integral representation \cite{Ponsot:1999uf,Ponsot:2000mt,Teschner:2012em,Collier:2018exn,Eberhardt:2023mrq}: 
\begin{multline}\label{eq:6jintegral}
    \left\{\begin{matrix}
        P_1 & P_2 & P_3 \\ P_4 & P_5 & P_6
    \end{matrix}\right\}_b = V_{123}V_{156}V_{246}V_{345} \times
\\ \int_{\mathcal{C}}\dd P
\frac{
S_b(i P-i P_{123})S_b(i P-i P_{156})S_b(i P-i P_{246})S_b(i P-i P_{345})
}{
S_b(\frac Q2+i P-i P_{1245})
S_b(\frac Q2+i P-i P_{1346})
S_b(\frac Q2+i P-i P_{2356})
S_b(\frac Q2+i P)}
\end{multline}
We follow the notation of \cite{Hartman:2025ula} (appendix~A)
with $P_{ijk}\coloneqq P_i+P_j+P_k$. The factors $V_{ijk}$ are products of double sine functions associated with the four vertices of the tetrahedron:
 \begin{equation}\label{eq:Vijk}
     V_{ijk} =\sqrt{S_b(\tfrac{Q}{2}+iP_k \pm i(P_i+P_j))S_b(\tfrac{Q}{2}-iP_k\pm i(P_i-P_j))}.
 \end{equation}
Here $\pm$ denotes a product over all choices of sign. The initial integration contour $\mathcal{C}$, before analytic continuation, is $P \in \mathcal{C} = -\frac{iQ}{4}+\mathbb{R}$.

For real $b\in (0,1)$, the double sine function has an integral representation that we review in appendix \ref{app:specialfunctions}. For $\mathrm{Im}(b^2)>0$, the double sine function 
admits a product representation in terms of the compact quantum dilogarithm function, 
\begin{equation}\label{eq:doublesinepochhammer}
   S_b(\tfrac{Q}{2}+z) =\frac{\psi_q(e^{2\pi i b z})}{\psi_{\tilde{q}}(e^{2\pi i z/b})} e^{-\frac{\pi i}{2}z^2+\frac{i\pi}{24}(Q^2-2)},\qquad \psi_q(x) 
\coloneqq \frac{1}{(-q x;q^2)_\infty}, 
\end{equation}
where $(x;q)_\infty = \prod_{n=0}^\infty(1-xq^n)$ is the $q$-Pochhammer symbol. Here we defined 
\begin{equation}
    q = e^{\pi i b^2}, \qquad \tilde{q} = e^{-\pi i /b^2}.
\end{equation}
This product representation converges for $\mathrm{Im}(b^2)>0$, which implies $|q|<1$ and $|\tilde{q}|<1$. It analytically continues to the lower half plane $\mathrm{Im}(b^2)<0$ (i.e. $q\to q^{-1}$) by the 
identity
\begin{equation}
    \psi_{q^{-1}}(x) = \psi_q(x)^{-1}.
\end{equation}
This defines $S_b$ in the spherical regime $b_\pm^2\in i\mathbb{R}$ that we need.
An alternative approach is to use the integral representation for the quantum dilogarithm at imaginary $b^2\in i\mathbb{R}$, with a deformed integration contour, as explained in \cite{Kim:2021nyy,Kim:2023wsq}.

Since the Liouville parameters $b_\pm^2$ are purely imaginary, the nomes $q_\pm, \tilde{q}_\pm$ are real. Without loss of generality, consider $b_+^2 = \frac{i}{\kappa}$, so $q_+ = e^{-\pi/ \kappa}$. We can use the known semiclassical limit of the compact quantum dilogarithm as $\kappa\to\infty$ to derive the following asymptotic behavior of the double sine (see \eqref{app:doublesineasymp}):
\begin{equation}\label{eq:doublesinekappa}
    \log S_b(b^{-1}z) \sim  -\frac{\kappa}{2\pi }\left( \Li_2(e^{-2\pi i z})-\pi^2(z^2-z)-\frac{\pi^2}{6}\right) + \mathcal{O}(1).
\end{equation}
Here $\mathrm{Li}_2(x)$ is the classical dilogarithm function, with a branch cut on the half line $(1,\infty)$. 
The above asymptotics hold for $0<\Re{\,z} <1$. To use the above asymptotic formula for $\Re(z)$ outside $(0,1)$, we first use the exact shift relation 
\begin{equation}\label{eq:shiftrelation}
    S_b\left(b^{-1}(z+1)\right) = 2\sin(\frac{\pi z}{b^2})S_b(b^{-1}z),
\end{equation}
to bring $\Re(z)$ inside the unit interval, 
and then take the semiclassical limit.

 With these preparations in place, let us take the semiclassical limit of the $6j$ symbol in the spherical regime. We focus on $b^2\equiv b_+^2 = \frac{i}{\kappa}$ (the other chiral half is analogous) and write $P_{j,+} =\frac{i\theta_j}{2\pi  b}$, with $\theta_j = \frac{\alpha_j}{2} \in (0,\pi)$ the six dihedral angles of a non-degenerate convex tetrahedron in $S^3$. At each vertex $(ijk)$, the dihedral angles must satisfy
 \begin{equation}\label{eq:angleinequalities}
    \pi<\theta_i+\theta_j+\theta_k < 3\pi, \qquad -\pi<\theta_i+\theta_j-\theta_k<\pi, 
 \end{equation}
 and all permutations thereof.
 Now consider the vertex factors $V_{ijk}$ in equation \eqref{eq:Vijk}. Each consists of a product of double sines, of the form
  \begin{equation}
    V_{ijk}^2\sim S_b\left(b^{-1}(\tfrac{1}{2}-\tfrac{\theta_k\pm(\theta_i+\theta_j)}{2\pi})\right)S_b\left(b^{-1}(\tfrac{1}{2}+\tfrac{\theta_k\pm(\theta_i-\theta_j)}{2\pi})\right).
 \end{equation}
 For three out of the four double sines, we can immediately use the asymptotic formula \eqref{eq:doublesinekappa}, because the argument is in $(0,1)$ by the angle inequalities \eqref{eq:angleinequalities}. Since the full partition function is the absolute value square, we only have to consider the real part of their exponents. But the real part of the dilogarithm on the unit circle satisfies
 \cite{murakami2012volume}
\begin{equation}
    \text{Re} \left(\text{Li}_2(e^{i\theta})\right)= \frac{\theta^2}{4} -  \frac{\pi\theta}{2} + \frac{\pi^2}{6}, 
\end{equation}
cancelling the other terms in \eqref{eq:doublesinekappa}.
Hence the contribution of these double sine factors to the classical limit of $|V_{ijk}|^2$ vanishes, to leading order in $\kappa$. The only non-zero contribution comes from the single double sine function in \eqref{eq:Vijk} whose argument is negative, $\frac{1}{2\pi}(\pi-\theta_i-\theta_j-\theta_k)<0$. We shift it back to $(0,1)$ using \eqref{eq:shiftrelation} and then take the classical limit. The $\sin(\pi z/b^2)$ factor then contributes non-trivially to the real part of the classical action. Thus we have shown that semiclassically
\begin{equation}
    V_{ijk} \sim e^{\frac{\kappa}{4}A(\theta_i,\theta_j,\theta_k)} \quad \text{with}\quad  \Re\, A(\theta_i,\theta_j,\theta_k) = \pi - \theta_i-\theta_j-\theta_k.
\end{equation}
This means that the real part of $A$ precisely cancels against the semiclassical limit of the corresponding $C_0$ normalization factor that we derived in section \ref{sec:observer2point}, cf.~\eqref{eq:C0_volume_result}:
\begin{equation}
   \Bigg| \frac{V_{ijk}}{\sqrt{\mathsf{C}_{ijk}}} \Bigg|^2 = \mathcal{O}(1), \qquad \kappa\to\infty.
\end{equation}

Now we move on to the $P$-integral \eqref{eq:6jintegral} defining the $6j$ symbol. To analyze its classical limit, we parametrize $P = -\frac{iQ}{2} -\frac{s}{4\pi b}$. If we rotate $b$ from the real axis to $b_+$ in the first quadrant, the $s$-contour gets transported too. Let us turn on the external angles as $t\theta_j$ with $t\in [0,1]$. Then a suitable transported contour for $s$ is  
\begin{equation}\label{eq:sixj-contour}
\Gamma_s(t)=\mathbb R-i\pi(1+b^2)
-it\left(\min_v\theta_v+\max_r\theta_r\right).
\end{equation}
 Here we introduced the notation $v\in\{123,156,246,345\}$, $r\in\{1245,1346,2356,\emptyset\}$ for the indices in respectively the numerator and denominator of the integrand in \eqref{eq:6jintegral}. This keeps the numerator poles
below $\Gamma_s(t)$ and the poles arising from denominator zeros above
$\Gamma_s(t)$; no poles are crossed in the analytic continuation of $b^2$. Moreover, the tails are exponentially convergent,
including at the endpoint $b^2=i/\kappa$, approached from the right half plane
$\text{Re}\,b^2\to 0^+$.

 Using the asymptotic formula \eqref{eq:doublesinekappa} to expand the integrand at large $\kappa$, we find
\begin{equation}\label{eq:sixj-saddle-integral}
    \left\{\begin{matrix}
        \frac{i\theta_1}{2\pi b_+} & \frac{i\theta_2}{2\pi b_+} & \frac{i\theta_3}{2\pi b_+} \\[0.5em] \frac{i\theta_4}{2\pi b_+} & \frac{i\theta_5}{2\pi b_+} & \frac{i\theta_6}{2\pi b_+}
    \end{matrix}\right\}_{b_+} \sim \,\prod_v e^{\frac{\kappa}{4} A_v} 
\int_{\Gamma_s(1)}\dd s\,
\exp\!\left[\frac{\kappa}{\pi}S(s;\theta_j)\right],
\end{equation}
where \vspace{-4mm}
\begin{align}\label{eq:classaction}
 S(s;\theta_j)
={}&
  U(e^{-s/2};e^{-i\theta_j})-\frac{1}{2} \sum_{j=1}^3 \theta_{j}\theta_{j+3}-\frac{\pi}{2}\sum_{j=1}^6\theta_j
 +\frac{i\pi s}{2}
 -\frac{ \pi^2}{2}.
\end{align}
Here $U(z;x_j)$ is precisely the dilogarithm combination \eqref{eq:U-action}, evaluated on $z = e^{-s/2}$ and $x_j = e^{-i\theta_j}$. To write the saddle-point equation $\partial_s S =0$, we use that 
\begin{equation}\label{eq:derivdilog}
    \frac{d}{du}\mathrm{Li}_2(u) = \frac{-\log(1-u)}{u}.
\end{equation}
Exponentiating both sides, the saddle-point equation becomes
\begin{align}
&(1-z)
(1-x_1x_2x_4x_5z)
(1-x_1x_3x_4x_6z)
(1-x_2x_3x_5x_6z)
\nonumber\\
&\hspace{1cm}=
(1+x_1x_2x_3z)
(1+x_1x_5x_6z)
(1+x_2x_4x_6z)
(1+x_3x_4x_5z).
\label{eq:saddle-equation}
\end{align}
Expanding the products and factoring out $z$, this is a quadratic equation for $z = e^{-s/2}$ with $x_j$-dependent coefficients. Let $z_0$ be the root with $|z_0|>1$, and $s_0$ its logarithmic lift. Its upward thimble intersects the transported contour $\Gamma_s(1)$ on the same branch, with signed intersection number 1 in the spherical configurations checked.\footnote{For example, at the right-angled point $\theta_j=\pi/2$, $z_0=1+i$ and
$s_0=-\log2-9\pi i/2$. The $\text{Li}_2(z_0)$ term in $U$ is
its principal value minus $2\pi i\log z_0$, with the logarithm principal;
the other seven dilogarithms are principal. Subsequent branches are
transported along $G(t)=(1-t)\mathbf1+tG$, which stays positive definite.} An explicit expression for $z_0$ is given in \eqref{eq:root}.

Interestingly, the discriminant of the quadratic polynomial in $z$ is proportional to the determinant of the angle Gram matrix of the tetrahedron \eqref{eq:grammatrix}. This was also observed in \cite{Hartman:2025ula}. It signals the hyperbolic-to-spherical transition when the discriminant changes sign (positive in the spherical regime). The transition occurs when $D = 0$, which for equal dihedral angles is at $\theta = \arccos(1/3)$.

As discussed above, the full partition function is a modulus squared, so only the real part of $S$ survives in the spherical regime (where $\kappa$ is real). So, evaluating $S$ on the saddle $s_0$ that can be picked up by $\Gamma_s$, and taking the real part, we precisely land on Murakami's expression for the volume of a spherical tetrahedron, recall \eqref{eq:murakamivolume}:
     \begin{equation}\label{eq:6jactionvol}
  \mathrm{Re}\,S(s_0;\theta_j)= \ell_{\rm dS}^{-3}\text{Vol}(T_{\vec{\alpha}}).
 \end{equation}
We have also verified this directly by showing that $\Re S(s_0,\theta_j)$ obeys the spherical Schl\"afli formula \eqref{eq:tetravolume}. Plugging this back into the saddle-point approximation of the $6j$ symbol \eqref{eq:sixj-saddle-integral}, and multiplying by the complex conjugate to get $Z_{\rm grav}^{\Lambda>0}$,
we have thus shown that 
\begin{equation}\label{eq:tetravolumeresult}
    Z_{\rm grav}^{\Lambda>0}\big[S^3 \setminus \tetra_{\vec{\alpha}} \big] \sim \exp[\frac{\text{Vol}\left(S^3 \setminus \tetra_{\vec{\alpha}}\right)}
{4\pi G_N\ell_{\mathrm{dS}}^2}]
\end{equation}
where the spherical cone manifold $S^3\setminus\tetra_{\vec\alpha}$ \;is the doubled tetrahedron whose volume is 
$2\text{Vol}(T_{\vec\alpha})$.
The exponent agrees with the Einstein-plus-worldline action of
subsection~\ref{sec:Schlafli}. This expansion holds at fixed nondegenerate angles. We can easily incorporate the one-loop determinant, by keeping the $\order{1}$ terms in the expansion of the double sine. Since the 1-loop correction is $b$-independent, we  can use the result of \cite{Meng:2026xlo}, who found $Z_{\rm 1-loop}(6j) \propto (\det G)^{-1/4}$ where $G$ is the angle Gram matrix of $T_{\alpha}$. This should be multiplied by the one-loop corrections to the $\mathsf{C}_{ijk}$ factors, which are also known.

\subsection*{Heavy--light limit of the spherical tetrahedron and the de Sitter OTOC}

Next we study a degeneration of the spherical tetrahedron in
which four dihedral angles correspond to a heavy observer and the remaining
two to light scalar probes $V,W$. The picture to have in mind is\footnote{To go from the planar $6j$ diagram to this `crossed' diagram, Virasoro TQFT gives two additional braiding phases $\mathbb{B}_{ij}^k$. However, as explained in the introduction, these cancel in the product $Z_+ Z_-$.}
\begin{equation}
    \vcenter{\hbox{
    \begin{tikzpicture}[scale=0.5]
    \draw[very thick, black, fill=black!5] (0,0) circle (2);
    \draw[very thick, green!60!black, dashed] (-2,0) 
    node[circle, fill=green!60!black, inner sep=1.5pt] {}
    node[left,black] {$W$}
    to[out=30, in=150] (2,0)
    node[right,black] {$W^{\dagger}$}
    node[circle, fill=green!60!black, inner sep=1.5pt] {};
    \draw[very thick, red] (0,2) 
    node[circle, fill=red, inner sep=1.5pt] {} 
    node[above,black] {$V$}
    to[out=-140,in=140] (0,-2)
    node[below,black] {$V^{\dagger}$}
    node[circle, fill=red, inner sep=1.5pt] {};
\end{tikzpicture}
    }}
\end{equation}
In this limit we can connect to the recent literature on the de Sitter OTOC \cite{Aalsma:2020aib,Cui:2026bcd,Harlow:2026pwe,Chen:2026boh,Milekhin:2026tbi}. In particular, we will derive the positive sign of the growing OTOC correction associated with
the negative de Sitter shock. This is what we will now show.

We choose the labeling of the tetrahedron such that the four heavy angles
are
\begin{align}
    \theta_1=\theta_4
    &=
    \theta_H
    +\frac{\pi}{2\kappa}\,\omega,
    \nonumber\\
    \theta_3=\theta_6
    &=
    \theta_H
    -\frac{\pi}{2\kappa}\,\omega,
    \label{eq:heavy-angles}
\end{align}
with $\kappa=\frac{\ell_{\rm dS}}{4G_N}$, while the two probe angles are
\begin{equation}
    \theta_2
    =
    \pi-\frac{\pi\nu'}{\kappa},
    \qquad
    \theta_5
    =
    \pi-\frac{\pi\nu}{\kappa}.
    \label{eq:light-angles}
\end{equation}
Using $\alpha_j =2\theta_j =  2\pi(1-4G_Nm_j)$, we see that $\omega = (m_3-m_1)\ell_{\rm dS}$ measures the energy transferred between the two pairs of
observer segments, while their average energy is held fixed. Reversing the
assignment of the two heavy pairs amounts to $\omega\rightarrow-\omega$. As in the previous section, $\nu,\nu'$ are the continuous principal series labels of the scalar probes $\Delta_V = 1+i\nu$, $\Delta_W = 1+i\nu'$. For the observer, we stay in the fully backreacted regime, so $\theta_H<\pi$ fixed. We work in the WKB regime for the probe particles,
\begin{equation}
    \nu,\nu' \gg 1,
    \qquad
    \frac{\nu}{\kappa},
    \frac{\nu'}{\kappa}\ll 1.
\end{equation}
So as $\kappa\to\infty$, the probe dihedral angles approach $\theta_{V,W}\to \pi$. This is a partially ideal limit of the tetrahedron. We will expand the spherical volume function \eqref{eq:classaction} of the semiclassical $6j$ symbol for large but finite $\kappa$. The analysis is very similar to that of \cite{Jackson:2014nla} for the hyperbolic tetrahedron in the AdS$_3$ shockwave calculation, but with important differences in the signs.

Consider the saddle-point equation \eqref{eq:saddle-equation} that arose from the $6j$ integral. In the strict $\kappa\to\infty$ limit, the saddle-point equation for $z=e^{-s/2}$ degenerates to
\begin{equation}
    z(1-x^2)^2(1-x^2z)^2=0,
\end{equation}
where we used that $x_1 = x_3 = x_4 = x_6 = e^{-i\theta_H} \equiv x$ and $x_2 = x_5 = -1$. Discarding the endpoint root $z=0$, the relevant saddle is the double root $
    z_0=x^{-2}=e^{2i\theta_H}.$ At finite but large $\kappa$, our OTOC limit \eqref{eq:heavy-angles} resolves this double root. Namely, using \eqref{eq:heavy-angles},
    \begin{equation}\label{eq:otocx}
        x_1=x_4
    =
    xe^{
       -\frac{i\pi\omega}{2\kappa}
    },\qquad x_3=x_6
    =
    xe^{
       +\frac{i\pi\omega}{2\kappa}
   },\qquad x_2=-e^{\frac{i\pi\nu'}{\kappa}}, \qquad x_5=-e^{\frac{i\pi\nu}{\kappa}},
    \end{equation}
where we parametrize the saddle near $z_0 =e^{2i\theta_H}$ by
\begin{equation}\label{eq:otocx2}
    z
    =
    z_0
    \exp\left[
       -\frac{i\pi}{\kappa}(\nu+\nu')
       +\delta
    \right].
\end{equation}
To resolve the double root, $\delta$ should be of order $\kappa^{-2}$. Expanding the saddle-point equation \eqref{eq:saddle-equation} to this order, we show in appendix \ref{app:saddlepoint6j} that it becomes 
\begin{equation}\label{eq:saddlepointdelta}
    \delta^2
    =
    \left(
       \frac{\pi}{\kappa}
    \right)^2
    \left(\omega_c^2-\omega^2\right), \qquad \omega_c \coloneqq \frac{\pi\nu\nu'}
         {2\kappa\sin\theta_H}\,.
\end{equation}
The $6j$ integration contour picks up the root with $|z|>1$, which corresponds to $\delta>0$.

Next, we expand the  $6j$-action $S(s;\theta_j)$ \eqref{eq:classaction} in the heavy-light limit. Recall that it contains the dilogarithm combination $U(z;x_j)$. We need the following expansion of the dilogarithm close to its branch point:
\begin{equation}
    \text{Li}_2(e^u)
    =
    \frac{\pi^2}{6}
    +
    u\bigl(1-\log(-u)\bigr)
    -\frac{u^2}{4}
    +\order{u^3},
    \qquad
    u\rightarrow0.
    \label{eq:dilog-expansion}
\end{equation}
The branch of the logarithm is inherited from the analytic continuation of
the original $6j$ contour. Carefully expanding the 8 dilogarithms in $U(z;x_j)$, we find that to $\order{\kappa^{-3}}$,
\begin{multline}
  S(\delta;\theta_j) =
 \frac12\Bigg[
 2\delta
 -
 \left(
    \delta+\frac{i\pi}{\kappa}\omega
 \right)
 \log\left(
 \frac{
    \delta+\frac{i\pi}{\kappa}\omega
 }{
    \frac{\pi}{\kappa}\omega_c
 }
 \right)
\\
 -
 \left(
    \delta-\frac{i\pi}{\kappa}\omega
 \right)
 \log\left(
 \frac{
    \delta-\frac{i\pi}{\kappa}\omega
 }{
    \frac{\pi}{\kappa}\omega_c
 }
 \right)
 \Bigg] + S_{\rm reg.}
 \label{eq:S-local-integrated}
\end{multline}
Here $S_{\rm reg}$ contains the $\delta$-independent pieces coming from the polynomial part of \eqref{eq:classaction}. As a check, we can take the $\delta$-derivative of $S$, which gives
\begin{equation}
 S'(\delta)
 =
 -\frac12
 \log\left(
 \frac{
   \delta^2+
   \left(
      \frac{\pi}{\kappa}\omega
   \right)^2
 }{
   \left(
      \frac{\pi}{\kappa}\omega_c
   \right)^2
 }
 \right).
\end{equation}
Indeed, setting $S'(\delta)=0$ precisely gives the saddle-point equation \eqref{eq:saddlepointdelta} for $\delta$. Evaluating the action \eqref{eq:S-local-integrated} on the saddle point $\delta = \frac{\pi}{\kappa}\sqrt{\omega_c^2-\omega^2}$, and using the identity
\begin{equation}
\log \frac{\sqrt{\omega_c^2-\omega^2}\pm i\omega}{\omega_c}
 =
    \pm i\arcsin\left(\frac{\omega}{\omega_c}\right),
\end{equation}
we have shown that
\begin{equation}
   \Re( S(\omega;\nu,\nu') - S(0;\nu,\nu')) = \frac{\pi}{\kappa}
 \left[
    \omega
    \arcsin\left(\frac{\omega}{\omega_c}\right)
    +
    \sqrt{\omega_c^2-\omega^2}
    -
    \omega_c
 \right].
\end{equation}
Here we normalized by the $\omega=0$ action, which has gotten rid of $S_{\rm reg}$; this is a convenient normalization choice for the OTOC. Recalling that the semiclassical $6j$ symbol scales as $ \exp(\frac{\kappa}{\pi}S)$, and multiplying by the complex conjugate chiral half, the normalized energy space OTOC takes the form 
\begin{equation}
   \log K^{\rm OTOC}_{\nu,\nu'}(\theta_H,\omega) \sim W(\omega) \equiv 2\left[\omega
    \arcsin\left(\frac{\omega}{\omega_c}\right)
    +
    \sqrt{\omega_c^2-\omega^2}
    -
    \omega_c\right].
\end{equation}

Since $W(\omega)$ was derived from the spherical volume function, it has an interesting geometric interpretation using the Schl\"afli formula. Recall that $d\text{Vol}(T) = \frac{\ell_{\rm dS}^2}{2}\sum_e L_e d\theta_e$, where $L_e$ is the geodesic length of the edge $e$. Using $\partial_\omega \theta_1 = \partial_\omega \theta_4 = \frac{\pi}{2\kappa}$ and $\partial_\omega \theta_3 = \partial_\omega \theta_6 = -\frac{\pi}{2\kappa}$, the variation of the volume with respect to $\omega$ satisfies
\begin{equation}
    \frac{\partial}{\partial\omega} \text{Vol}(T_{\vec{\alpha}}) = \frac{\pi\ell_{\rm dS}^2}{2\kappa}(L_1 -L_3).
\end{equation}
Since the on-shell action is proportional to $\text{Vol}(T_{\vec{\alpha}})$ by \eqref{eq:tetravolumeresult}, we obtain a relation between $W'(\omega)$ and the difference in observer geodesic lengths in the heavy-light limit:
\begin{equation}\label{eq:Wprime}
    W'(\omega) = 2\arcsin(\frac{\omega}{\omega_c}) = \ell_{\rm dS}^{-1}(L_1-L_3).
\end{equation}

Let us now study the Lorentzian continuation of the tetrahedral
saddle. We define its contribution by continuing the Euclidean
steepest-descent cycle through $\omega=0$ at $t=0$:
\begin{equation}
G^{\mathrm{OTOC}}_{\nu,\nu'}(t)
\sim
\int_{i\mathbb{R}}\frac{d\omega}{i}\,
e^{W(\omega)-2i\omega t/\ell_{\mathrm{dS}}}\, .
\end{equation}
The branches of $W$ are fixed by continuation from $\omega=0$,
where $\sqrt{\omega_c^2-\omega^2}=\omega_c$, and the contour is
oriented from $-i\infty$ to $+i\infty$. In the following,
$G^{\mathrm{OTOC}}_{\nu,\nu'}$ denotes this tetrahedral
contribution.\footnote{This prescription specifies the inverse
transform of the reduced semiclassical action. Identifying it
with the inverse transform of the full finite-$\kappa$
energy-space amplitude requires additional contour information;
see the related discussion in Sec.~2.6 of \cite{Chen:2026boh}.}
The factor of two in the exponent is appropriate to the symmetric energy-transfer
variable used in our antipodal configuration \eqref{eq:heavy-angles}. The saddle-point equation is
$W'(\omega_*) = 2it/\ell_{\rm dS}$, which we can solve using \eqref{eq:Wprime} as\footnote{To check that this saddle contributes, write $\omega=iy$ with
$y\in\mathbb R$. The exponent is real on the imaginary axis,
has strictly negative second derivative
$-2/\sqrt{\omega_c^2+y^2}$ with respect to $y$, and tends to
$-\infty$ at both ends. Its unique maximum is therefore at
$y=\omega_c\sinh(t/\ell_{\mathrm{dS}})$.
Thus the imaginary axis is the descending cycle through
$\omega_*$, and this saddle contributes with intersection
number one.}
\begin{equation}
    \omega_*(t) = i\omega_c
    \sinh\left(\frac{t}{\ell_{\rm dS}}\right).
\end{equation}
So in real Lorentzian time, the energy saddle has moved to the imaginary axis. Moreover, the energy-transfer saddle $\omega_*$
grows exponentially with static-patch time. The geometric interpretation can be read off from \eqref{eq:Wprime}: the saddle-point obeys $L_1-L_3 = 2it$, in other words the spherical length difference is rotated to an imaginary value, whose imaginary part is the Lorentzian proper-time separation.

Evaluating the Fourier exponent on this saddle gives
\begin{align}
   \log G^{\rm OTOC}_{\nu,\nu'}(t) \sim  W(\omega_*)-2i\omega_*t/\ell_{\rm dS}
    &=
    \frac{\pi\nu\nu'}
         {\kappa\sin\theta_H}
    \left[
       \cosh\left(\frac{t}{\ell_{\rm dS}}\right)-1
    \right],
\end{align}
where we substituted in the definition of $\omega_c$. In particular, we confirm the positive sign for the growing term at late times, agreeing 
with the classical calculations in \cite{Aalsma:2020aib,Aalsma:2021kle,Cui:2026bcd,Harlow:2026pwe,Chen:2026boh,Milekhin:2026tbi} of the time advance produced by a shockwave in de Sitter space.\footnote{For example, the classical impact parameter $h(\phi)$ at the antipodal point $\phi = \pi$ is minus our $\omega_c$, \begin{equation}
    \nu\nu' h(\pi)
    =
    -\frac{2\pi \nu\nu' G_N}
           {\ell_{\rm dS}\sin(\alpha/2)} = -\omega_c.
\end{equation}} Here we derived it purely from a semiclassical heavy-light limit of the analytically continued Virasoro $6j$ symbol. 

Moreover, we can take a further pure recoil limit where we set $\theta_H = \pi(1-\frac{m\ell_{\rm dS}}{\kappa})$, keeping the observer mass $m$ fixed as we take $\kappa\to \infty$. Then $\sin\theta_H \approx \frac{\pi m\ell_{\rm dS}}{\kappa}$, so the classical action scales in the pure recoil limit as
\begin{equation}\label{eq:6jpurerecoil}
    \log G^{\rm OTOC}_{\nu,\nu'}(t) \sim \frac{\nu\nu'}{2m\ell_{\rm dS}}e^{t/\ell_{\rm dS}}
\end{equation}
at late times, in agreement with the pure recoil regime of \cite{Chen:2026boh}. In appendix \ref{sec:tetra-recoil}, we further show that the tetrahedron volume function \eqref{eq:6jactionvol} that arose from the semiclassical $6j$ symbol reduces in the pure recoil limit to the tetrahedral graph action of \cite{Chen:2026boh}. This reduction is a simple consequence of the Schl\"afli formula. 

We emphasize that the above tetrahedral regime is not the Lyapunov regime of the OTOC. Schematically, the former has $G \sim \exp \left(\omega_c e^{t/\ell_{\rm dS}}\right)$ whereas the latter has $G \sim 1+ \omega_c e^{t/\ell_{\rm dS}}$. Naively expanding the exponential in the tetrahedral regime is not allowed, because the Fourier saddle is only sharply peaked if the following ratio is small:
\begin{equation}
    \frac{\Delta \omega}{|\omega_*|}\sim \left (\frac{\nu \nu'}{m\ell_{\rm dS}}e^{t/\ell_{\rm dS}}\right)^{-\frac{1}{2}},
\end{equation}
meaning that the exponent is large.
To probe the scrambling regime, we have to take a different limit of the $6j$ symbol where the Liouville momenta $P_j$ scale as $b$, analogous to the Schwarzian limit of the $6j$ symbol studied in \cite{Mertens:2017mtv}.

Lastly, let us comment on the puzzle raised in \cite{Chen:2026boh} for interpreting the OTOC as a (microcanonical) trace. Our proposal for the OTOC in terms of the analytically continued Virasoro $6j$ symbol is non-perturbative in $G_N$, but not `microscopic'. Nonetheless, it resolves at least one of the puzzles of \cite{Chen:2026boh}, namely cyclicity. A cyclic shift of the four insertions together with $\nu\leftrightarrow\nu'$ is just a relabeling of the tetrahedron's vertices, which is an exact symmetry of the $6j$ symbol. Before taking the semiclassical limit, $\{6j\}_b$ is an analytic function of $\omega$, with a well-defined integration contour. It is only in the semiclassical approximation that the square root branch cut $\sqrt{\omega_c^2-\omega^2}$ arises. 
The two eikonal resummations $F_{12}, F_{14}$ found in \cite{Chen:2026boh} are then two Stokes sectors of the same analytic function of $\omega$, as was already intuited there.

The second puzzle, that of positivity, is not resolved by our proposal. One suggestion \cite{Harlow:2026pwe} is that the quantum mechanical observer OTOC is not a microscopic trace, but instead a
Euclidean-folded correlator rather than an ordinary thermal correlator. Since this paper is mainly focused on the gravitational calculations, we leave the microscopic interpretation in observer holography as a major open problem.

\section{Hyperbolic-to-spherical transitions in cone manifolds}\label{sec:hyptosph}

In this section we compute the partition functions of knot- and link complements with a hyperbolic-to-spherical transition: the figure-eight knot, the three-twist knot and the Whitehead link. In each case, the starting point is the exact Virasoro TQFT partition function, which we analytically continue to the de Sitter regime following our general prescription \eqref{eq:zgrav}. We will show that the semiclassical limit of the resulting partition function reproduces the transition to the spherical regime at the critical cone angle $\alpha_*$, and we will demonstrate how the semiclassical actions encode the cone manifold volumes on both sides of the transition. This follows from a non-trivial Picard-Lefschetz analysis of the relevant integration contours. We hence confirm the spherical volume conjecture \eqref{eq:volconj} in these more complicated examples too.

Physically, we may think of the knot partition functions studied in this section as non-perturbative corrections to the observer partition function (viewed as a Euclidean gravitational path integral with a sum over topologies), where the observer worldline is embedded in $S^3$ in a topologically non-trivial way and follows a knotted geodesic trajectory. The classical volume of any knotted embedding of the worldline is always smaller than that of the unknot, whose conical dS$_3$ partition function we computed in section \ref{sec:observerpartitionfunction}. Hence semiclassically, these corrections to the observer partition function are non-perturbatively suppressed in $G_N$. 

An interesting feature of such an interpretation is that the existence of a spherical saddle,\footnote{This is the so-called Porti condition for any non-toric 2-bridge knot \cite{Porti2004}.} $\alpha_* < \alpha < 2\pi - \alpha_*$, translates to a condition on the observer mass through $\alpha = 2\pi(1-4G_Nm)$. In other words, depending on the observer mass, different saddles may contribute to the observer partition function. For the figure-eight knot, this gives the spherical mass interval\footnote{For every other hyperbolic two-bridge knot, the critical angle
satisfies $\alpha_*>2\pi/3$ \cite{Porti2004, Porti2022Cone}, so its spherical mass interval is
strictly contained in the figure-eight interval (\ref{eq:massrange}).}
\begin{equation} \label{eq:massrange}
    \frac{1}{12}<G_Nm< \frac{1}{6}\,.
\end{equation}
At the lower endpoint for $m$, i.e. when $\alpha$ reaches $\alpha_{\rm max}$, the knot degenerates into a tetravalent theta graph which we also discuss in section \ref{sec:Tetravalent}. At this angle, the geometry becomes time-reflection symmetric and admits a Lorentzian continuation, which we can interpret as a closed de Sitter cosmology with heavy massive particles.

\subsection{The figure-eight knot cone manifolds}

The first knot for which we study the spherical volume conjecture using analytically continued Virasoro TQFT is the figure-eight knot. 
The figure-eight knot is an amphichiral knot with 4 crossings denoted $\mathbf{4}_1$ in the Alexander-Briggs notation,
\begin{equation}
   \vcenter{\hbox{ \begin{tikzpicture} [scale=0.7,xscale=0.75]      
        \draw[very thick, out = 90, in = 180, black] (-5/2,-3/2) to (-1,0);
        \draw[very thick, out = 0, in = 180, black] (-1,0) to (0,0);
        \draw[fill=white,draw=white] (-1,0) circle (1/10);
        \draw[very thick, out = 180, in = 90, black] (0,1) to (-1,0);
\draw[very thick, out = 270, in = 150, black] (-1,0) to (0,-1);
        \draw[very thick, out = 330, in = 90, black] (0,-1) to (1,-2);
        \draw[fill=white,draw=white] (0,-1) circle (1/10);
\draw[very thick, out = 90, in = 270, black] (-1,-2) to (1,0);
        \draw[very thick, out = 90, in = 0, black] (1,0) to (0,1) ;
        \draw[fill=white,draw=white] (1,0) circle (1/10);
        \draw[very thick, out = 0, in = 180, black] (0,0) to (1,0) node[right=10pt,black] {$P_0$};
        \draw[very thick, out = 0, in = 90, black] (1,0) to (5/2,-3/2);
        \draw[very thick, out = 270, in = 330, black] (5/2,-3/2) to (0,-3);
\draw[very thick, out = 150, in = 270, black] (0,-3) to (-1,-2);
        \draw[fill=white,draw=white] (0,-3) circle (1/5);
\draw[very thick, out = 270, in= 0, black] (1,-2) to (0,-3);
        \draw[very thick, out = 180, in = 270, black] (0,-3) to (-5/2,-3/2);
\end{tikzpicture}}}
\end{equation}
Let us parametrize the meridian holonomy by $P_0 = \frac{i\alpha}{4\pi b}$. Then it is known that $\mathbf{4}_1$ has a hyperbolic-to-spherical transition \cite{Mednykh2003}: for $\alpha = 0$, there is a unique complete hyperbolic metric with finite volume (the boundary is a torus cusp). For $0<\alpha < \frac{2\pi}{3}$, there is a hyperbolic cone structure; for $\alpha = \frac{2\pi}{3}$ a flat cone structure; and for $\frac{2\pi}{3}<\alpha<\frac{4\pi}{3}$ a spherical cone metric with a conical singularity along the figure-eight knot.

Here we will study this transition on the level of the full quantum partition function. Our starting point is the exact Virasoro TQFT partition function, which was shown to be \cite{Collier:2024mgv}
\begin{equation}
    Z^{(b)}_{\rm Vir}[\mathbf{4}_1;P_0]= \frac{1}{\rho_0(P_0)} \int_0^\infty \dd P\,\rho_0(P)\, \mathbb{S}^*_{P_0P_0}[P]\left(\mathbb{B}_{P_0P_0}^P\right)^2\ .
\end{equation}
The braiding phase is defined as $\mathbb{B}_{P_1P_2}^P = e^{\pi i (h_1+h_2 - h_P)}$ and $\sker{P_1}{P_2}{P}$ is the modular kernel of the torus one-point conformal block in 2D CFT. Without going into detail, we can simplify this expression considerably by using a non-standard integral representation for the modular kernel (eq.~(3.89) in \cite{Eberhardt:2023mrq}) and performing the $P$-integral in closed form using the two-term Euler integral (eq.~(C.26) in \cite{Eberhardt:2023mrq}). The exact result is
\begin{equation} \label{eq:figure-eight_part_func}
    Z^{(b)}_{\rm Vir}[\mathbf{4}_1;P_0]=\sqrt{2}\,e^{2\pi QP_0}\int_{\mathcal{C}}\dd y \,S_b \left(\tfrac{Q}{2}-iy+iP_0\right) S_b \left(\tfrac{Q}{2}-iy-iP_0\right)e^{-6\pi i P_0 y}
\end{equation}
where $S_b(z)$ is the double sine function that we introduced in section \ref{sec:observerOTOC}.
We will discuss the integration contour $\mathcal{C}$ momentarily. This integral expression coincides with the Andersen-Kashaev Teichm\"uller TQFT partition function of the figure-eight knot complement \cite{EllegaardAndersen:2011vps}, upon translating conventions.

Taking \eqref{eq:figure-eight_part_func} as our starting point, we will show that
the semiclassical limit of $Z_{\rm Vir}$ reproduces
the hyperbolic-to-spherical transition at cone angle $\alpha_*=2\pi/3$ as a
collision of its two saddles. Transporting the integration contour and the
logarithmic sheets identifies the contributing geometric saddle on each
side of the transition. Its semiclassical action reproduces the hyperbolic and spherical
cone manifold volumes throughout the
angle ranges studied below.

First, we take $b$ real and positive, so we want to relate the above partition function (\ref{eq:figure-eight_part_func}) to the volume of the hyperbolic branch in the semiclassical limit. 
To determine the integration contour $\mathcal{C}$, note the asymptotics of the integrand,
\begin{equation}
    I(y;P_0)=s_b(y-P_0)s_b(y+P_0)e^{-6\pi iP_0y}\,,
\end{equation}
where we have introduced the shifted double-sine function,
\begin{equation}
    s_b(x)\coloneqq S_b\left(\frac{Q}{2}-ix\right)\,.
\end{equation}
The large-$x$ asymptotics of $s_b(x)$ are given by
\begin{equation}\label{eq:sbasymptotics}
    s_b(x)\sim \begin{cases}
        \exp[-\frac{i\pi x^2}{2}], \qquad & \Re(x)\to \infty,\\
        \exp[\frac{i\pi x^2}{2}], \qquad & \Re(x)\to -\infty\,.
    \end{cases}
\end{equation}
Using this, we see that the asymptotics of the integrand are
\begin{equation}
    I(y;P_0)\sim \begin{cases}
        \exp[-i\pi y^2-6\pi i P_0 y], \qquad & \Re(y) \to \infty,\\
        \exp[i\pi y^2 - 6\pi i P_0 y], \qquad & \Re(y) \to -\infty\,.
    \end{cases}
\end{equation}
From these asymptotics, we see that the horizontal contour $y=\mathbb{R}-i\epsilon$ is convergent only for $P_0 \in \mathbb{R}$. Since we need $\text{Im}P_0>0$, we need to deform the contour near infinity. A convenient choice is the following bent contour:
\begin{equation}
    \mathcal{C}: \infty e^{i(\pi+\phi)}\to -R-i\epsilon \to R-i\epsilon \to \infty e^{-i\phi}, \qquad 0<\phi<\frac{\pi}{2}\,.
\end{equation}
Here, $R$ can be conveniently chosen to be any positive value greater than $|P_0|$ which ensures that no poles cross the integration contour. To see this, note that the two families of poles coming from the two $S_b$ functions are at 
\begin{equation}
    y_{m,n}=\pm P_0-iA_{m,n}, \qquad A_{m,n}\equiv \frac{Q}{2}+mb+nb^{-1}\,.
\end{equation}
Since the two pole lattices are in the lower half plane and contained within the strip $|\Re \,y_{m,n}|<|P_0|$, choosing $R>|P_0|$ ensures that no pole crosses the integration contour.
In the spherical branch, since we continue $b^2$ to be imaginary, the poles of the integrand are no longer restricted to the strip, rather they spread into a wedge/cone given by
\begin{equation}
    -\frac{\pi}{2}-|\arg b|  \leq \text{arg}(\Delta y) \leq -\frac{\pi}{2}+|\arg b|\,.
\end{equation}
Here $\Delta y$ is the difference in the location of a pole in the family and $m=n=0$ pole in the same family. So the poles are located in a downward pointing wedge below the real axis. The bent portion of the integration contour should be such that it encloses this wedge. 

\begin{figure}
    \centering
    \includegraphics[width=0.6\linewidth]{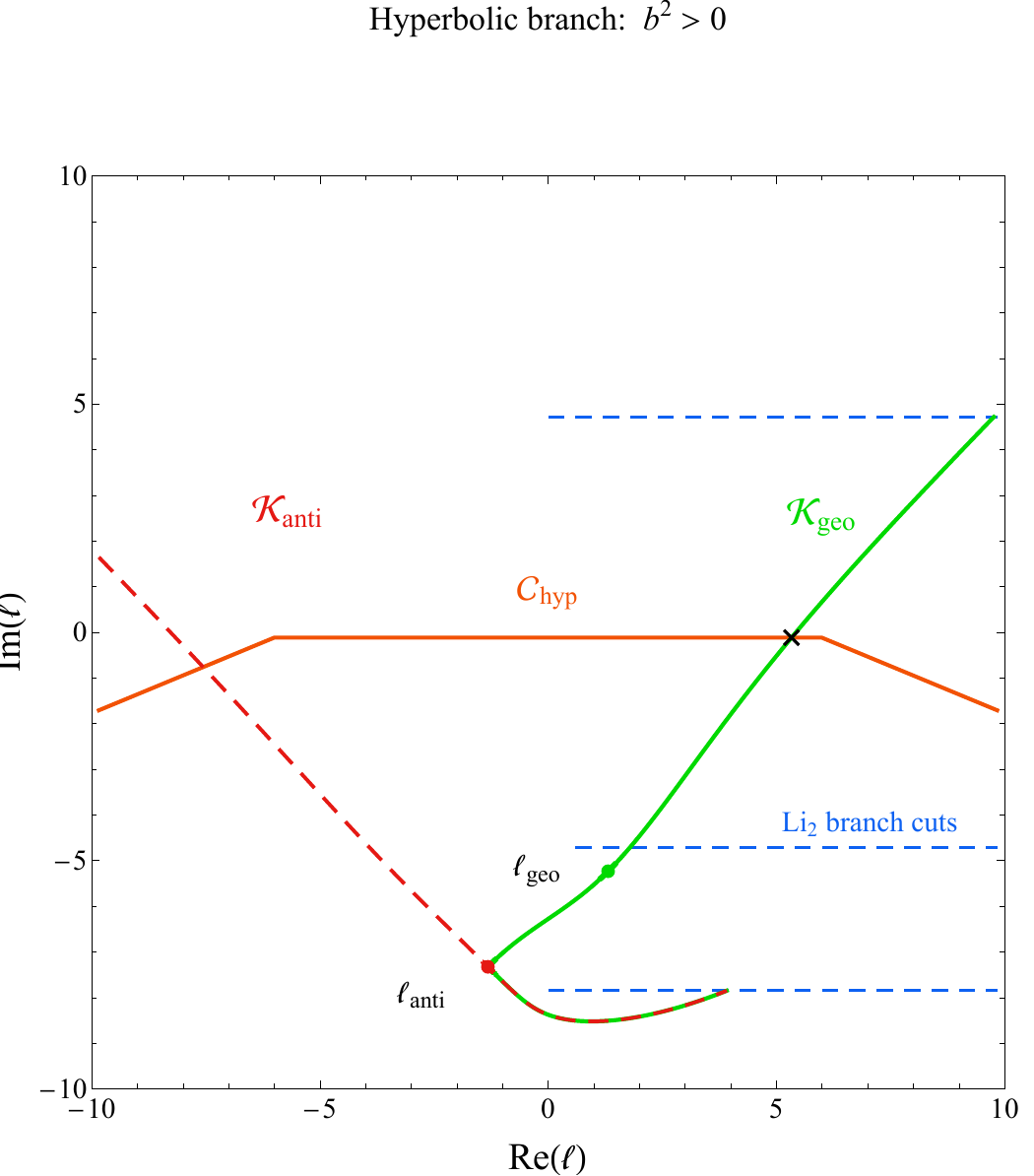}
    \caption{Lefschetz-thimble structure in the hyperbolic branch at $\alpha=\pi/2$. The orange curve is the integration contour $\mathcal C_{\rm hyp}$, while the green solid and red dashed curves are the upward thimbles $\mathcal K_{\rm geo}$ and $\mathcal K_{\rm anti}$ through the geometric and anti-geometric saddles. The blue dashed lines denote the $\text{Li}_2$ branch cuts at $\text{Im}(\ell)=\frac{3\pi}{2}, -\frac{3\pi}{2}, -\frac{5\pi}{2}$, $\text{Re}(\ell)>0$. Only $\mathcal K_{\rm geo}$ intersects the contour on the transported physical sheet, giving $n_{\rm geo}=1$ and $n_{\rm anti}=0$.}
    \label{fig:fig8_hyp_thimble}
\end{figure}

\subsubsection*{Saddlepoint analysis and Lefschetz thimbles}
Now, we turn to the saddlepoint analysis. Introduce the semiclassical variable, $y=\frac{\ell}{4\pi b}$ in terms of which the partition function becomes
\begin{equation}
    Z^{(b)}_{\rm Vir}[\mathbf{4}_1;\tfrac{i\alpha}{4\pi b}]\sim \int_{\mathcal{C}}\dd\ell \exp\left[-\frac{1}{\pi b^2}Y(\ell,\alpha)\right]\,.
\end{equation}
Keeping all contributions of $\order{b^{-2}}$ in the exponent and using the semiclassical expansion of the double sine given in appendix \ref{app:specialfunctions}, the full semiclassical potential is
\begin{equation}\label{eq:figure8-full-potential}
\begin{aligned}
Y(\ell,\alpha)
=-\frac{i}{4}\Big[&
2\text{Li}_2\!\left(-e^{\ell/2-i\alpha/2}\right)
+2\text{Li}_2\!\left(-e^{\ell/2+i\alpha/2}\right)\\
&\hspace{2cm}+\frac{\pi^2}{3}
+\frac{\ell^2}{4}
-\frac{\alpha^2}{4}
-\frac{3i\alpha\ell}{2}
+2\pi\alpha
\Big].
\end{aligned}
\end{equation}
In this potential, 
the dilogarithms (and the logarithms obtained by differentiating them) are evaluated
on the sheets transported along the analytic continuation, rather than reset to
their principal branches.

The saddle-point equation $\partial_\ell Y=0$ is obtained using the derivative of the dilogarithm given in \eqref{eq:derivdilog}. Exponentiating the saddle equation and
introducing $z=e^{\ell/2}$  gives the following quadratic equation: 
\begin{equation}
    z^2+B(\alpha)z+1=0, \qquad B(\alpha)=2\cos (\tfrac{\alpha}{2})- e^{-\frac{3i\alpha}{2}}\,.
\end{equation}
Its roots $z_\pm$ must be
assigned logarithmic lifts $\ell_\pm$ that satisfy the un-exponentiated equation on the
transported sheets.
The two solutions of the quadratic equation are
\begin{equation}
z_{\pm}(\alpha)
=
\frac{-B(\alpha)\pm\sqrt{B(\alpha)^2-4}}{2},
\qquad
z_{+}z_{-}=1.
\end{equation}
At the critical angle $\alpha_*=2\pi/3$, one has $B=2$, so the discriminant vanishes and the two saddles collide at $z=-1$. This is the degeneration to the Euclidean cone structure that separates the hyperbolic and spherical branches. The ordinary Gaussian saddle approximation also breaks down at this point.

At $\alpha =0$, we get $B=1$ and hence $z_\pm = e^{\pm 2\pi i/3}$. Then the on-shell action reduces to the well-known dilogarithm formula for the volume of the figure 8 knot complement in the complete hyperbolic metric \cite{Hikami_2007,Dijkgraaf:2010ur}.

For $0<\alpha<2\pi/3$ and $b^2>0$, the principal logarithmic lift of the algebraic root $z_-$ is what we will call the
`geometric' hyperbolic saddle,
\begin{equation}
    \ell_{\rm geo}(\alpha)=\ell_{-,\rm pr}(\alpha)\,.
\end{equation}
The companion `anti-geometric' saddle is not
the principal lift of the second algebraic root $z_+$. Since
$z=e^{\ell/2}$, logarithmic lifts are related by
$\ell\rightarrow\ell+4\pi i$, and the lift that is continuously connected
to the geometric saddle at the Euclidean degeneration $\alpha_*=\frac{2\pi}{3}$ is
\begin{equation}
    \ell_{\rm anti}(\alpha)
    =
    \ell_{+,\rm pr}(\alpha)-4\pi i .
\end{equation}
Indeed, as $\alpha\rightarrow(2\pi/3)^\pm$, the geometric and
anti-geometric lifts coalesce at $\ell=-2\pi i$.

To determine which of the saddles gets `picked up' by the $\ell$ contour integral, we perform a thimble analysis using Picard-Lefschetz theory. In the hyperbolic branch, illustrated in Figure \ref{fig:fig8_hyp_thimble}, the upward
thimble through $\ell_{\rm geo}$ intersects the transported contour once,
while that through $\ell_{\rm anti}$ does not. Thus the Stokes constants are
\begin{equation}
    n_{\rm geo}=1,\qquad n_{\rm anti}=0 .
\end{equation}
To evaluate the on-shell action $Y(\ell_{\rm geo},\alpha)$ for the geometric saddle, we need to be careful about the choice of branches of the dilogarithms in the semiclassical potential \eqref{eq:figure8-full-potential}. The branch cuts of the two dilogarithms are chosen to be along
\begin{equation}
    \text{Im}(\ell)=\pm \alpha+2(2k+1)\pi, \quad \text{Re}(\ell)>0, \quad k \in \mathbb{Z}\,.
\end{equation} 
As we increase $\alpha$ from zero, the geometric saddle crosses the branch cut at $\text{Im}(\ell)=\alpha-2\pi$, $\text{Re}(\ell)>0$ when the angle becomes $\alpha=\alpha_c\equiv \arccos(1/4)$. 
Continuing the geometric saddle from its cusp value ($\alpha=0$) of $\ell_{\rm geo}(0)=-\frac{4\pi i}{3}$ satisfied by the principal branches of both the dilogarithms,
one has 
\begin{equation} \label{eq:figure8_saddleaction_hyp}
    Y_{\rm geo}(\alpha)=\begin{cases}
        Y_{\rm pr}(\ell_{\rm geo}(\alpha),\alpha), \qquad & 0\leq\alpha<\alpha_c,\\
        Y_{\rm pr}(\ell_{\rm geo}(\alpha),\alpha)+ \frac{\pi}{2}\ell_{\rm geo}(\alpha)+i\pi^2-\frac{i\pi\alpha}{2}, \qquad &\alpha_c<\alpha < \frac{2\pi}{3}\,,
    \end{cases}
\end{equation}
where $Y_{\rm pr}$ denotes the potential evaluated with principal-branch dilogarithms. Similarly, for the antigeometric saddle, we get
\begin{equation}
    Y_{\rm anti}(\alpha)= Y_{\rm pr}(\ell_{\rm anti}(\alpha),\alpha)+ \frac{\pi}{2}\ell_{\rm anti}(\alpha)+i\pi^2-\frac{i\pi\alpha}{2}, \qquad 0<\alpha < \frac{2\pi}{3}\,.
\end{equation}

\begin{figure}
    \centering

    \begin{subfigure}[t]{0.49\textwidth}
        \centering
        \includegraphics[width=\linewidth]{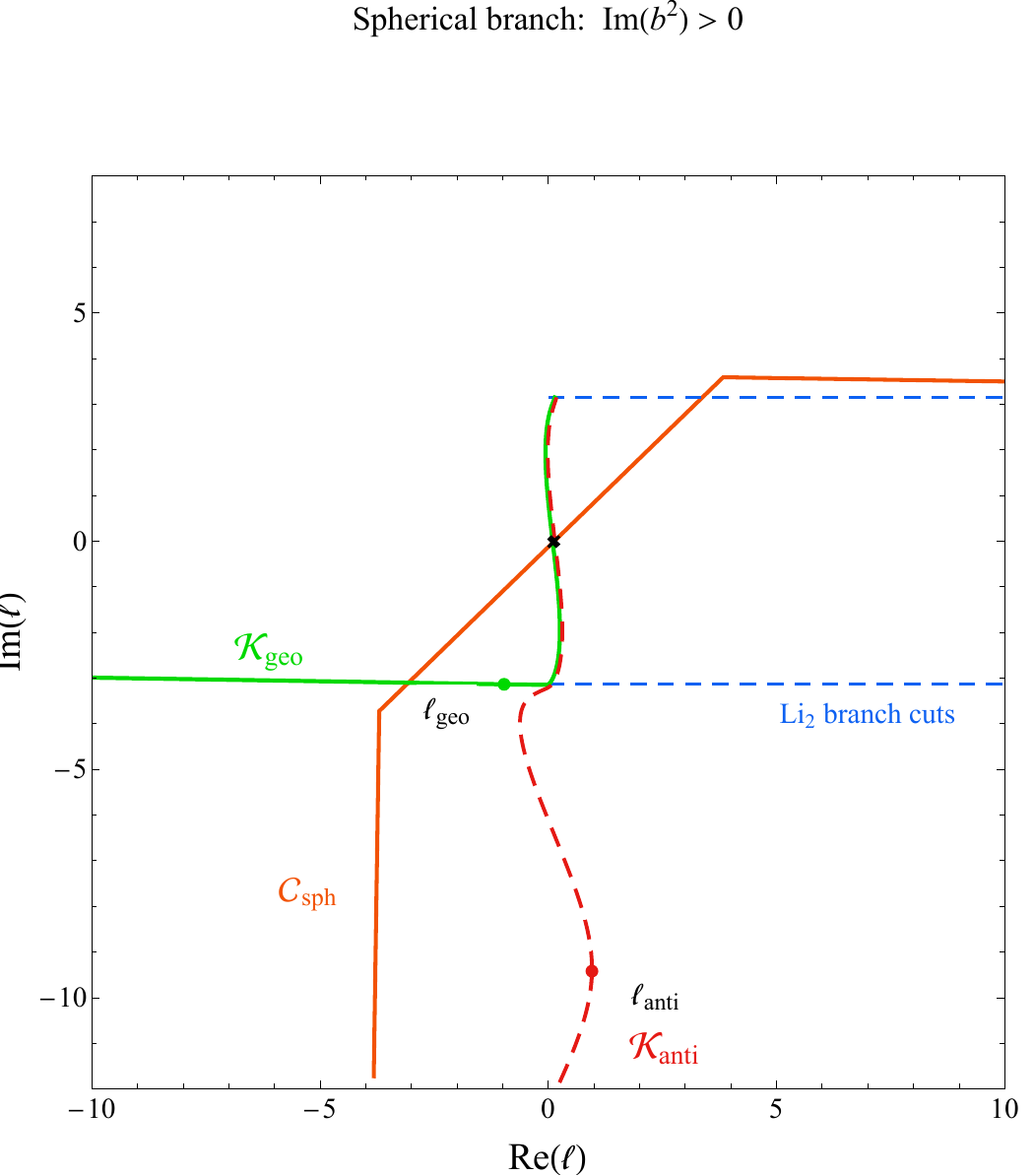}
        \caption{}
        \label{fig:fig8-thimble-spherical1}
    \end{subfigure}
    \hfill
    \begin{subfigure}[t]{0.49\textwidth}
        \centering
        \includegraphics[width=\linewidth]{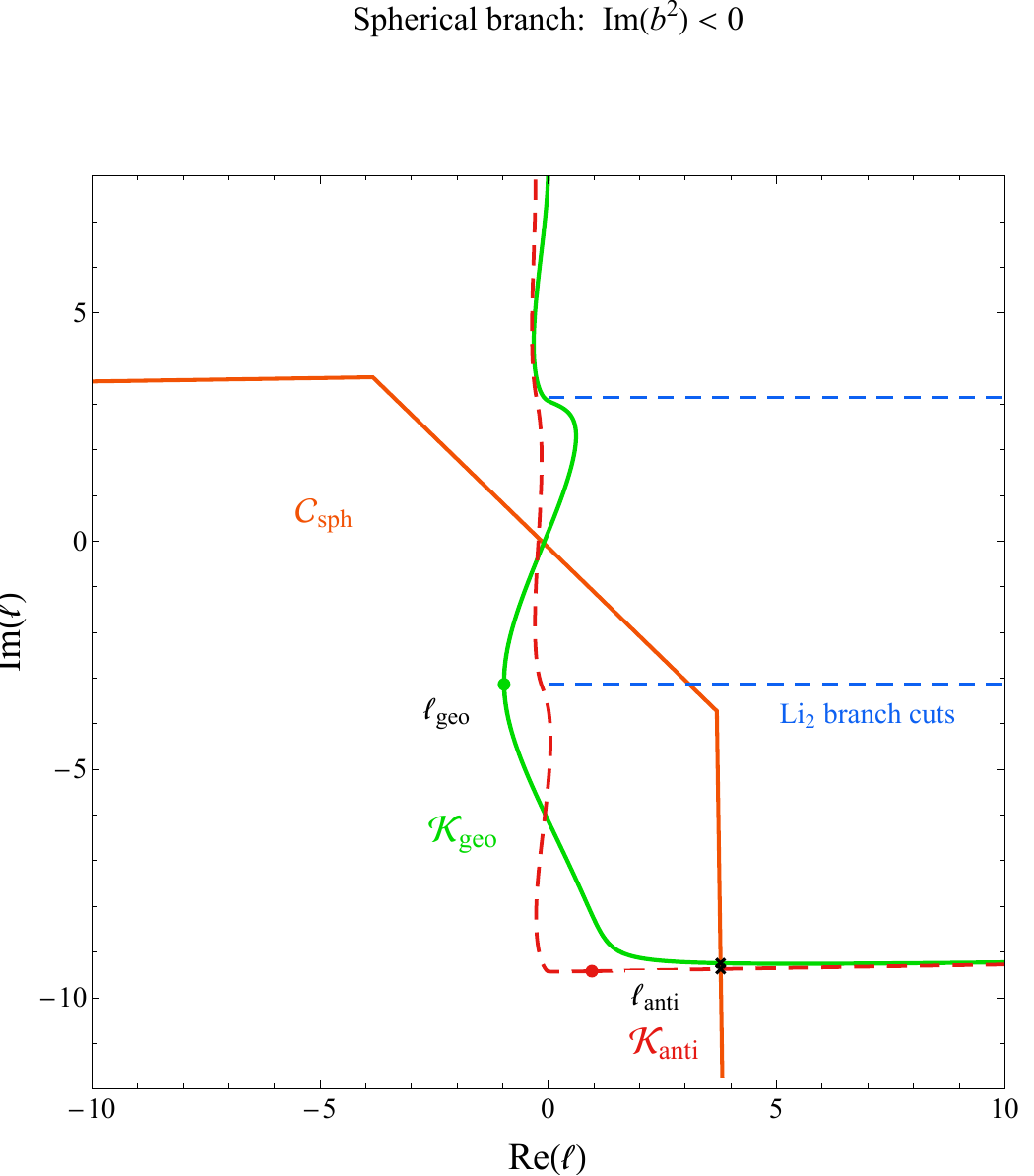}
        \caption{}
        \label{fig:fig8-thimble-spherical2}
    \end{subfigure}

    \caption{Lefschetz-thimble structure in the spherical branch at
$\alpha=\pi$ for (a) $\text{Im}(b^2)>0$ and
(b) $\text{Im}(b^2)<0$. The orange curves are the transported
contours $\mathcal C_{\rm sph}$, while the green solid and red dashed
curves are projections of the upward thimbles through the geometric and anti-geometric
lifts. The blue dashed lines denote the $\text{Li}_2$ branch
cuts at $\text{Im}(\ell)=\pm \pi$, $\text{Re}(\ell)>0$. In both lateral continuations, each thimble has one same-sheet
intersection with the contour, giving $n_{\rm geo}=n_{\rm anti}=1$. In plot (a), the nearly overlapping curves remain separated at finite lateral regulator. The two projected crossings in plot (b) occur on distinct logarithmic sheets, differing by \(2\pi i\) in \(\log(1+e^{\ell/2+i\alpha/2})\), and are not intersections of the lifted thimbles.}
    \label{fig:fig8-spherical-thimbles}
\end{figure}

The analysis readily generalizes for cone angles $\frac{2\pi}{3}<\alpha <\frac{4\pi}{3}$ in the spherical branch. Continuing the geometric saddle through the Euclidean transition ($|z_{\rm geo}|<1$) gives
\begin{equation} \label{eq:figure8_saddleaction_sph_geo}
    Y_{\rm geo}(\alpha)=
        Y_{\rm pr}(\ell_{\rm geo}(\alpha),\alpha)+ \frac{\pi}{2}\ell_{\rm geo}(\alpha)+i\pi^2-\frac{i\pi\alpha}{2}, \qquad \frac{2\pi}{3}<\alpha < \frac{4\pi}{3}\,.
\end{equation}
The anti-geometric saddle crosses a branch cut of the dilogarithm at $\alpha=\pi$ so the action has a piecewise form,
\begin{equation} \label{eq:figure8_saddleaction_sph_antigeo}
    Y_{\rm anti}(\alpha)=\begin{cases}
        Y_{\rm pr}(\ell_{\rm anti}(\alpha),\alpha)+\frac{\pi}{2}\ell_{\rm anti}(\alpha)+i\pi^2-\frac{i\pi\alpha}{2}, \qquad & \frac{2\pi}{3}<\alpha<\pi,\\
        Y_{\rm pr}(\ell_{\rm anti}(\alpha),\alpha)+\pi \ell_{\rm anti}(\alpha) + 2i\pi^2, \qquad &\pi<\alpha < \frac{4\pi}{3}\,.
    \end{cases}
\end{equation}
The above saddlepoint actions for the geometric and anti-geometric saddles imply the simple relation for all the angles in the spherical branch,
\begin{equation}
    Y_{\rm geo}(\alpha)=-Y_{\rm anti}(\alpha), \qquad \frac{2\pi}{3}<\alpha<\frac{4\pi}{3}\,.
\end{equation}

The integration contour, logarithmic sheets, and thimbles must be
transported together. As shown in Fig.~\ref{fig:fig8-spherical-thimbles}, for both
$\text{Im}(b^2)>0$ and $\text{Im}(b^2)<0$, the upward
thimbles through both $\ell_{\rm geo}$ and $\ell_{\rm anti}$ have one
same-sheet intersection with the integration contour. Thus the intersection numbers for the two saddles are\footnote{Projected intersections occurring after crossing onto a different logarithmic
sheet do not contribute.}
\begin{equation}
    n_{\rm geo}=1,\qquad n_{\rm anti}=1 .
\end{equation}

For illustration, consider
$\alpha=\pi$, for which the two algebraic roots 
are
\begin{equation}
    z_{\rm geo}=-\frac{i}{2}(\sqrt{5}-1),
    \qquad
    z_{\rm anti}=\frac{i}{2}(\sqrt{5}+1) .
\end{equation}
The relevant logarithmic lifts are 
\begin{equation}
    \ell_{\rm geo}
    =
    -2\log\left(\frac{\sqrt{5}+1}{2}\right)-i\pi,
    \qquad
    \ell_{\rm anti}
    =
    2\log\left(\frac{\sqrt{5}+1}{2}\right)-3i\pi .
\end{equation} 
Evaluating the full potential on the two transported
geometric sheets gives
\begin{equation}
    Y_{\rm geo}(\pi)
    =
    -\frac{i\pi^2}{10},
    \qquad
    Y_{\rm anti}(\pi)
    =
    +\frac{i\pi^2}{10},
\end{equation}
where we used the special values of the dilogarithm 
\begin{equation}
    \mathrm{Li}_2\left(\mfrac{\sqrt{5}-1}{2}\right) = \mfrac{\pi^2}{10}-\log^2\left(\mfrac{\sqrt{5}+1}{2}\right), \quad \mathrm{Li}_2\left(-\mfrac{\sqrt{5}-1}{2}\right) =- \mfrac{\pi^2}{15}+\frac{1}{2}\log^2\left(\mfrac{\sqrt{5}+1}{2}\right).
\end{equation}

Although both thimbles have nonzero intersection number in each lateral continuation, their real exponential weights are opposite. The geometric
saddle is exponentially growing for $\text{Im}(b^2)>0$, whereas
the anti-geometric saddle is exponentially growing for
$\text{Im}(b^2)<0$. Since $b_-^2 = -b_+^2$, with $\kappa = \frac{i}{b_+^2} >0$, the dominant growing contribution to the gravitational partition
function $ Z_{\rm grav}^{\Lambda>0} = Z_{+} Z_{-}$ therefore pairs the geometric saddle on the first ray with the
anti-geometric saddle on the conjugate ray,
\begin{equation}
    Z_{\rm grav}^{\Lambda>0}[\mathbf{4}_1;\alpha] 
    \sim
    \exp\left[
        -\frac{Y_{\rm geo}(\alpha)}{\pi b_+^2}
        -\frac{Y_{\rm anti}(\alpha)}{\pi b_-^2}
    \right]
    =
    \exp\left[
        \frac{\kappa}{\pi}
        \left(
            iY_{\rm geo}(\alpha)-iY_{\rm anti}(\alpha)
        \right)
    \right].
\end{equation}
At $\alpha=\pi$ this gives 
\begin{equation}
    Z^{\Lambda>0}_{\rm grav}[\mathbf{4}_1;\pi]
    \sim
    \exp\left[
        \frac{\text{Vol}(M_\pi)}
        {4\pi G_N\ell_{\rm dS}^2}
    \right], \qquad \text{Vol}(M_\pi)
    =
    \frac{\pi^2}{5}\ell_{\rm dS}^3.
\end{equation}
The volume agrees with the spherical volume computed in \cite{MednykhRasskazov:2006} for the figure 8 knot at $\alpha=\pi$, denoted by $M_\pi$. To see why this is the volume, use the fact that $M_\pi$ at this angle has a 2-fold branched cover that is the Lens space $L(5,2)$, whose universal cover (with 5 sheets) is $S^3$. Since the volume of $S^3$ is $2\pi^2\ell_{\rm dS}^3$, we get $\text{Vol}(M_\pi) = 2\pi^2\ell_{\rm dS}^3/10$.

For $\alpha\neq \pi$, we can also show that the on-shell action agrees with the figure-eight knot volume. The figure-eight cone manifold volumes (both hyperbolic and spherical) were obtained from length-angle identities and the Schläfli formula in~\cite{MednykhRasskazov:2006, Porti2004, tran2026exact}. In our cone-angle convention, these results read
\begin{equation}\label{eq:figure8-volumes-schlafli}
\frac{\text{Vol}_{\rm hyp}(\alpha)}{\ell_{\rm AdS}^{3}}
=
\int_{\alpha}^{2\pi/3}
d\theta\,
\cosh^{-1}\left(1+\cos\theta-\cos2\theta\right),
\qquad 0\leq\alpha<\frac{2\pi}{3},
\end{equation}
for the cone angles in the hyperbolic branch and 
\begin{equation} \label{eq:sphericalvolumes_figureeight}
    \frac{\text{Vol}_{\rm sph}(\alpha)}{\ell_{\rm dS}^{3}}
= \begin{cases}
\displaystyle\int_{2\pi/3}^{\alpha}
d\theta\,
\cos^{-1}\left(1+\cos\theta-\cos2\theta\right),\quad 
&\frac{2\pi}{3}<\alpha\leq\pi,\\[15pt]
\displaystyle 2\pi(\alpha-\pi)+\int_{2\pi/3}^{2\pi-\alpha}
d\theta\,
\cos^{-1}\left(1+\cos\theta-\cos2\theta\right),\quad &\pi<\alpha <\frac{4\pi}{3}\,,
\end{cases}
\end{equation}
for cone angles in the spherical branch.
Differentiating the saddle action (\ref{eq:figure8-full-potential}) for the geometric saddle on the transported sheets reproduces
the derivatives of these integrals,
\begin{equation}
    \begin{split}
         \frac{d Y_{\rm geo}}{d\alpha} = -\frac{1}{2}\cosh^{-1}\left(1+\cos\alpha-\cos2\alpha\right), \qquad 0\leq \alpha <\frac{2\pi}{3}\,,
    \end{split}
\end{equation}
in the hyperbolic branch and 
\begin{equation}
    \frac{d Y_{\rm geo}}{d\alpha} = \begin{cases}
        -\frac{i}{2}\cos^{-1}\left(1+\cos\alpha-\cos2\alpha\right), \qquad
&  \frac{2\pi}{3}<\alpha < \pi,\\
-i\pi + \frac{i}{2}\cos^{-1}\left(1+\cos\alpha-\cos2\alpha\right), \qquad & \pi <\alpha<\frac{4\pi}{3}\,,
    \end{cases}
\end{equation}
in the spherical branch.
The integration constant is fixed by setting the value of the potential at the Euclidean transition to be zero, $Y_{\rm geo}(\frac{2\pi}{3})=0$. 

In summary, we have shown that the hyperbolic and spherical volumes obtained from the saddlepoint analysis of the Virasoro TQFT partition function on the figure-eight knot cone manifolds are given by
\begin{equation} \label{eq:volumematchaction}
    \frac{\text{Vol}_{\rm hyp}(\alpha)}{\ell_{\rm AdS}^3}
    =
    2\,\text{Re}Y_{\rm geo}(\alpha),
    \qquad
    \frac{\text{Vol}_{\rm sph}(\alpha)}{\ell_{\rm dS}^3}
    =
       i Y_{\rm geo}(\alpha)-iY_{\rm anti}(\alpha)
    =
    -2\,\text{Im}Y_{\rm geo}(\alpha)\,,
\end{equation}
where $Y_{\rm geo}(\alpha)\equiv Y(\ell_{\rm geo}(\alpha),\alpha)$ is the saddlepoint action for the geometric saddle.
The agreement holds throughout these angle ranges. We have therefore verified the volume conjecture for the figure-eight knot cone manifolds.

\subsubsection*{A tetravalent graph from the figure-eight knot} \label{sec:Tetravalent}

We now analyze the upper endpoint $\alpha\to(4\pi/3)^-$ of the spherical branch of the figure-eight knot. Denoting the
limiting cone manifold by $M_{\Theta_4}$, our proposal becomes
\begin{align}
 Z_{\mathrm{grav}}^{\Lambda>0}[M_{\Theta_4}]
 &=Z_{\mathrm{Vir}}^{(b_+)}\!\left[\mathbf{4}_1;\tfrac{i}{3b_+}\right]
   Z_{\mathrm{Vir}}^{(b_-)}\!\left[\mathbf{4}_1;\tfrac{-i}{3b_-}\right] \\[1em]
 &=\left|\sqrt{2} e^{\frac{2\pi iQ_+}{3b_+}}
 \int_{\mathcal C_+}\!\dd y\,
 S_{b_+}\!\left(\tfrac{Q_+}{2}-iy\pm\tfrac{1}{3b_+}\right)
 e^{\frac{2\pi y}{b_+}}\right|^2 
 \label{eq:figure-eight-theta4-exact}
\end{align}
where $Q_+=b_++b_+^{-1}$. The modulus square follows from the reality property of the
figure-eight amplitude in the chosen framing~\cite{Collier:2024kmo}.
At this endpoint, the expressions in (\ref{eq:sphericalvolumes_figureeight}) and (\ref{eq:volumematchaction}) for the spherical volume give
\begin{equation} \label{eq:tetravol}
 \frac{\text{Vol}(M_{\Theta_4})}{\ell_{\mathrm{dS}}^3}
 =\lim_{\alpha\to \frac{4\pi}{3}^-}\bigl[-2\text{Im}\,Y_{\mathrm{geo}}(\alpha)\bigr]
=\frac{2\pi^2}{3}.
\end{equation}
In this limit, the `unknotting tunnels' between the two linked portions of the knot shrink to zero and the singular locus
becomes an unknotted tetravalent theta graph~\cite{Porti2004} (which is referred to as a `spherical suspension of a sphere with 4 cone points' in \cite{Porti2004}),
\begin{equation}
     \vcenter{\hbox{ \begin{tikzpicture} [scale=0.7,xscale=0.75]      
        \draw[very thick, out = 90, in = 180, black] (-5/2,-3/2) to (-1,0);
        \draw[very thick, out = 0, in = 180, black] (-1,0) to (0,0);
        \draw[fill=white,draw=white] (-1,0) circle (1/10);
        \draw[very thick, out = 180, in = 90, black] (0,1) to (-1,0);
\draw[very thick, out = 270, in = 150, black] (-1,0) to (0,-1);
        \draw[very thick, out = 330, in = 90, black] (0,-1) to (1,-2);
        \draw[fill=white,draw=white] (0,-1) circle (1/10);
\draw[very thick, out = 90, in = 270, black] (-1,-2) to (1,0);
        \draw[very thick, out = 90, in = 0, black] (1,0) to (0,1) ;
        \draw[fill=white,draw=white] (1,0) circle (1/10);
        \draw[very thick, out = 0, in = 180, black] (0,0) to (1,0) ;
        \draw[very thick, out = 0, in = 90, black] (1,0) to (5/2,-3/2);
        \draw[very thick, out = 270, in = 330, black] (5/2,-3/2) to (0,-3);
\draw[very thick, out = 150, in = 270, black] (0,-3) to (-1,-2);
        \draw[fill=white,draw=white] (0,-3) circle (1/5);
\draw[very thick, out = 270, in= 0, black] (1,-2) to (0,-3);
        \draw[very thick, out = 180, in = 270, black] (0,-3) to (-5/2,-3/2);
\end{tikzpicture}}} \qquad 
\xrightarrow[]{\;\;\alpha \to \frac{4\pi}{3}\;\;} \qquad
\vcenter{\hbox{
\begin{tikzpicture}[scale=0.8]
    \draw[very thick, black] (-2,0)
    node[circle, fill=black, inner sep=1.5pt]{}
    to[bend left = 70,looseness=1.4] (2,0) 
    node[circle, fill=black, inner sep=1.5pt]{};
    \draw[very thick, black] (-2,0) to[bend right = 70,looseness=1.4] (2,0);
    \draw[very thick, black] (-2,0) to[bend left = 40] (2,0);
    \draw[very thick, black] (-2,0) to[bend right = 40] (2,0);
\end{tikzpicture}
}}
\end{equation}
The limiting
geometry can be explicitly described by the metric,
\begin{equation}
 ds_E^2=\ell_{\mathrm{dS}}^2
 \left(d\tau^2+\cos^2\tau\,d\widetilde\Sigma_{0,4}^2\right),
 \qquad -\frac{\pi}{2}\leq\tau\leq\frac{\pi}{2},
\end{equation}
where $d\widetilde\Sigma_{0,4}^2$ is the unit-curvature metric on the
four-punctured sphere selected by this degeneration, with cone angle
$4\pi/3$ at every puncture. Gauss--Bonnet therefore gives
\begin{equation}
 \text{Vol}(M_{\Theta_4})
 =\frac{\pi\ell_{\mathrm{dS}}^3}{2}
 \text{Area}(\widetilde\Sigma_{0,4})
 =\frac{2\pi^2\ell_{\mathrm{dS}}^3}{3}\,,
\end{equation}
in agreement with (\ref{eq:tetravol}). To completely determine the geometry of the tetravalent theta graph, we also need to specify the cross-ratio between the 4 cone points on $\Tilde{\Sigma}_{0,4}$. To this end, note that the rotational holonomy group around the cone points is the tetrahedral group $A_4$ which is a finite group of order $12$. This follows from the tetrahedral rotational holonomy of the Euclidean figure-eight
cone manifold ($\alpha=\frac{2\pi}{3}$) shown in \cite{Hilden1995Polyhedron}, together with the
reflected-angle continuation ($\alpha \to 2\pi-\alpha$) of \cite{Porti2004} to the spherical endpoint at $\alpha=\frac{4\pi}{3}$. This means we can write a single-valued map composing the multi-valued developing map into the round $S^2$ with the tetrahedral quotient, 
\begin{equation}
    F: \widetilde\Sigma_{0,4} \to S^2/A_4 \,.
\end{equation}
The quotient is isometric to a thrice-punctured sphere $\Tilde{\Sigma}_{0,3}$ with cone angles $\{\frac{2\pi}{3}, \frac{2\pi}{3}, \pi\}$. Since $F$ pulls back the quotient metric on $\Tilde{\Sigma}_{0,3}$ to the metric on
$\widetilde\Sigma_{0,4}$, its degree equals the ratio of their areas,
$(4\pi/3)/(\pi/3)=4$. Thus $F$ is a degree-four branched covering. The locations of the three cone points on the auxiliary $\Tilde{\Sigma}_{0,3}$ can be chosen to be at $0,1,\infty$ without loss of generality, with the cone angles at $0,1$ being $\frac{2\pi}{3}$ and the cone angle at $\infty$ being $\pi$. Since all the cone angles in $\Tilde{\Sigma}_{0,4}$ are $\frac{4\pi}{3}$, the cone points at $0,1$ in $\Tilde{\Sigma}_{0,3}$ should each have two preimages under $F$ of local degree $2$ into two cone points in $\Tilde{\Sigma}_{0,4}$. The cone point at $\infty$ in $\Tilde{\Sigma}_{0,3}$ has a smooth preimage (since the cone angle of $\pi$ doubles to $2\pi$) of local degree 2 at two points in $\Tilde{\Sigma}_{0,4}$ which we place at $z=0,\infty$ using a M\"obius transformation. These conditions fix the map $F$ to be
\begin{equation}
    F(z)=\frac{1}{4}\left(z+\frac{1}{z}\right)^2\,.
\end{equation}
We can now deduce the locations of the 4 cone points on $\Tilde{\Sigma}_{0,4}$ by solving $F(z)=0$ and $F(z)=1$ giving $\{\pm 1, \pm i\}$. The cross-ratio between these points is $1/2$ upto relabelling of the four cone points. We have thereby determined the complete geometry of the tetravalent theta graph.

The geometry of the tetravalent theta graph described above is reflection-symmetric under $\tau \to -\tau$. We can therefore continue it to Lorentzian time. Cutting along the totally geodesic slice $\tau=0$ and continuing to Lorentzian time $\tau=\frac{it}{\ell_{\mathrm{dS}}}$ gives
\begin{equation}
 ds_L^2=-dt^2+\ell_{\mathrm{dS}}^2
 \cosh^2\!\left(\frac{t}{\ell_{\mathrm{dS}}}\right)
 d\widetilde\Sigma_{0,4}^2.
\end{equation}
For $t\geq0$, this describes an expanding closed de Sitter cosmology
sourced by four comoving particles of equal proper mass
$m=\frac{1}{12G_N}$.
Their relative configuration is determined by the cross-ratio determined above and their proper separations
on these geometrically selected slices are coordinate-independent and
grow as $\cosh(t/\ell_{\mathrm{dS}})$. On the $t=0$ slice, there are four extremal closed curves separating each pair of cone points. Two of the curves have proper lengths $\pi \ell_{\rm dS}$ and the other two have proper lengths $\frac{4\pi \ell_{\rm dS}}{3}$. These curves are maximal under uniform normal displacements within the spatial slice and are minimal under uniform temporal displacements.
Reflection-symmetry also suggests interpreting
$Z_{\mathrm{grav}}^{\Lambda>0}[M_{\Theta_4}]$ as the norm-squared of a state
on the four-punctured sphere prepared by one Euclidean half.
This is consistent with \eqref{eq:figure-eight-theta4-exact} being manifestly non-negative.

\subsection{The three-twist knot cone manifolds}\label{sec:threetwist}

The next example that we will study is the three-twist knot.
The three-twist knot, denoted $\textbf{5}_2$ in the Alexander-Briggs notation, is a chiral knot
with 5 crossings:
\begin{equation}
\vcenter{\hbox{
\begin{tikzpicture}[scale=0.75,xscale=-1]
\draw[very thick] (-1.41,0.73)
    to[out=145.16,in=-161.33,out distance=0.37cm,in distance=0.5cm] (-1.1,1.78)
    to[out=18.68,in=136.63,out distance=0.5cm,in distance=1.28cm] (1.04,0.83)
    ;
\draw[very thick] (-1.08,0.48)
    to[out=-22.13,in=39.18,out distance=1.1cm,in distance=0.89cm] (0.09,-1.54)
    ;
\draw[very thick] (-0.15,1.46)
    to[out=-146.63,in=101.96,out distance=2.05cm,in distance=0.59cm] (-1.91,-1.01)
    to[out=-78,in=-162.36,out distance=0.59cm,in distance=0.9cm] (-0.21,-1.78)
    ;
\draw[very thick] (0.32,1.7)
    to[out=26.06,in=96.99,out distance=0.76cm,in distance=0.5cm] (1.59,1.28)
    to[out=-82.97,in=38.75,out distance=0.5cm,in distance=0.92cm] (0.15,-0.18);
\draw[very thick] (1.39,0.53)
    to[out=-46.98,in=66.2,out distance=0.98cm,in distance=0.3cm] (1.85,-1.22)
    to[out=-113.77,in=4.28,out distance=0.3cm,in distance=0.64cm] (0.86,-1.9)
    to[out=-175.72,in=-123.5,out distance=0.64cm,in distance=1.12cm] (-0.14,-0.43)
    node[right=41pt,black] {$P_0$};
\end{tikzpicture}}}
\end{equation}
Let $P_0 = \frac{i\alpha}{4\pi b}$. The Virasoro TQFT partition function computed in \cite{Chandra:2025fef} is\footnote{The knot shown above is the mirror image of the corresponding knot in \cite{Chandra:2025fef}. The two differ by a complex conjugation of the Virasoro TQFT partition function.} 
\begin{equation}\label{eq:VTQFT1}
    Z^{(b)}_{\rm Vir}[\mathbf{5}_2;P_0] = \frac{e^{6\pi i h_0}}{\rho_0(P_0)}\int_0^{\infty}\dd P \rho_0(P)e^{-3\pi i h_P}\mathbb{S}^*_{P_0,P_0}[P].
\end{equation}
We can insert the Teschner-Vartanov integral representation of the S-kernel \cite{Teschner:2012em} (with the correct phase $e^{\frac{i\pi}{2}h_P}$ clarified in \cite{Eberhardt:2023mrq}), and use double sine integral identities to arrive at the following form (we have verified this numerically):
\begin{equation}\label{eq:52integral}
     Z^{(b)}_{\rm Vir}[\mathbf{5}_2;P_0] = \sqrt{2}e^{\frac{\pi i}{24}(7Q^2-2)}e^{2\pi i P_0^2}\int_{\mathbb{R}-i\epsilon}\dd z\,S_b(\tfrac{Q}{2}-iz\pm 2iP_0)S_b(\tfrac{Q}{2}-iz)e^{-\frac{\pi i}{2}z^2}.
\end{equation}
Up to an overall framing factor,
this agrees with the Teichm\"uller TQFT partition function on the $\mathbf{5}_2$ complement \cite{EllegaardAndersen:2011vps}, if we translate notations.\footnote{In particular, use the relation between $S_b$ and the quantum dilogarithm \eqref{eq:quantumdilog}, and set $x=2P_0$. The relative framing factor $e^{\frac{5\pi i}{12}Q^2+10\pi i P_0^2}$ drops out of the chiral-antichiral product defining $Z_{\rm grav}$.} 

We will now show that the integral \eqref{eq:52integral} realizes the hyperbolic-to-spherical transition of the $\mathbf{5}_2$ knot in the semiclassical approximation. The novelty compared to the previous section is that multiple saddle-points contribute: a Lefschetz-thimble analysis of the two
conjugate analytic continuations shows that the $Z^{(b_+)}_{\rm Vir}$
factor contains the geometric saddle together with an additional non-geometric
saddle, while the $Z^{(b_-)}_{\rm Vir}$ factor contains the anti-geometric
saddle only. The non-geometric contribution is exponentially subleading in the
semiclassical limit, so the leading product is controlled by the difference
between the geometric and anti-geometric actions, which reproduces the spherical volume computed using the Schl\"afli formula.

Let us first work with the hyperbolic branch so we take $b$ real and positive. To check that the horizontal contour for $z$ is convergent, write $s_b(z) = S_b(\tfrac{Q}{2}-iz)$ as before and use the asymptotics \eqref{eq:sbasymptotics} to derive the large $z$ behaviour of the integrand $I(z;P_0)$ of \eqref{eq:52integral},
\begin{equation}
    I(z;P_0) \sim \begin{cases}
        e^{i\pi z^2}, \quad & \text{Re}\,z \to -\infty,\\
        e^{-2\pi  iz^2}, \quad & \text{Re}\,z \to \infty\,.
    \end{cases}
\end{equation}
Unlike the figure-eight case, there is no term linear in \(z\), so the tails of \(\mathbb R-i\epsilon\) remain convergent at fixed \(P_0\), including along \(|\arg b|\leq\pi/4\). Pole avoidance needs to be checked separately. There are three families of poles coming from the three double sine functions,
\begin{equation}
    \begin{split}
        z^{+}_{m,n} &=-2P_0- \frac{iQ}{2}-imb-inb^{-1},\\
        z_{m,n} &=-\frac{iQ}{2}-imb-inb^{-1},\\
        z^{-}_{m,n} &=2P_0-\frac{iQ}{2}-imb-inb^{-1}\,
    \end{split}
\end{equation}
where $m,n \in \mathbb{Z}_{\geq 0}$. For real \(b\), these pole families are downward vertical arrays. For complex \(b\), their increments \(-imb-inb^{-1}\) lie in a downward wedge. The origins of the arrays move with \(P_0\) and \(b\), so the downward orientation alone does not guarantee that the poles remain below the real axis. The horizontal contour \(\mathbb R-i\epsilon\) is admissible whenever
\[
0<\epsilon<\frac{\text{Re}Q}{2}-2|\text{Im}P_0|.
\]
Along \(b=\kappa^{-1/2}e^{i\eta\phi}\), \(P_0=\eta i\alpha/(4\pi b)\), with \(\eta=\pm1\) and \(0\leq\phi\leq\pi/4\), this condition can always be satisfied for \(0\leq\alpha\leq\pi\). At larger angles where it fails, we transport \(\mathcal C_{52}\) with compact upward deformations that keep the moving poles below it, while retaining the asymptotic tails \(\mathbb R-i\epsilon\). No poles are crossed by the transported contour itself.

Having determined the integration contour, we will now perform the saddlepoint analysis. We take the semiclassical limit of \eqref{eq:52integral} with the following scaling,
\begin{equation}
    z=\frac{Z}{2\pi b}, \quad P_0=\frac{i\alpha}{4\pi b}\,.
\end{equation}
Using the semiclassical limit of the double sine function \eqref{eq:form2}, 
 we find
\begin{equation}\label{eq:semiclassical52}
     Z_{\rm Vir}^{(b)}[\mathbf{5}_2]\sim \int \dd Z \exp\left[\frac{i}{2\pi b^2}W(Z,\alpha)\right]\,,
\end{equation}
with the semiclassical potential given by\footnote{We have removed the terms arising from the framing factors since they cancel from the chiral-antichiral pairing.}
\begin{equation}
    W(Z,\alpha) = \frac{1}{2}(Z^2+\alpha^2)+\text{Li}_2(-e^{Z+i\alpha})+\text{Li}_2(-e^Z)+\text{Li}_2(-e^{Z-i\alpha})\,.
\end{equation}
To determine the saddlepoint equation, we use $
    \frac{d}{du}\text{Li}_2(-e^u)=-\log(1+e^u) \,,
$
giving
\begin{equation}
    e^Z=(1+e^{Z+i\alpha})(1+e^Z)(1+e^{Z-i\alpha})\,.
\end{equation}
Defining $r=e^Z$, we get the following cubic equation,
\begin{equation} \label{eq:AKcubic}
    r^3+(1+2\cos\alpha)r^2+2r\cos \alpha +1=0\,.
\end{equation}
The discriminant of the above cubic equation is given by
\begin{equation}
    \Delta (A)=A^4-6A^3 + 7A^2 + 6A -31, \qquad A=2\cos \alpha \,.
\end{equation}
The hyperbolic-to-spherical transition occurs when the discriminant vanishes.  In the range $A\in [-2,2]$, the only cone angle at which this happens is
\begin{equation}
    \alpha_* =\cos^{-1}\left(\frac{3-\sqrt{13+16\sqrt{2}}}{4}\right)\approx 2.40717\,.
\end{equation}
This agrees with the known geometric transition angle \cite{Mednykh2021,tran2026exact}.
For angles $0\leq \alpha <\alpha_*$, the cone manifold is hyperbolic; at $\alpha=\alpha_*$, it is Euclidean; and for $\alpha>\alpha_*$, it is spherical. In the hyperbolic branch, $\Delta(A)<0$ so there is one real and one pair of complex conjugate roots. The root with $\text{Im}(r_{\rm geo})<0$ will be called the geometric saddle, and its complex conjugate the anti-geometric saddle. At $\alpha=\alpha_*$, the complex conjugate pair collides and $\Delta(A)=0$. For $\alpha>\alpha_*$, $\Delta(A)>0$ so all three roots are real. After the continuation into the spherical branch, the root corresponding to the geometric saddle becomes the larger positive root.

\begin{figure}
    \centering
    \includegraphics[width=0.6\linewidth]{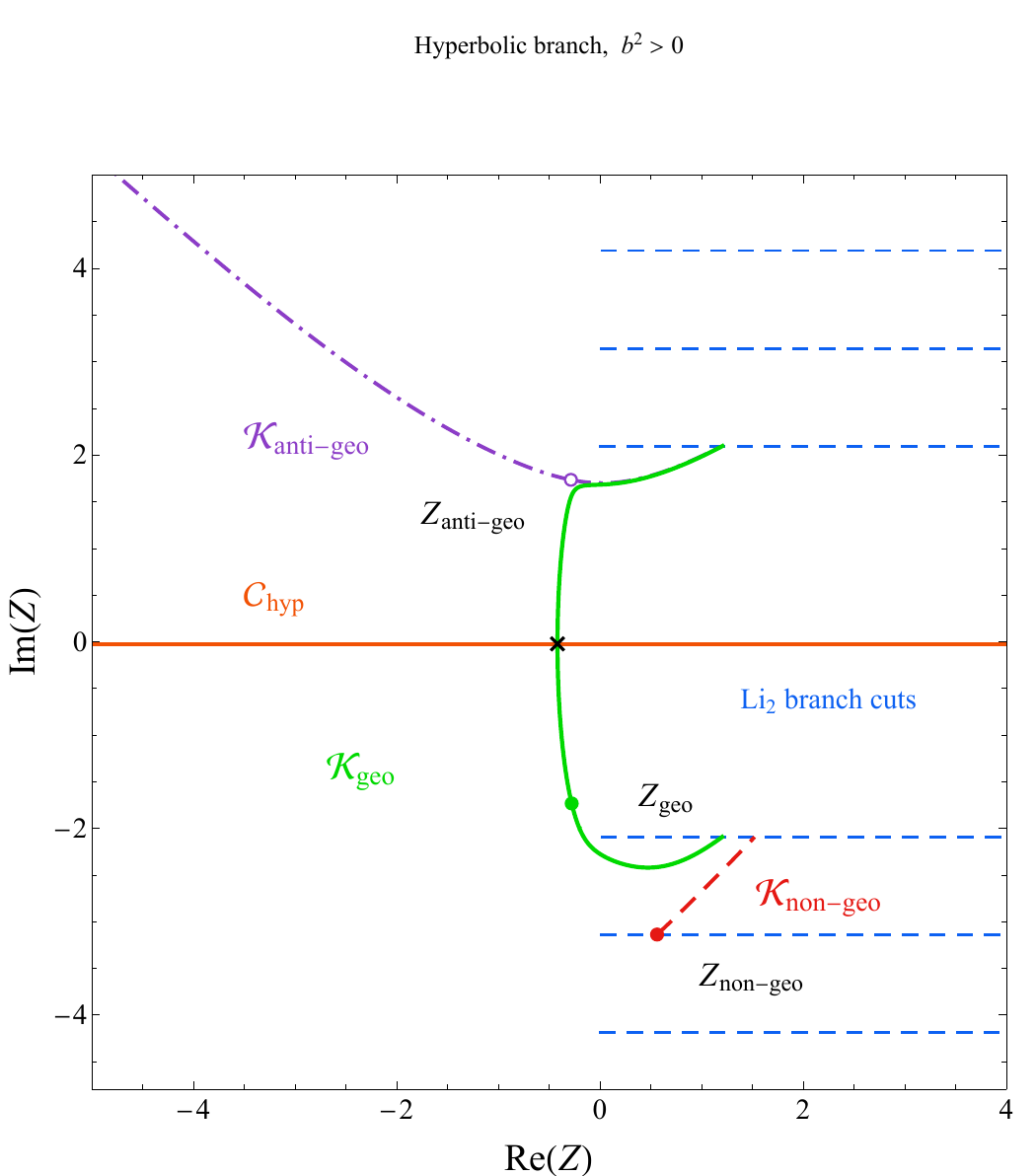}
    \caption{Lefschetz-thimble structure of the semiclassical integral \eqref{eq:semiclassical52} for the $\mathbf{5}_2$ knot in the hyperbolic branch, shown at
$\alpha=\pi/3$ and $b^2>0$. The orange curve is the physical integration
contour, while the green, purple, and red curves are the upward thimbles through
the geometric, anti-geometric, and non-geometric saddles, respectively. The blue
dashed lines denote the $\Li_2$ branch cuts at $\text{Im}(Z)=\pm \frac{2\pi}{3}, \pm \pi, \pm \frac{4\pi}{3}$, $\text{Re}(Z)>0$. Only the geometric upward thimble
has a same-sheet intersection with the integration contour, giving
$(n_{\rm geo},n_{\rm anti},n_{\rm non})=(1,0,0)$.}
    \label{fig:tt_hyperbolic}
\end{figure}

As a sanity check, we reproduce the known volume of the $\mathbf{5}_2$-knot complement when $\alpha=0$ i.e., in the cusp limit. The cubic saddlepoint equation in this limit becomes
\begin{equation}
    r^3+3r^2+2r+1=0 \,.
\end{equation}
The real root is not geometric.
The two complex roots given by
\begin{equation}
    r_{\rm geo} \approx -0.34 - 0.56 i, \quad r_{\rm anti} \approx -0.34 + 0.56 i
\end{equation}
correspond to the pair of geometric saddles related to each other by orientation reversal (since the knot is chiral). The saddlepoint action for these two roots is 
\begin{equation}
    W_{\rm cusp}(Z=\log r,\alpha=0) = \frac{1}{2}(\log r)^2+ 3 \text{Li}_2(-r)\approx-1.38 \pm 2.83i\,.
\end{equation}
The volume of the knot complement is the magnitude of the imaginary part of either saddle,
\begin{equation}
    \text{Vol}_{\text{hyp}}(\alpha=0)=\left|\Im\, W_{\rm cusp}\right|\approx 2.83
\end{equation}
which is in agreement with the known value from the literature \cite{Hikami_2007}. Equivalently, the volume can be read off from the difference in the geometric and anti-geometric actions $\frac{1}{2i}\left(W_{\rm geo}-W_{\rm anti}\right)$. This is important because once we cross the transition point, the actions for the (anti-)geometric saddles will become real and we will have to read off the spherical volume of the cone manifold at that angle using the difference in their actions, as dictated by the Lefschetz-thimble analysis.

A Lefschetz-thimble analysis (illustrated in figure \ref{fig:tt_hyperbolic} for $\alpha=\pi/3$) gives the following intersection numbers for the upward thimbles of the three saddles with the integration contour in the hyperbolic branch,
\begin{equation}
\bigl(n_{\rm geo},n_{\rm anti},n_{\rm non}\bigr)_{\rm hyp}
=(1,0,0),
\qquad
0\leq\alpha<\alpha_* .
\end{equation}
Thus only the geometric saddle contributes throughout the hyperbolic branch.
Although the geometric and non-geometric saddles undergo a Stokes
alignment at an intermediate angle, the intersection number of the thimble for the non-geometric saddle remains zero; the Stokes connection
therefore does not change the decomposition of the physical integration cycle. The semiclassical potential on the geometric saddle is to be evaluated on the principal branches of the three dilogarithms,
\begin{equation}
    W_{\rm geo}(\alpha)=W_{\rm pr}(Z_{\rm geo}(\alpha),\alpha), \quad 0\leq \alpha <\alpha_*.
\end{equation}
The hyperbolic volume of the cone manifold can be read off from the imaginary part of the above semiclassical potential at the geometric saddle.

Now, we turn to the spherical branch. Although the spherical branch covers the angle range $\alpha_*<\alpha<2\pi-\alpha_*$, we restrict to $\alpha_*<\alpha<\pi$ for the present discussion. In this restricted range, there is no crossing of the saddles across the branch cuts of the dilogarithms. The analysis for $\alpha>\pi$ can be done similarly accounting for the branch-cut crossing. As a first special example, let us choose the cone angle to be $\alpha=\pi^-$. The 3 real roots of the cubic saddle-point equation \eqref{eq:AKcubic} at this angle are 
\begin{equation}
    r_{\rm geo}=2\cos(\frac{\pi}{7}),\quad  r_{\rm anti}=2\cos(\frac{3\pi}{7}), \quad r_{\rm non} =2\cos(\frac{5\pi}{7})\,.
\end{equation}
The first two roots correspond to the geometric and anti-geometric saddles analytically continued past the Euclidean transition point, while the third root corresponds to the non-geometric saddle.
The corresponding classical actions at these values are
\begin{equation}
    W = \frac{37\pi^2}{42}, \quad \frac{25\pi^2}{42}, \quad \frac{\pi^2}{42}\,.
\end{equation}
The spherical volume of the cone manifold at this angle is given by the difference in actions of the two geometric saddles,
\begin{equation}
    \text{Vol}_{\rm sph}(\alpha=\pi)=\frac{1}{2}\left(\frac{37\pi^2}{42}-\frac{25\pi^2}{42}\right)=\frac{\pi^2}{7}\,,
\end{equation}
which is in agreement with the volume of the cone manifold whose two-fold branched cover is the lens space $L(7,3)$. 

\begin{figure}
    \centering

    \begin{subfigure}[t]{0.45\textwidth}
        \centering
        \includegraphics[width=\linewidth]{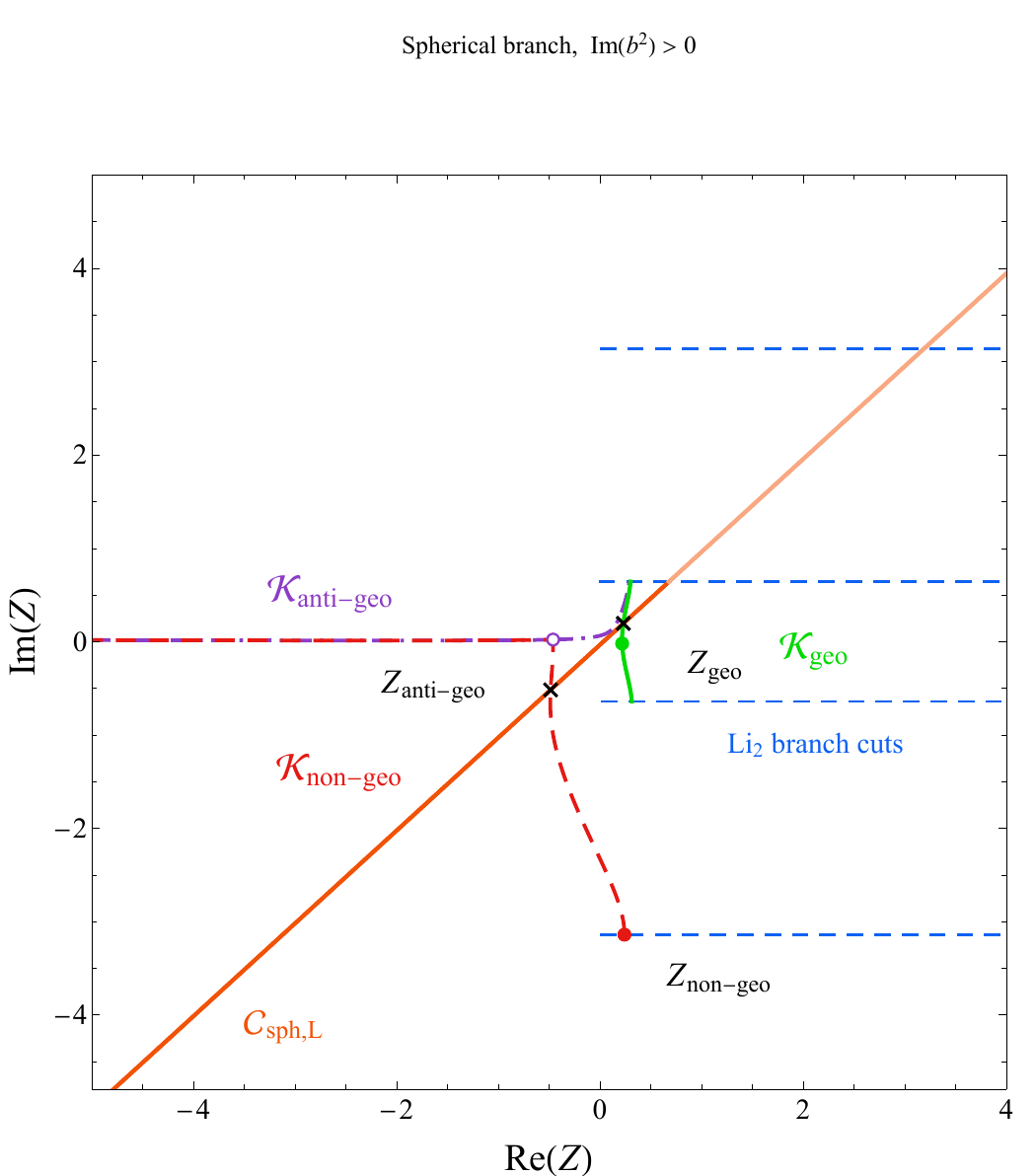}
        \caption{}
        \label{fig:tt-thimble-hyperbolic}
    \end{subfigure}
    \hfill
    \begin{subfigure}[t]{0.45\textwidth}
        \centering
        \includegraphics[width=\linewidth]{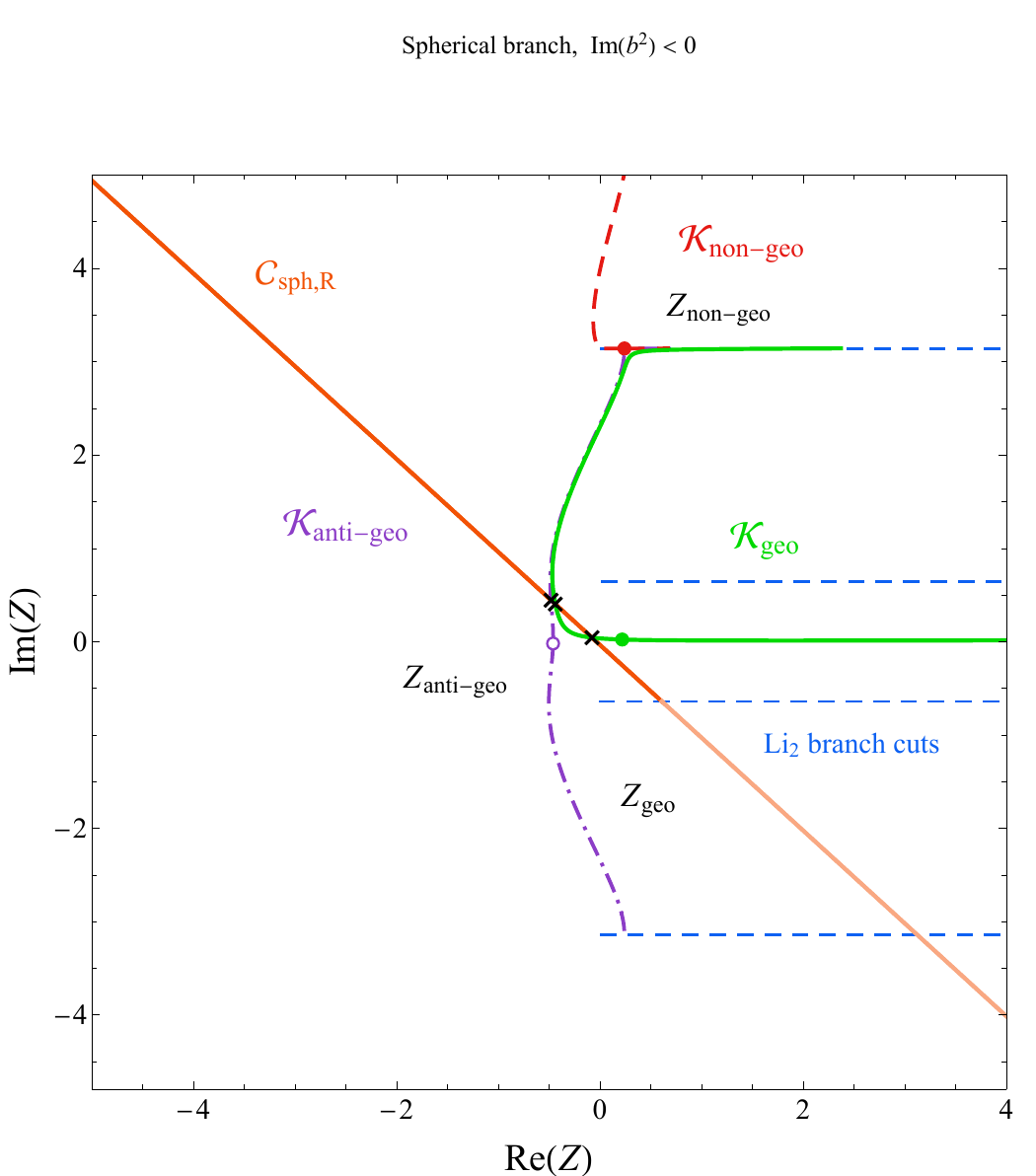}
        \caption{}
        \label{fig:tt-thimble-spherical}
    \end{subfigure}
\caption{Lefschetz-thimble structure of the semiclassical Virasoro TQFT partition function for the $\mathbf{5}_2$ knot in the spherical branch at $\alpha=2.5$.
The left and right panels show the conjugate lateral continuations with
$\text{Im}(b^2)>0$ and $\text{Im}(b^2)<0$, respectively.
The orange curves are the transported integration contours; the green, purple,
and red curves are the upward thimbles through the geometric, anti-geometric,
and non-geometric saddles; and the blue dashed lines denote the $\Li_2$ branch
cuts at $\text{Im}(Z)=\pm (\pi-2.5), \pm \pi$, $\text{Re}(Z)>0$. Black crosses mark same-sheet intersections with the physical contour.
The signed Lefschetz coefficients are
$(n_{\rm geo},n_{\rm anti},n_{\rm non})_L=(1,0,1)$ and
$(n_{\rm geo},n_{\rm anti},n_{\rm non})_R=(0,1,0)$; additional visible
intersections may occur in cancelling pairs and therefore do not change these
coefficients.}
    \label{fig:tt-thimbles}
\end{figure}

The origin of this difference is most transparent in the two chiral factors.
Let $g_{\rm geo}$ and $g_{\rm anti}$ denote the geometric and
orientation-reversed geometric saddles, and let $g_{\rm non}$ denote the third,
non-geometric saddle of the cubic equation. Their logarithmic and dilogarithmic sheets
are fixed by continuation from the hyperbolic branch. We transport the
integration contours and the full lifted actions along a pair of conjugate
lateral paths ending at
$\arg b_+=\pi/4-0^+$ and $\arg b_-=-\pi/4+0^+$, so that
\begin{equation}
 \frac{1}{b_-^2}=i\kappa,
 \qquad
 \frac{1}{b_+^2}=-i\kappa,
 \qquad
 \kappa\sim\frac{\ell_{\rm dS}}{4G_N}.
\end{equation}
For the transported cycles, the signed Lefschetz intersection numbers in the
spherical branch are
\begin{equation}
 \bigl(n_{\rm geo},n_{\rm anti},n_{\rm non}\bigr)_+=(1,0,1),
 \qquad
 \bigl(n_{\rm geo},n_{\rm anti},n_{\rm non}\bigr)_-=(0,1,0).
\end{equation}
We have checked this pattern numerically at representative angles throughout
$\alpha_*<\alpha<\pi$, while transporting the logarithmic sheets and the full
dilogarithmic action. See the plot in Fig.~\ref{fig:tt-thimble-spherical} for illustration of this pattern with the thimbles drawn for the saddles at the cone angle $\alpha=2.5$. These are signed intersection numbers: additional
same-sheet crossings can occur in cancelling pairs without changing the
coefficients above. Apparent $2\pi i$ translates of the cubic roots do not
supply further contributions to the transported contour once the full action
sheet is taken into account.

It follows that, including only the exponential dependence explicitly, the two
chiral factors have the semiclassical form
\begin{equation}
\begin{split}
 Z_{\text{Vir}}^{(b_+)}[\mathbf{5}_2] &\sim
 A_{{\rm geo}}^{(b_+)}
 \exp\!\left[\frac{\kappa}{2\pi}W_{\rm geo}\right]
 +
 A_{{\rm non}}^{(b_+)}
 \exp\!\left[\frac{\kappa}{2\pi}W_{\rm non}\right],
 \\
 Z_{\text{Vir}}^{(b_-)}[\mathbf{5}_2] &\sim
 A_{{\rm anti}}^{(b_-)}
 \exp\!\left[-\frac{\kappa}{2\pi}W_{\rm anti}\right],
 \end{split}
\end{equation}
where the $A_{\sigma}^{(b)}$ contain the fluctuation determinants and other
subleading prefactors. On the transported spherical sheets we have checked that
\begin{equation}
 \Delta_{\rm non}(\alpha)
 \equiv W_{\rm geo}(\alpha)-W_{\rm non}(\alpha)>0,
 \qquad
 \alpha_*<\alpha<\pi .
\end{equation}
Hence the non-geometric term in $Z_{\text{Vir}}^{(b_+)}$ is exponentially suppressed relative to
the geometric one, 
\begin{equation}
 \frac{Z_{{\rm non}}^{(b_+)}}{Z_{{\rm geo}}^{(b_+)}}
 \sim
 \frac{A_{{\rm non}}^{(b_+)}}{A_{{\rm geo}}^{(b_+)}}
 \exp\!\left[
 -\frac{\kappa}{2\pi}\Delta_{\rm non}(\alpha)
 \right].
\end{equation}
The leading semiclassical product is therefore
\begin{align}
Z_{\text{Vir}}^{(b_-)}[\mathbf{5}_2] Z_{\text{Vir}}^{(b_+)} [\mathbf{5}_2]
 &=
 A_{{\rm geo}}^{(b_+)}A_{{\rm anti}}^{(b_-)}
 \exp\!\left[
 \frac{\kappa}{2\pi}
 \bigl(W_{\rm geo}-W_{\rm anti}\bigr)
 \right]
 \left(
 1+
 \frac{A_{{\rm non}}^{(b_+)}}{A_{{\rm geo}}^{(b_+)}}
 e^{-\frac{\kappa}{2\pi}\Delta_{\rm non}}
 +\cdots
 \right)
 \nonumber\\
 &\sim
 \exp\!\left[
 \frac{\kappa}{2\pi}
 \bigl(W_{\rm geo}-W_{\rm anti}\bigr)
 \right].
\end{align}

The volumes extracted from the saddle actions agree with the
geometric volumes obtained by integrating the Schl\"afli formula
\cite{Mednykh2021,tran2026exact}.
Indeed, the change of variables $\xi=1-r^{-1}$ maps the holonomy
equation in the literature to the Virasoro TQFT saddle cubic
derived in \eqref{eq:AKcubic}. In the spherical range $\alpha_*<\alpha<\pi$,
the geometric volume integral becomes
\begin{equation}\label{eq:three-twist-schlafli-match}
\begin{aligned}
\frac{\text{Vol}(M_\alpha)}{\ell_{\rm dS}^{\,3}}
&=
-\frac12\int_{r_{\rm anti}}^{r_{\rm geo}}
\frac{dt}{t}\,
\log\!\left[
\frac{(1+t)(1+2t\cos\alpha+t^2)}{t}
\right]
\\
&=
\frac12\left(W_{\rm geo}(\alpha)-W_{\rm anti}(\alpha)\right),
\end{aligned}
\end{equation}
where $r_{\rm geo}>r_{\rm anti}>0$ are the two continued (anti-)geometric
roots, and the endpoint $\alpha=\pi$ is understood by continuity
from below. The corresponding hyperbolic formula gives
$(W_{\rm geo}-W_{\rm anti})/(2i)$ on the continued logarithmic sheets.
Both expressions vanish as the geometric saddles coalesce at
$\alpha_*$.
At $\alpha=\pi$, the spherical volume is
$\pi^2\ell_{\rm dS}^{\,3}/7$, also obtained from the two-fold
branched cover $L(7,3)$, whose universal cover is $S^3$.
Thus the semiclassical partition function reproduces the spherical cone manifold volume and hence
the on-shell gravitational action,
\begin{equation}
   Z_{\rm grav}^{\Lambda>0}[\mathbf{5}_2]= Z_{\text{Vir}}^{(b_-)}[\mathbf{5}_2] Z_{\text{Vir}}^{(b_+)}[\mathbf{5}_2] \sim \exp\left[\frac{\kappa}{2\pi}(W_{\rm geo}-W_{\rm anti})\right]=\exp\left[\frac{\text{Vol}(M_\alpha)}{4\pi G_N \ell^2_{\rm dS}}\right]\,.
\end{equation}
This provides a verification of the spherical volume conjecture for the spherical cone manifolds of the three-twist knot.

\subsection{The Whitehead link cone manifolds}

Our last example will be that of the Whitehead link, also known as $\mathbf{5}_1^2$. This is a two-component alternating link which can be drawn with only 5 crossings. For this paper, it will be convenient to work with the following presentation:\\[-11pt]
\begin{equation}
\mathbf{5}^2_1 \;\;=\!\!\! \vcenter{\hbox{
\begin{tikzpicture}[scale=0.76,yscale=0.79]
\draw[very thick,black]
    (-0.22,-2.10)
    to[out=197.39,in=182.12,out distance=2.28cm,in distance=2.76cm]
    (-0.07,1.98);
\draw[very thick,black]
    (0.08,-1.96)
    to[out=32.01,in=287.10,out distance=0.32cm,in distance=0.46cm]
    (0.54,-0.81) node[right, black]{$P_2$}
    to[out=107.10,in=292.48,out distance=0.46cm,in distance=0.46cm]
    (-0.46,0.52)
    to[out=112.47,in=186.91,out distance=0.46cm,in distance=0.40cm]
    (-0.05,1.65);
\draw[very thick,black]
    (-0.09,-0.12)
    to[out=223.78,in=89.63,out distance=0.47cm,in distance=0.26cm]
    (-0.56,-1.10)
    to[out=269.70,in=149.31,out distance=0.26cm,in distance=0.52cm]
    (-0.05,-2.03)
    to[out=329.29,in=255.96,out distance=0.51cm,in distance=0.84cm]
    (1.91,-0.91)
    to[out=75.96,in=0.97,out distance=0.84cm,in distance=2.01cm]
    (0.13,1.99);
\draw[very thick,black]
    (0.06,-0.01)
    to[out=39.97,in=258.21,out distance=0.47cm,in distance=0.24cm]
    (0.57,0.83)
    to[out=78.31,in=4.60,out distance=0.24cm,in distance=0.49cm]
    (0.14,1.65);
\draw[very thick,black]
    (0.31,1.56)
    to[out=245.69,in=268.06,out distance=0.22cm,in distance=0.41cm]
    (-0.01,1.82)
    to[out=88.01,in=102.71,out distance=0.41cm,in distance=0.20cm]
    (0.34,2.04);
\draw[very thick,black]
    (0.36,1.92)
    to[out=274.86,in=75.96,out distance=0.07cm,in distance=0.07cm] node[above=3pt,black]{$P_1$}
    (0.35,1.69);
\end{tikzpicture}}}.
\end{equation}
One useful property of this link is that the components 1 and 2 can be exchanged using only Reidemeister moves so, up to a factor of the framing anomaly, the Virasoro TQFT amplitude is a symmetric function of $P_1$ and $P_2$. To evaluate this amplitude, we can insert an identity line in the second link component and apply an $\mathbb{F}$ transform. After reducing the diagram, the result is:\footnote{This expression matches other known formulas for the Virasoro TQFT amplitude of the Whitehead link, up to framing factors. For example, to match with equation (3.24) in \cite{Post:2024itb}, we simply have to substitute $\sker{P_2}{P_2}{P}$ with their formula (3.20), identify the F-kernel with their fusion density (2.19), and finally apply their relation (3.22). Equivalence with the Whitehead-link amplitude in \cite{Chandra:2025fef} follows by applying the hexagon identity and the braiding-weighted $\mathbb{S}-\mathbb{F}$ relation \cite{Collier:2024mgv}, which rewrite that amplitude as a bilinear in $\mathbb{S}$-kernels. The same bilinear follows from (\ref{eq:Whitehead_VTQFT}) using the inverse Virasoro–Verlinde transform \cite{Post:2024itb}, with the remaining difference accounted for by external framing phases.}
\begin{equation} \label{eq:Whitehead_VTQFT}
Z_{\rm Vir}^{(b)}\left[\!\!\!\!\!\!\vcenter{\hbox{
\begin{tikzpicture}[yscale=0.6,xscale=0.75,scale=0.9]
\draw[very thick,black]
    (-0.22,-2.10)
    to[out=197.39,in=182.12,out distance=2.28cm,in distance=2.76cm]
    (-0.07,1.98);
\draw[very thick,black]
    (0.08,-1.96)
    to[out=32.01,in=287.10,out distance=0.32cm,in distance=0.46cm]
    (0.54,-0.81) node[right, black]{$P_2$}
    to[out=107.10,in=292.48,out distance=0.46cm,in distance=0.46cm]
    (-0.46,0.52)
    to[out=112.47,in=186.91,out distance=0.46cm,in distance=0.40cm]
    (-0.05,1.65);
\draw[very thick,black]
    (-0.09,-0.12)
    to[out=223.78,in=89.63,out distance=0.47cm,in distance=0.26cm]
    (-0.56,-1.10)
    to[out=269.70,in=149.31,out distance=0.26cm,in distance=0.52cm]
    (-0.05,-2.03)
    to[out=329.29,in=255.96,out distance=0.51cm,in distance=0.84cm]
    (1.91,-0.91)
    to[out=75.96,in=0.97,out distance=0.84cm,in distance=2.01cm]
    (0.13,1.99);
\draw[very thick,black]
    (0.06,-0.01)
    to[out=39.97,in=258.21,out distance=0.47cm,in distance=0.24cm]
    (0.57,0.83)
    to[out=78.31,in=4.60,out distance=0.24cm,in distance=0.49cm]
    (0.14,1.65);
\draw[very thick,black]
    (0.31,1.56)
    to[out=245.69,in=268.06,out distance=0.22cm,in distance=0.41cm]
    (-0.01,1.82)
    to[out=88.01,in=102.71,out distance=0.41cm,in distance=0.20cm]
    (0.34,2.04);
\draw[very thick,black]
    (0.36,1.92)
    to[out=274.86,in=75.96,out distance=0.07cm,in distance=0.07cm] node[above=7pt,black]{$P_1$}
    (0.35,1.69);
\draw[very thick, black,dashed] (-1.7,0.9)  -- node[below] {$\bbi$} (-0.5,0.9);
\end{tikzpicture}}}    
    \right] =
    \int_0^\infty \dd P\;
    \sker{P_1}{P}{\bbi}\;
    \frac{\sker{P_2}{P_2}{P}}{\rho_0(P_2)}.
\end{equation} 
As in the example of the three-twist knot, the combination involving the $\mathbb{S}$ kernel can be written as an integral which only involves double-sine functions:
\begin{equation}
\frac{\sker{P_2}{P_2}{P}}{\rho_0(P_2)} = \frac{\e^{\frac{i\pi}{2}h_P}}{
S_b\left(\frac{Q}{2}+iP\right)}\int_{\mathbb{R}} \dd x
\,\e^{-4\pi i x P_2}
S_b\left(
\frac{Q}{4}+\frac{i P}{2} \pm i P_2\pm ix
\right).
\end{equation}
Here the notation $\pm$ corresponds to a product over four different factors. We pick the following scalings for our variables 
\begin{equation}
 x = \frac{u}{4\pi b},\quad P = \frac{v}{4\pi b},\quad P_{1,2} = \frac{i\alpha_{1,2}}{4\pi b}.
\end{equation}
In the semiclassical limit, where $b\rightarrow0$, the Virasoro TQFT amplitude becomes
\begin{equation}
    Z_{\text{Vir}}^{(b)}[\mathbf{5}_1^2] \sim 
    \int
    \dd u\dd v\;
    \exp[\frac{i}{2\pi b^2} W_{\mathbf{5}_1^2}(\alpha_1,\alpha_2,u,v)],
\end{equation}
where the classical action takes the form
\begin{multline}
     W_{\mathbf{5}_1^2}(\alpha_1,\alpha_2,u,v) =
\frac{\pi^2}{4}
-\frac{i\alpha_1v}{2}
+\frac{v^2}{16}
-\frac{i\alpha_2u}{2}
+\mathcal{F}\left(\frac{v}{2}\right)
\\ -
\sum_{\tau=\pm1}\sum_{\sigma=\pm1}
\mathcal{F}\!\left(
  \frac{v+2\pi i+2i\tau \alpha_2+2\sigma u}{4}
\right),
\end{multline}
and $\mathcal{F}$ is given by 
\begin{equation}
\label{eq:defF}
    \log S_b\left(\frac{Q}{2}+i\frac{z}{2\pi b}\right)\sim \frac{1}{2\pi i b^2}\mathcal{F}(z), \qquad \mathcal{F}(z)=\text{Li}_2(-e^z)+\frac{z^2}{4}+\frac{\pi^2}{12}\,.
\end{equation}
The integral defining this amplitude is over two complex variables. In the previous sections, we reduced the partition function to a single-integral representation. For the $\mathbf{5}_1^2$ knot, we do not know whether a similar single-integral representation exists. 
The fact that the integral is over two complex variables is unfortunate because the multivariable generalization of Picard--Lefschetz theory is significantly more difficult than its single-variable counterpart, see e.g. \cite{Aniceto:2026oxu}. A full thimble analysis of this contour is therefore beyond the scope of this paper. 
However, the shape of the potential $W_{\mathbf{5}_1^2}$ suggests an elegant way to check whether the de Sitter amplitude reproduces the volume of the spherical Whitehead link in the semiclassical limit. 

Consider the saddle-point approximation of the integral
\begin{equation}
    \log Z_{\text{Vir}}^{(b)}[\mathbf{5}_1^2] =  \frac{i}{2\pi b^2} W_{\mathbf{5}_1^2}(\alpha_1,\alpha_2,u^\star,v^\star) +\order{1},
\end{equation}
where $v^\star(\alpha_1,\alpha_2)$ and $u^\star(\alpha_1,\alpha_2)$ are the solutions to the saddle-point equations that approximate the $u$ and $v$ integrals. If $\ell_{\text{dS}}^3W_{\mathbf{5}_1^2}$ is the volume of the spherical cone manifold, then it must obey the Schläfli formula. Moreover, since $W_{\mathbf{5}_1^2}$ comes from a saddle-point approximation, we have a slightly more useful set of equations that we can check, namely:
\begin{equation}
\frac{d}{d\alpha_1} W_{\mathbf{5}_1^2}(\alpha_1,\alpha_2,u^\star,v^{\star}) =\left(
\partial_{v}W_{\mathbf{5}_1^2}\; \frac{dv^\star}{d\alpha_1} + 
\partial_{u}W_{\mathbf{5}_1^2} \;\frac{du^\star}{d\alpha_1} + \partial_{\alpha_1}W_{\mathbf{5}_1^2}\right)_{u=u^{\star},v=v^{\star}}
\end{equation}
and the analogue equation for $\alpha_2$. At the saddle-point, the first two terms vanish and we are left with the following relation to check:
\begin{equation}
\label{eq:schlafliwhite}
  \frac{d}{d\alpha_1} W_{\mathbf{5}_1^2}(\alpha_1,\alpha_2,u^\star,v^{\star}) = \partial_{\alpha_1}W_{\mathbf{5}_1^2}\Big\lvert_{u=u^{\star},v=v^{\star}} = \frac{\ell_1}{2}.
\end{equation}
Here $\ell_1$ is the geodesic length of the singular geodesic with conical opening angle $\alpha_1$. The second equality follows from the Schläfli formula. 

Formula \eqref{eq:schlafliwhite} is especially useful for the example of the Whitehead link, because the potential $W_{\mathbf{5}_1^2}$ is of the form 
\begin{equation}
    W_{\mathbf{5}_1^2} = -\frac{1}{2}\big(
    i \alpha_1 v
    -V_0(\alpha_2,v,u)
    \big).
\end{equation}
Since $V_0$ is independent of $\alpha_1$, we have the following simple result: 
\begin{equation}
    v^{\star} = i\ell_1.
\end{equation}
That is, the saddle-point equation for $v$ should correspond to the algebraic equation whose solution yields the length of the geodesic $\ell_1$. 

There is one important subtlety, however: as with the example of the three-twist knot, we do not expect $Z_{\text{Vir}}^{(b)}[\mathbf{5}_1^2]$ to exponentiate to the spherical volume. The total partition function is a combination of left and right movers, one whose dominant term is the geometric saddle and one whose dominant term is the anti-geometric saddle. The appropriate equation to check is therefore 
\begin{equation}
   \left(v^{\star}\right)_\text{geo}-  \left(v^{\star}\right)_\text{anti} = 2i\ell_1,
\end{equation}
where $\left(v^{\star}\right)_\text{geo}$ and $\left(v^{\star}\right)_\text{anti}$ are the two geometric roots of the saddle-point equation.

\begin{figure}[t]
    \centering
\begin{tikzpicture}
    \node at (0,0) {    \includegraphics[width=0.5\linewidth]{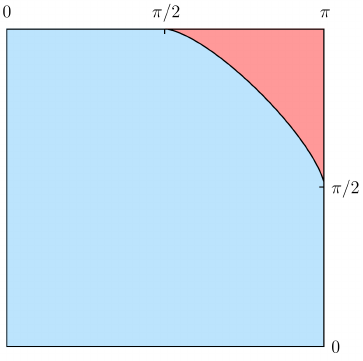}};
    \node at (-4.2,-0.2) {$\alpha_2$};
    \node at (-0.2,-4) {$\alpha_1$};
    \node at (-0.6,-0.8) {Hyperbolic};
    \node[rotate=-45] at (1.4,1.4) {Euclidean};
    \node[rotate=-45] at (2.2,2.2) {Spherical};
\end{tikzpicture}
    \caption{Geometric structures on the Whitehead link cone manifold as a function of the cone angles $\alpha_1$ and $\alpha_2$.}
    \label{fig:geometric_structures_whitehead}
\end{figure}

The analysis now is straightforward: the first saddle point equation $\partial_u W_{\mathbf{5}_1^2} = 0$ leads to the following quadratic equation:
\begin{equation}
\label{eq1}
    q^2\cos\left(\frac{\alpha_2}{2}-\frac{\alpha_v}{4}\right)-q\sin\left(
    \alpha_2\right)
    -\cos\left(\frac{\alpha_2}{2}+\frac{\alpha_v}{4}\right) =0,
\end{equation}
where $q=\e^{\frac{u}{2}}$, and $v = i \alpha_v$. The second saddle-point equation is equivalent to
\begin{equation}
\label{eq2}
\exp[-2i\left(\alpha_1 - \frac{\alpha_v}{4}\right)] = 
\frac{
\cos^2\left(\frac{\alpha_v}{4}\right)
}{
\left[
\sin(\frac{\alpha_v}{4})
-\cos(\frac{iu-\alpha_2}{2})
\right]
\left[
\sin(\frac{\alpha_v}{4})
-\cos(\frac{i u+\alpha_2}{2})
\right]
}.
\end{equation}
Using \eqref{eq1} and \eqref{eq2}, one can eliminate the variable $iu$ without having to solve explicitly for $q$. The resulting saddle-point equation for $\alpha_v$ is 
\begin{equation}
\cos(\alpha_1-\frac{\alpha_v}{4})
=
\frac{
\cos^2\left(\frac{\alpha_2}{2}\right)
\cos(\frac{\alpha_v}{4})
}{
\cos^2\left(\frac{\alpha_2}{2}\right)
-
\sin^2\left(\frac{\alpha_v}{4}\right)
}.
\end{equation}
The following change of variables,
\begin{equation}
 A=\cot\frac{\alpha_1}{2},\quad B=\cot\frac{\alpha_2}{2},\quad y = \cot\frac{\alpha_2}{2}\; \cot\frac{\alpha_v}{4} 
\end{equation}
leads to the cubic equation 
\begin{equation}
\label{eq:cubic-whitehead}
       y^3 - A B y^2 +\frac{1}{2}\left(A^2B^2 +A^2 + B^2-1\right)y+AB =0,
\end{equation}
which is precisely the equation that determines the lengths $\ell_1,\ell_2$ of the Whitehead link cone manifold, see e.g.~\cite{Abrosimov2008}.
So we have demonstrated the spherical volume conjecture for the Whitehead link cone manifolds at the level of the geometric-saddle action. 

The cubic equation \eqref{eq:cubic-whitehead}, which was purely derived from the saddle-point analysis of the Virasoro TQFT amplitude, contains a considerable amount of geometric information. In particular the locus where the discriminant of \eqref{eq:cubic-whitehead} vanishes defines a curve in the $(\alpha_1,\alpha_2)$-plane, shown in figure \ref{fig:geometric_structures_whitehead}, which separates the regions corresponding to spherical and hyperbolic structures. The locus itself corresponds to Euclidean structures, which is where the saddle-point approximation breaks down.

\section{Seifert manifolds from spherical Dehn fillings} \label{sec:Seifert}

In the previous section, we studied cone manifolds whose singular loci are knots
and links with hyperbolic complements, and followed their geometric structures
from the hyperbolic to the spherical regime. In this section, we instead consider
torus knots and links. Their complements are Seifert-fibered rather than
hyperbolic, but they admit spherical cone structures for suitable cone angles
\cite{KolpakovMednykhSpherical}. We will also study Dehn fillings for which the
singular locus is filled in and the resulting closed manifold admits a smooth
spherical metric. We illustrate these constructions with two basic examples.
The Hopf link, the simplest nontrivial two-component link, admits Dehn fillings
that produce lens spaces, while the trefoil, the simplest nontrivial torus knot,
admits particular Dehn fillings that produce spherical Seifert manifolds
\cite{Bojkovic:2023Hopf,MoserTorusKnotSurgery}. These examples provide a class
of smooth spherical topologies that is qualitatively distinct from the
hyperbolic knot and link complements considered in the previous section.

\subsection{Lens spaces from the Hopf link}

The Hopf link, denoted by $\mathbf{2}^2_1$ in the Alexander–Briggs notation, is the simplest nontrivial link with more than one component:
\begin{equation}
    \begin{tikzpicture}
        \draw[very thick] (0.5,0) node[right=8pt,black] {$P_2$}  [partial ellipse=-110:230:1 and 1];
        \draw[very thick] (-0.5,0) node[left=8pt,black] {$P_1$} [partial ellipse=70:410:1 and 1];
    \end{tikzpicture}
\end{equation}
If the two link components correspond to conical defects, the complement in $S^3$ admits a cone metric with positive curvature.
As with the unlink and the theta graph, we can write down an explicit  metric for its corresponding  spherical cone manifold:
\begin{equation}
    \dd s^2_{E} = \ell_{\rm dS}^2 (\dd r^2 + \cos^2 r \dd \phi^2 + \sin^2 r \dd \theta^2), 
\end{equation}
where $r \in [0,\pi/2]$, $\phi \in [0,\alpha_1]$ and $\theta \in [0,\alpha_2]$. The two conical defects have opening angles $\alpha_1$ and $\alpha_2$. Comparing to \eqref{eq:conicaldSmetric}, the geometry is that of the static patch in Euclidean signature, with one conical defect line along the observer worldline and one at the cosmological horizon. The spherical volume of the manifold is given by:
\begin{equation}
    \text{Vol}(\mathbf{2}_1^2) = \ell_{\rm dS}^3\frac{\alpha_1 \alpha_2}{2}.
\end{equation}
The Virasoro TQFT amplitude of this manifold is also very simple,
\begin{equation}
    Z_{\rm Vir}^{(b)}[\mathbf{2}_1^2] = \sker{P_1}{P_2}{\bbi} = 2\sqrt{2}\cos(4\pi P_1P_2).
\end{equation}
In particular, it is independent of $b$.
Identifying the angles $\alpha_j$ with the momenta $P_j$ via the relation $i\alpha_j = 4 \pi b P_j$, $j=1,2$ and substituting $\kappa = \frac{i}{b^2}$, we find that 
\begin{equation}
    Z_{\rm Vir}^{(b)}[\mathbf{2}_1^2] = \sqrt{2} \left(
    \e^{\frac{\kappa}{2\pi \ell_{\rm dS}^3} \text{Vol}(\mathbf{2}_1^2)}
    + \e^{-\frac{\kappa}{2\pi \ell_{\rm dS}^3} \text{Vol}(\mathbf{2}_1^2)}
    \right).
\end{equation}
Thus the growing real exponential confirms the general form of the spherical volume conjecture. This is interesting because the topology is off-shell in AdS$_3$, whereas it is an on-shell saddle in the spherical regime.

A simple family of smooth spherical manifolds (i.e.~without defects) is obtained by Dehn filling both components of the Hopf link. The resulting manifolds are lens spaces, which admit the quotient description
\begin{equation}
L(N,\mathcal{Q})=S^3/\mathbb{Z}_N,\qquad
(z_1,z_2)\sim
\left(e^{2\pi i/N}z_1,e^{2\pi i\mathcal{Q}/N}z_2\right),
\qquad \gcd(N,\mathcal{Q})=1.
\end{equation}
Here $S^3=\{(z_1,z_2)\in\mathbb{C}^2:|z_1|^2+|z_2|^2=1\}$. If the two Hopf-link components are filled along slopes $p/q$ and $r/s$, respectively, then $N=|pr-qs|$, while $\mathcal{Q}$ is determined modulo $N$ by the corresponding gluing matrices; this reproduces the standard classification of rational surgeries on the Hopf link \cite{Bojkovic:2023Hopf}. In Virasoro TQFT the Hopf-link state is simply the modular $\mathbb{S}$-kernel as we discussed above. Smooth filling of each component is implemented by the vacuum Dehn filling kernel (see e.g. \cite{Benjamin:2020mfz, Gang:2017cwq,Collier:2024mgv}),
\begin{equation}
\mathcal{K}_{p/q}(P)=\epsilon(p,q)\sqrt{\frac{8}{|q|}}
\,e^{-\frac{2\pi i }{q}\left(\frac{p^*Q^2}{4}-pP^2\right)}\left[\cosh\left(\frac{2\pi QP}{q}\right)-e^{2\pi i \frac{p^*}{q}}\cosh\left(\frac{2\pi \Hat{Q}P}{q}\right)\right]
\end{equation}
Here, $p^*$ is the modular inverse of $p$, $pp^*\equiv 1\, \text{mod}\, q$. Moreover $\epsilon(p,q)$ is a phase and $Q =b+b^{-1}$ and $\hat{Q} = b-b^{-1}$.
In terms of the above Dehn filling kernels, one obtains the integral expression for the partition function on the Lens space $L(N,\mathcal{Q})$ as
\begin{equation} \label{eq:VTQFTLenspace}
Z_{L(N,\mathcal{Q})}=
\int_0^\infty \dd P_1\dd P_2\,
\mathcal{K}_{p/q}(P_1)
\sker{P_1}{P_2}{\bbi}
\mathcal{K}_{r/s}(P_2).
\end{equation}
The double integral in (\ref{eq:VTQFTLenspace}) can be evaluated exactly, giving
\begin{equation} \label{eq:Lensexact}
Z_{L(N,\mathcal{Q})}
=
\epsilon_G\sqrt{\frac{8}{N}}
\sum_{\sigma,\tau=\pm1}
\sigma\tau\,
\exp\!\left[
\frac{2\pi i}{N}
\left(A P_\sigma^2+\mathcal{Q} P_\tau^2\right)
\right]
\cos\!\left(\frac{4\pi P_\sigma P_\tau}{N}\right),
\end{equation}
where we defined $P_\sigma=\frac{i}{2}\left(b^{-1}+\sigma b\right)$. Here $\epsilon_G$ is an
overall framing phase and $A$ is the modular inverse of $\mathcal{Q}$, $A\mathcal{Q}\equiv 1 \,\text{mod}\,N$. The semiclassical $|b|\to 0$ expansion of $Z$ is\footnote{We consider the generic case \(\mathcal Q\not\equiv\pm1\pmod N\). The exceptional cases require retaining the first nonzero terms in the expansion of (\ref{eq:Lensexact}).}
\begin{equation}
Z_{L(N,\mathcal{Q})}
\sim
\sum_{\eta=\pm1}
Z_{\rm 1\mbox{-}loop}^{(\eta)}
\exp\!\left[
-\frac{i\pi}{2Nb^2}(2\eta+A+\mathcal{Q})
\right],
\end{equation}
with
\begin{equation}
Z_{\rm 1\mbox{-}loop}^{(\eta)}
=
-4\epsilon_G\sqrt{\frac{2}{N}}\,
\sin\!\left(\frac{\pi(A+\eta)}{N}\right)
\sin\!\left(\frac{\pi(\mathcal{Q}+\eta)}{N}\right).
\end{equation}
Now, pairing the growing exponentials of the left-moving and the right-moving sectors after analytically continuing $b^2$ to $b_\pm^2$, with $\kappa = \pm i/b_\pm^2 = \frac{\ell_{\rm dS}}{4G_N}$, we arrive at the following semiclassical expansion of the dS$_3$ partition function on the lens space,
\begin{multline}
    Z^{\Lambda>0}_{\rm grav} \sim \frac{32}{N}\sin\left(\frac{\pi(A-1)}{N}\right)\sin\left(\frac{\pi(\mathcal{Q}-1)}{N}\right)\\ \times \sin\left(\frac{\pi(A+1)}{N}\right)\sin\left(\frac{\pi(\mathcal{Q}+1)}{N}\right)\exp\left[\frac{\pi \ell_{\rm dS}}{2G_N N}\right]\,.
\end{multline}
The exponential factor is consistent with the known volume of the lens space,
\begin{equation}
    \exp\left[\frac{\pi \ell_{\rm dS}}{2G_N N}\right] =\exp\left[\frac{\text{Vol}(L(N, \mathcal{Q}))}{4\pi G_N \ell^2_{\rm dS}}\right], \qquad \text{Vol}(L(N, \mathcal{Q}))=\frac{2\pi^2}{N}\ell_{\rm dS}^3\,.
\end{equation}
The 1-loop determinant also has a natural interpretation in terms of Reidemeister torsion. The two semiclassical sectors have adjoint holonomies labelled by $\eta=\pm1$, and the corresponding reduced adjoint Reidemeister torsion obeys 
\begin{equation}
    \mathcal{T}_{\eta}^{1/2}=\frac{4}{\sqrt{N}}\left|\sin\left(\frac{\pi(A+\eta)}{N}\right)\sin\left(\frac{\pi(Q+\eta)}{N}\right)\right|
\end{equation}
This is consistent with the semiclassical description of Chern--Simons theory on lens spaces \cite{Jeffrey:1992Lens}.
So, the gravitational 1-loop determinant can be expressed in terms of the above Reidemeister torsion as,
\begin{equation}
    Z_{\rm grav}^{1-\rm loop}=2\sqrt{\mathcal{T}_+\mathcal{T}_-}\,.
\end{equation}
Thus, apart from an $N,\mathcal{Q}$-independent normalization associated with the Virasoro momentum basis, the one-loop coefficient obtained from the double-filling integral is precisely the geometric mean of the reduced Reidemeister torsion.

Up to an overall normalization, the spherical saddle contribution
obtained above, including its perturbative fluctuations, agrees with
the Chern-Simons calculation of~\cite{Castro:2011xb}.\footnote{The comparison uses Eq. (5.1) of \cite{Castro:2011xb} with the geometric holonomies and continuation in their Eqs. (5.4)–(5.5). In their Eq. (5.6), the cosines involving \(q+q^*\) and \(q-q^*\) in the last two terms appear to be interchanged. Note also that the exact fixed-topology amplitudes obtained here agree, up to normalization and framing conventions, with the vacuum-to-vacuum modular-matrix elements of \cite{Blommaert:2026ofx}. Our double-filling calculation evaluates these amplitudes explicitly and identifies their semiclassical coefficients with reduced Reidemeister torsions.}
We can therefore form the de Sitter Farey-tail sum over lens spaces,
\begin{equation}\label{eq:lens_space_farey_sum}
Z_{\mathrm{lens}}
=
Z_{\mathrm{grav}}(S^3)
+
\sum_{N=2}^{\infty}
\sum_{\substack{1\leq \mathcal Q<N\\
                \gcd(N,\mathcal Q)=1}}
Z_{\mathrm{grav}}\!\left(L(N,\mathcal Q)\right),
\end{equation}
where the superscript denotes the spherical Einstein saddle together
with its perturbative corrections, and we use the same counting
conventions as~\cite{Castro:2011xb}.
This sum has the same residual divergence as the de Sitter Farey tail:
it remains divergent after the zeta-function regularization employed
there, and higher-loop corrections do not remove the divergence.

\subsection{Spherical Seifert manifolds from the trefoil knot}\label{sec:trefoil}

The trefoil knot is a torus knot with 3 crossings denoted $\mathbf{3}_1$,\\[-15pt]
\begin{equation}\label{eq:picutretrefoil}
\vcenter{\hbox{
        \begin{tikzpicture}[scale=0.75]
          \draw[very thick, out=-30, in =-30, looseness=2.5] (1,0) to (0,-1.732);
           \draw[fill=white, draw=white] (-1,0) circle (1/10);
           \draw[very thick, out=30, in =150, looseness=1] (-1,0) to (1,0);
           \draw[very thick, out=-150, in =-150, looseness=2.5] (-1,0) to (0-0.1,-1.732-0.1);
            \draw[fill=white, draw=white] (1,0) circle (1/10);
             \draw[very thick, out=90, in =90, looseness=2.5] (-1,0.15) to (1,0);
              \draw[very thick, out=-90, in =30, looseness=1] (1,0) to (0.1,-1.732+0.1);
              \draw[very thick, out=-90, in =150, looseness=1] node[right=30pt,black] {$P_1$}(-1,-0.15) to (0,-1.732);
        \end{tikzpicture}}}
\end{equation}\\[-15pt]
\noindent It is well-known that the trefoil is homeomorphic to the $T(2,3)$ torus knot \cite{rolfsen2003knots}, which can be presented as a Dehn filling of two of the links in the three-component Hopf chain with slopes $(2,1)$ and $(-3,1)$, respectively (we have verified this in SnapPy \cite{SnapPy}). Since we know how to perform Dehn filling in Virasoro TQFT, we can easily write down the partition function on the trefoil knot complement:
\begin{align}
    Z^{(b)}_{\rm Vir}[\mathbf{3}_1;P_1] &=  \int_{\mathbb{R}+i\epsilon} \!\frac{\dd P}{2}\,\mathbb{S}_{P_1P}[\mathds{1}]Z_{\rm Vir}\left[
       \vcenter{\hbox{\begin{tikzpicture}[scale=0.9]
        \draw[very thick] (0,0.2) node[below=4pt]{$P$} ellipse (1.4 and 0.8);
        \draw[line width=4pt, white] (0.7,0.9) [partial ellipse=150:-170:0.3 and 0.4];
        \draw[very thick] (0.7,0.9) [partial ellipse=150:-170:0.3 and 0.4];
        \draw[line width=4pt, white] (-0.7,0.9) [partial ellipse=185:-145:0.3 and 0.4];
        \draw[very thick] (-0.7,0.9) [partial ellipse=185:-145:0.3 and 0.4];
        \node at (-1.4,1.1) {$-3$};
        \node at (1.2,1.1) {$2$};
    \end{tikzpicture}}}
\,\right] \\[1em]
&= \int_{\mathbb{R}+i\epsilon} \!\frac{\dd P}{2}\,\mathbb{S}_{P_1P}[\mathds{1}] \frac{(ST^{-3} S)_{\mathds{1}P} (ST^2 S)_{\mathds{1}P}}{\rho_0(P)}.
\end{align}
To arrive at this answer, we used standard TQFT technology to unlink the top loops and integrate against the appropriate Dehn filling kernels (for a similar calculation, see e.g. \cite{deBoer:2025oge}). The additional S-transform is necessary because the momentum $P$ parametrizes the holonomy around a regular Seifert fiber of the trefoil complement, whereas $P_1$ parametrizes the holonomy around the meridian cycle as illustrated in \eqref{eq:picutretrefoil}. The $P$-contour is shifted by $i\epsilon$ to avoid the double pole at $P=0$ coming from $\rho_0(P)$.\footnote{As an aside, we remark that one can arrive at the same integral expression by a careful analysis of the $P_a\to\mathds{1}$ limit of the trefoil `fragmentation' analyzed in \cite{Chandra:2025fef}. This requires using the Virasoro-Verlinde formula \cite{Post:2024itb} and a change of variables that we will not discuss here.}

The relevant integer Dehn filling kernels are given by 
\begin{equation} \label{eq:Dehn_kernels}
    (ST^{n}S)_{\mathds{1}P} = e^{\pi i\theta_n} e^{-\frac{2\pi i}{n}(-\frac{Q^2}{4}+P^2)}\sqrt{\frac{8}{|n|}}\left(\cosh(\frac{2\pi Q}{n}P)-e^{-\frac{2\pi i}{n}}\cosh(\frac{2\pi \hat{Q}}{n}P)\right)
\end{equation}
where $\theta_n = \frac{\text{sgn}(n)}{4}-\frac{n}{12}$ and we used the convention for the Dehn twist $\mathbb{T}_P = e^{2\pi i (P^2-\frac{1}{24})}$.
The $ST^2S$ kernel simplifies together with the $\rho_0(P)$ factor in the denominator, using that $\sinh(2x) = 2\sinh x\cosh x$, to
\begin{align}
   \frac{(ST^2S)_{\mathds{1}P}}{\rho_0(P)} &= \frac{4 \cosh(\pi b P)\cosh(\pi P/b)e^{\pi i\theta_2} e^{\frac{\pi iQ^2}{4}-\pi i P^2}}{4\sqrt{2}\sinh(2\pi bP)\sinh(2\pi P/b)} \\[0.5em]
   &=\frac{e^{\pi i\theta_2} e^{\frac{\pi iQ^2}{4}-\pi i P^2} }{4\sqrt{2}\sinh(\pi bP)\sinh(\pi P/b)}\,.
\end{align}
Hence the full trefoil partition function becomes
\begin{equation} \label{eq:Trefoil_partition_function}
    Z^{(b)}_{\rm Vir}[\mathbf{3}_1;P_1] = \frac{e^{\frac{\pi i}{12} (1+Q^2)}}{\sqrt{6}} \int_{\mathbb{R}+i\epsilon} \!\!\dd P \cos(4\pi PP_1) e^{-\frac{i\pi}{3} P^2} \frac{\cosh(\tfrac{2\pi Q}{3}P)-e^{\frac{2\pi i}{3}}\cosh(\tfrac{2\pi\hat{Q}}{3}P)}{\sinh(\pi b P)\sinh(\pi P/b)}.
\end{equation}
Let us study the analytic continuation of this integral from real $b^2$ to $b_\pm^2 = \pm \frac{i}{\kappa}$, with $\kappa >0$. We write $P_1^\pm =\pm \frac{i\alpha}{4\pi b_\pm}$ as before for the external momentum, with cone angle $\alpha$. 

We begin with $b_-$. Parametrizing $P = -is/b_-$, the integral can be written as
\begin{equation}
    Z^{(b_-)}_{\rm Vir}[\mathbf{3}_1;P_1^-] = \frac{e^{\frac{\pi i}{12} (2+Q_-^2)}}{\sqrt{6}} I_\kappa(\alpha) 
\end{equation}
where the integral $I_\kappa$ has the following large $\kappa$ asymptotics:
\begin{align}
    I_\kappa(\alpha) &\coloneqq \sqrt{\kappa}\int_{\mathcal{C}}\dd s \,e^{-\frac{\kappa \pi}{3}s^2} \cosh(\kappa \alpha s) \frac{e^{2\pi \kappa s/3}\sin \frac{\pi(2s+1)}{3}-e^{-2\pi \kappa s/3}\sin \frac{\pi(2s-1)}{3}}{\sin(\pi s)\sinh(\pi \kappa s)} \\[1em]
    &= 2\sqrt{\kappa}\int_{\mathcal{C}}\dd s \,e^{\kappa \left(-\frac{\pi }{3}s^2+\left(\alpha - \frac{\pi}{3}\right)s\right)}\,\frac{\sin\frac{\pi(2s+1)}{3}}{\sin(\pi s)}\left[1+\order{\tfrac{1}{\kappa}}\right].
\end{align}
The prefactor of $2$ in the second line arises because the integrand in the first line is even in $s$. The saddle-point is at $s_* = \frac{3}{2\pi}(\alpha-\frac{\pi}{3})$. This requires the cone angle to satisfy
\begin{equation}
    \frac{\pi}{3}<\alpha <\frac{5\pi}{3},
\end{equation}
otherwise $s_*$ hits a pole of $1/\sin(\pi s)$.\footnote{The apparent pole at $s_* = 1$ is canceled by a zero of the numerator.} Evaluating the integrand on the saddle-point and including the 1-loop fluctuation, we have thus shown that
\begin{equation}
    I_\kappa(\alpha) = \frac{2\sqrt{3}\sin(\alpha)}{\sin(\frac{3\alpha-\pi}{2})} \exp\left(\frac{\kappa}{\pi}\frac{(3\alpha-\pi)^2}{12}\right)\left[1+\order{\tfrac{1}{\kappa}}\right].
\end{equation}

Let us now turn to $b_+$. The steepest descent path of the $P$-integral is $\arg P = -\pi/4$. Hence along the steepest descent contour $PP_1^+$ is real, so the cosine $\cos(4\pi PP_1)$ is oscillating. The $P$-integral then does not have a saddle-point: it only gives at most a power law in $P_1$. By analogy with the Chern-Simons analysis of the trefoil \cite{Witten:2010cx,Dimofte:2011gm}, we can call the $b_+$ sector the abelian branch, and the $b_-$ sector the non-abelian geometric branch. The only exponentially growing contribution comes from $e^{\frac{\pi i}{12}Q_+^2}\sim e^{\frac{\pi \kappa}{12}}$, which cancels the corresponding factor in $Z^{(b_-)}_{\rm Vir}$. 

Combining the large $\kappa$ asymptotics of the two factors $Z^{(b_\pm)}_{\rm Vir}$ we have shown that
\begin{equation}
    Z_{\rm grav}^{\Lambda>0}[\mathbf{3}_1;\alpha] \sim \exp(\frac{\text{Vol}(\mathbf{3}_1;\alpha)}{4\pi G_N \ell_{\rm dS}^2}), \quad \text{Vol}(\mathbf{3}_1;\alpha) =\ell_{\rm dS}^3\frac{(3\alpha-\pi)^2}{12}. 
\end{equation}
This agrees with the spherical volume of the trefoil cone manifold computed in \cite{KolpakovMednykhSpherical}, confirming our spherical volume conjecture for the trefoil knot complement. This is interesting, because the trefoil has no hyperbolic saddle for any $\alpha$: nevertheless, there is a spherical regime $\frac{\pi}{3}<\alpha<\frac{5\pi}{3}$ in which the analytically continued Virasoro TQFT partition function is well-defined and produces the correct spherical saddle.

\subsubsection*{Spherical Dehn fillings}
The trefoil also provides a source of examples of smooth spherical manifolds. Rational Dehn surgery on the trefoil generates, up to orientation,
every spherical three-manifold of tetrahedral, octahedral, or
icosahedral type
\cite[proof of Corollary~8, p.~674]{DoigFiniteKnotSurgeries}.
It also generates a subfamily of dihedral (prism) manifolds,
but not every spherical quotient. We focus here on the four
integer slopes $n=1,2,3,4$, which give the quotients by the
binary icosahedral, octahedral, tetrahedral, and dihedral group, respectively. These provide one representative
of each non-cyclic spherical family. 

Let $Y_n$ denote integer surgery of the trefoil, with the usual slope $n\mu+\lambda$, and set $k=6-n$. For the four cases $n=1,2,3,4$, the resulting spherical manifolds are Seifert fibrations over $S^2(2,3,k)$ with $k=6-n$. In standard Seifert notation, their invariants are $(-1;(2,1),(3,1),(k,1))$, up to orientation \cite{MoserTorusKnotSurgery}. They are quotients $S^3/\Gamma_n$, where the binary polyhedral groups $\Gamma_n\subset SU(2)$ act freely by multiplication on $S^3\simeq SU(2)$. Consequently,
\begin{equation}\label{eq:trefoil-surgery-quotient-volume}
|\Gamma_n|=\frac{4}{\frac12+\frac13+\frac1k-1}
=\frac{24k}{n},
\qquad
V_n\equiv\frac{\text{Vol}(Y_n)}{\ell_{\mathrm{dS}}^3}
=\frac{2\pi^2}{|\Gamma_n|}
=\frac{\pi^2 n}{12k}.
\end{equation}
Explicitly, the discrete groups and (smooth spherical) volumes are \cite{HikamiSphericalSeifert}
\begin{equation}\label{eq:trefoil-surgery-examples}
\begin{array}{c|c|c|c|c}
n&k&\Gamma_n&|\Gamma_n|&V_n\\ \hline
1&5&\text{binary icosahedral}&120&\pi^2/60\\
2&4&\text{binary octahedral}&48&\pi^2/24\\
3&3&\text{binary tetrahedral}&24&\pi^2/12\\
4&2&\text{binary dihedral }&12&\pi^2/6
\end{array}
\end{equation}
To implement the Dehn filling on the level of Virasoro TQFT, we integrate against the kernel $(ST^{-k})_{\mathds{1}P_1} = \rho_0(P_1)e^{-2\pi i  k(P_1^2-\frac{1}{24})}$, giving 
\begin{equation}
    Z^{(b)}_{\rm Vir}[Y_n] = \int_{\mathbb{R}+i\epsilon} \!\frac{\dd P}{2}\, \frac{(ST^{-k}S)_{\mathds{1}P}(ST^{-3} S)_{\mathds{1}P} (ST^2 S)_{\mathds{1}P}}{\rho_0(P)}
\end{equation}
Substituting equation (\ref{eq:Dehn_kernels}) and keeping the projective phase explicit, we obtain
\begin{align} \label{eq:trefoil-Seifert-VTQFT}
&Z_{\rm Vir}^{(b)}[Y_n]
=\frac{1}{\sqrt{6k}}
\exp\left[\frac{\pi i}{12}\left(k-5\right)
-\frac{\pi i nQ^2}{12k}\right]\times\nonumber\\
&\int_{\mathcal C}\dd P\,e^{\frac{\pi i nP^2}{3k}}
\frac{
\left(\cosh\left(\frac{2\pi QP}{3}\right)
-e^{2\pi i/3}\cosh\left(\frac{2\pi \Hat{Q}P}{3}\right)\right)
\left(\cosh\left(\frac{2\pi QP}{k}\right)
-e^{2\pi i/k}\cosh\left(\frac{2\pi \Hat{Q}P}{k}\right)\right)}{
\sinh(\pi bP)\sinh(\pi P/b)
} 
\end{align}
where $k=6-n.$
We now continue to $b_\pm^2=\pm i/\kappa$, with the integration
contours deformed accordingly.
Evaluating the integral in the $b_-$ sector in the semiclassical limit $\kappa \to \infty$, we obtain (see Appendix \ref{app:Trefoil} for details):
\begin{align} \label{eq:trefoil-Seifert-bminus}
Z_{\rm Vir}^{(b_-)}[Y_n]
={}&\frac{e^{\frac{\pi i}{12}(k-2)}}{\pi}
\sqrt{\frac{2\kappa}{k}}\sin\frac{\pi}{k}
\left[i\pi-\log\left(1-2\cos\frac{2\pi}{k}\right)\right]
\nonumber\\
&\times
e^{\frac{\pi\kappa n}{12k}}
\left[1+O(\kappa^{-1})\right].
\end{align}
In the $b_+$ sector, the exponential contribution from the
Gaussian saddle cancels the decaying prefactor, yielding
\begin{equation} \label{eq:b+semiclassics}
Z_{\rm Vir}^{(b_+)}[Y_n]
=\frac{4\sqrt{2}\pi}{n^{3/2}\kappa}
e^{\frac{\pi i}{12}(k+13)}
\left[1+\order{\kappa^{-1}}\right].
\end{equation}
Multiplying the two chiral contributions, we obtain\footnote{The net factor $\kappa^{-1/2}$ is consistent with relative
zero-mode counting. The manifolds $Y_n$ have three Killing vectors,
whereas a generic lens space has two \cite{McCullough2002}.
For these rigid saddles, each additional residual gauge generator
contributes a factor $\kappa^{-1/2}$ at one loop \cite{Adams1998}.
Thus the relative power agrees with our lens-space result (\ref{eq:VTQFTLenspace}), although
this comparison does not fix the absolute path-integral normalization. In the calculation leading to the result (\ref{eq:b+semiclassics}), the $\frac{1}{\kappa}$ factor shows up because after rescaling the integration momentum, one can show that the relevant portion of the integral takes the form $\sqrt{\kappa}\int_{-\infty}^{\infty}du\, u^2 e^{-a\kappa u^2}\sim \frac{1}{\kappa}$.}
\begin{align}
Z^{\Lambda>0}_{\rm grav}[Y_n]
&=Z_{\rm Vir}^{(b_+)}[Y_n]Z_{\rm Vir}^{(b_-)}[Y_n]
\\[0.7em]
&=\frac{8e^{\frac{\pi i}{6}(k+4)}
\sin(\pi/k)}{n^{3/2}\sqrt{k\kappa}}
\left[i\pi-\log\left(1-2\cos\frac{2\pi}{k}\right)\right]
e^{\frac{\pi\kappa n}{12k}}
\left[1+\order{\kappa^{-1}}\right].\nonumber
\end{align}
Thus, in the semiclassical limit,
\begin{equation}
\log|Z^{\Lambda>0}_{\rm grav}[Y_n]|
\sim \frac{\kappa}{\pi\ell_{\rm dS}^3}
\text{Vol}(Y_n),
\qquad
\text{Vol}(Y_n)
=\frac{\pi^2 n}{12k}\ell_{\rm dS}^3,
\end{equation}
in agreement with (\ref{eq:trefoil-surgery-quotient-volume}) and the spherical volume conjecture.

\section{Discussion}
\label{sec:discussion}

In this paper, we proposed a prescription \eqref{eq:zgrav} for computing the exact Euclidean gravitational partition function with positive cosmological constant on a large class of 3-dimensional spherical topologies with conical defects (interpreted as observer worldlines). This class was defined such that in one regime of cone angles $\alpha$ there is an AdS$_3$ saddle, for which the exact quantization is well-defined through Virasoro TQFT \cite{Collier:2023fwi}. Importantly, the Virasoro TQFT partition is an analytic function of $b^2\propto G_N/\ell_{\rm AdS}$ and meromorphic in $P=\frac{i\alpha}{4\pi b}$, which allowed us to analytically continue to the dS$_3$ regime at the level of the exact partition function. (Note that this is different from trying to continue  the $\Lambda<0$ on-shell action directly, which generically does \emph{not} reproduce the $\Lambda>0$ on-shell action.) 

We then showed that the semiclassical limit of $Z_{\rm grav}^{\Lambda>0}[M_{\vec{\alpha}}]$ reproduces the on-shell action of gravity coupled to a worldline, which is proportional to the spherical volume, in all the examples we checked. This provides non-trivial evidence for the validity of our proposal, at least in the semiclassical limit. In particular, the analytic continuation starting from the hyperbolic branch provided the non-perturbative integration contour for the saddle-point analysis, which non-trivially selects the geometric saddles needed to obtain the correct spherical volume.

To see why this is non-trivial, compare to Chern-Simons theory with purely imaginary level $k_\pm = \pm i\kappa$, with $\kappa = \frac{\ell_{\rm dS}}{4G_N}$, which is classically equivalent to 3D gravity with $\Lambda>0$, with gauge group $SL(2,\mathbb{C})$ in Lorentzian signature and $(SU(2)_L\times SU(2)_R) / \mathbb{Z}_2$ in Euclidean signature \cite{Witten:1988hc,Witten:1989ip,Hikida:2021ese,Castro:2023dxp}. The Chern-Simons path integral has saddle points (flat connections) that do not correspond to non-degenerate metrics (e.g.~the trivial connection $A = \bar A = 0$) so a naive sum over Chern-Simons saddles does not reproduce the spherical volume on-shell. Restricting to the dominant saddle does not help either: the geometric volume term need not be the dominant Chern-Simons saddle. For example, for the three-twist knot studied in section \ref{sec:threetwist}, the spherical saddle was the pairing of the geometric with the anti-geometric saddle, whereas the dominant saddle in complex Chern-Simons theory on the same topology would have been the pairing of the geometric with the non-geometric saddle. That is, in Chern-Simons theory the correct `gravitational' flat connection has to be picked by hand, whereas in our proposal the correct saddle points are uniquely determined by the Lefschetz thimble analysis. This gives us confidence that Virasoro TQFT gives the right prescription for 3D de Sitter.

  Despite the evidence gathered in this paper, we do not yet have a first-principles derivation of a quantization of 3D de Sitter with conical defects. We emphasize that the complex Virasoro TQFT amplitudes in our proposal are computed in the AdS$_3$ regime first, where they are well-defined, and then analytically continued to dS$_3$; but we have not derived a new `de Sitter TQFT' that can be used to assemble dS$_3$ partition functions from arbitrary spherical building blocks. We will now comment on the problem of quantizing dS$_3$ from first principles (closely related comments can be found in \cite{Collier:2025lux}).

  \paragraph{Quantizing the dS$_3$ phase space.} In Lorentzian signature, on manifolds $\Sigma_{g,n} \times \mathbb{R}$ where $\Sigma_{g,n}$ has $n$ punctures, it is known that the phase space of 3D gravity with $\Lambda>0$ is the space of complex projective structures $\mathcal{CP}(\Sigma_{g,n})$ \cite{andersson2007notes}. A useful set of coordinates are Thurston's shear-bend coordinates $z^\pm_j = x_j \pm i y_j$ associated to an ideal triangulation of $\Sigma_{g,n}$. In \cite{Kim:2021nyy,Kim:2023wsq}, the classical Poisson brackets of $z^\pm_j$ are promoted to canonical commutators of operators $\hat{z}^\pm_j$ satisfying $(\hat{z}^\pm_j)^\dagger = \hat{z}^\mp_j$,\footnote{The papers \cite{Kim:2021nyy,Kim:2023wsq} are more generally applicable to $\Lambda=-1,0$ and $1$; we only briefly summarized the $\Lambda>0$ case.} with quantization parameters $\pm i\hbar$. Changes of triangulation are represented by operator-valued Faddeev's quantum dilogarithms, like in ordinary quantum Teichm\"uller theory \cite{Chekhov:1999tn}, but analytically continued to real $q_\pm = e^{\mp \pi \hbar}$. Interestingly, this is the same analytic continuation as the double-sine function \eqref{eq:doublesinepochhammer} that appears in our analytically continued Virasoro TQFT partition functions; the dictionary is $b_\pm^2 = \pm i\hbar$. It would be interesting to understand how their quantization is related to our proposed quantization using Virasoro TQFT. One drawback of the above formalism is that it only works for $\Sigma_{g,n}\times \mathbb{R}$, whereas the topologies we have studied are more complicated. One would need to further develop the dS$_3$ quantization of \cite{Kim:2021nyy,Kim:2023wsq} into a version of Anderson-Kashaev Teichm\"uller TQFT \cite{EllegaardAndersen:2011vps}, which so far has not been done.  
  
A second approach is to use the fact that $\mathcal{CP}(\Sigma_{g,n})$ is symplectomorphic to the cotangent bundle of Teichm\"uller space, $T^*\mathcal{T}_{g,n}$, and do geometric quantization \cite{Verlinde:1989ua}. As explained e.g.~in \cite{Verlinde:2024zrh}, the physical state conditions on the wavefunctions become the Virasoro Ward identities with a pair of complex conjugate central charges $c_\pm \in 13+i\mathbb{R}$, so the gravitational Hilbert space is spanned by products of Virasoro conformal blocks with central charges $c_\pm$. As a vector space, this is the same as that of Virasoro TQFT. Thus, our product of analytically continued partition functions $Z_{\rm Vir}^{(b_+)}Z_{\rm Vir}^{(b_-)}$ defines a particular wavefunction in the gravitational Hilbert space. 

However, as emphasized in \cite{Collier:2025lux}, the natural inner product on this Hilbert space is not diagonalized by conformal blocks. A suitable basis should therefore be special (and so-far unknown) linear combinations of Virasoro conformal blocks that have been termed ``Virasoro partial waves'' in \cite{Collier:2025lux}. Making this concrete, and testing whether the resulting TQFT partition functions agree with the factorized form of our proposal \eqref{eq:zgrav}, is an  important future direction.

 \paragraph{Knotted contributions to the observer worldline correlators.} Another direction is to investigate possible non-perturbative corrections to observer worldline correlators from configurations in which probe worldlines are knotted around the observer. The unknotted theta graph studied in section \ref{sec:observer2point} suggests approaching these contributions through analytic continuation of the corresponding Virasoro TQFT amplitudes. Concrete starting points are the single-Wilson-line fragmentations of genus-one handlebody-knots studied in \cite{Chandra:2025fef}. Further examples are the theta-graph spines of genus-two handlebody-knots \cite{Ishii:2012HandlebodyTable}, with related Virasoro TQFT calculations appearing in \cite{Belin:2026pko}. Knot-fragmentation amplitudes contributing to the $6j$-contraction of OPE coefficients in \cite{Chandra:2025fef} may also provide  non-perturbative contributions to the observer four-point kernel beyond the spherical tetrahedron contribution considered in this paper. It would be interesting to investigate their effect on the OTOC and their possible relevance to the positivity puzzle of \cite{Chen:2026boh, Harlow:2026pwe, Cui:2026bcd, Milekhin:2026tbi}.

\paragraph{Extending our proposal to other spherical manifolds?} Given that our proposal appears to work for certain examples of Seifert manifolds, it is natural to consider extending the prescription to all spherical manifolds, provided that the underlying Virasoro TQFT amplitude remains finite. However, more care is needed when considering Seifert-fibred spherical cone manifolds. (Note that spherical cone manifolds admitting a hyperbolic transition are not Seifert fibred.)

When $M_{\vec{\alpha}}$ is not Seifert-fibred, and all its cone angles are smaller than $\pi$, global rigidity guarantees that if a spherical metric exists, it is unique up to isometry \cite{Weiss2007}. In particular, there are no continuous fixed-angle moduli, and geometric quantities such as the volume are determined by topological and angular data. This is analogous to Mostow rigidity in the hyperbolic setting, which states that a complete, finite-volume hyperbolic manifold of dimension greater than two is determined up to isometry by its fundamental group. 
For Seifert-fibred spherical cone manifolds, such rigidity cannot be assumed in general. Indeed, there are explicit examples that admit continuous families of non-isometric metrics with fixed cone angles. One of these examples is the torus link $T(4,4)$ examined in \cite{kolpakov2013examples}:
\begin{equation}
\label{eq:counterEx}
\vcenter{\hbox{
    \begin{tikzpicture}[scale=0.7]
\draw[very thick] (0.6,0.5) -- (0.65,0.5)
    to[out=0,in=90] (1,0.15) -- (1,-0.4);
\draw[very thick] (1,-0.6) -- (1,-1.15)
    to[out=-90,in=0] (0.65,-1.5) -- (-0.4,-1.5);
\draw[very thick] (-0.6,-1.5) -- (-1.65,-1.5)
    to[out=180,in=-90] (-2,-1.15) -- (-2,0.15)
    to[out=90,in=180] (-1.65,0.5) -- (0.4,0.5);

\draw[very thick] (-1.5,0.6) -- (-1.5,1.65)
    to[out=90,in=180] (-1.15,2) -- (0.15,2)
    to[out=0,in=90] (0.5,1.65) -- (0.5,-0.4);
\draw[very thick] (0.5,-0.6) -- (0.5,-0.65)
    to[out=-90,in=0] (0.15,-1) -- (-0.4,-1);
\draw[very thick] (-0.6,-1) -- (-1.15,-1)
    to[out=180,in=-90] (-1.5,-0.65) -- (-1.5,0.4);

\draw[very thick] (-1,0.6) -- (-1,1.15)
    to[out=90,in=180] (-0.65,1.5) -- (0.4,1.5);
\draw[very thick] (0.6,1.5) -- (1.65,1.5)
    to[out=0,in=90] (2,1.15) -- (2,-0.15)
    to[out=-90,in=0] (1.65,-0.5) -- (-0.4,-0.5);
\draw[very thick] (-0.6,-0.5) -- (-0.65,-0.5)
    to[out=180,in=-90] (-1,-0.15) -- (-1,0.4);

\draw[very thick] (-0.5,0.6) -- (-0.5,0.65)
    to[out=90,in=180] (-0.15,1) -- (0.4,1);
\draw[very thick] (0.6,1) -- (1.15,1)
    to[out=0,in=90] (1.5,0.65) -- (1.5,-0.4);
\draw[very thick] (1.5,-0.6) -- (1.5,-1.65)
    to[out=-90,in=0] (1.15,-2) -- (-0.15,-2)
    to[out=180,in=-90] (-0.5,-1.65) -- (-0.5,0.4);
\end{tikzpicture}}}\;,
\end{equation}
which admits a continuous family of spherical metrics labelled by a parameter $\tau$ which determines the overall shape of the cone manifold. 
Although the spherical volume is the same for every member of this family, the existence of these deformations means that, if these geometries contribute as saddles, an integral over their moduli must be accounted for in the gravitational path integral. It is not clear if the Virasoro TQFT prescription already incorporates this integration or if it requires an additional ingredient.

\paragraph{The mapping class group.} Alongside rigidity, spherical cone manifolds admitting a hyperbolic transition inherit another important property: a finite mapping class group. Finiteness in the hyperbolic setting, proved in \cite{kojima1998deformations}, persists across the transition, since the mapping class group depends only on the underlying topology. For Seifert-fibred spherical cone manifolds, however, the mapping class group can be infinite. The four-component link \eqref{eq:counterEx} again gives an example. The link consists of four fibers in the Hopf fibration, so its complement is an $S^1$-bundle over the four-punctured sphere $\Sigma_{0,4}$. Since the mapping class group of $\Sigma_{0,4}$ injects into that of the cone manifold, the mapping class group of the cone manifold is also infinite. 

Let us close this discussion with a related subtlety: the distinction between knotted conical defects and torus boundaries in the definition of the mapping class group. Here, we define the \emph{bulk} mapping class group using isotopies that preserve cone singularities setwise and fix boundaries pointwise. Fixing a singular core allows for rotations in its transverse and longitudinal directions, while fixing the boundary of a tubular neighbourhood pointwise excludes these isotopies. 
To illustrate this distinction, we can compare the mapping class group of the annulus with that of the once-punctured disk. With both boundaries fixed pointwise, the mapping class group of the annulus is $\mathbb{Z}$; it is generated by a Dehn twist. In the once-punctured disk, however, this twist can be continuously undone by rotating a neighbourhood of the puncture while keeping the puncture fixed. The mapping class group of the one-punctured disk is therefore trivial. This distinction  is relevant in the example of the Hopf link. If its components represent asymptotic boundaries, then the manifold (which has the topology of $T^2\times [0,1]$) has a $\mathbb{Z}\times \mathbb{Z}$ bulk mapping class group. If instead the components represent conical defects, these Dehn twists become isotopic to the identity, and the bulk mapping class group is finite.

\section*{Acknowledgments}

We thank Ahmed Abdalla, Jan de Boer, Scott Collier, Lorenz Eberhardt, Tom Hartman, Luca Iliesiu, Juan Maldacena, Tommaso Marini, Victor Rodriguez, Martin Sasieta, Douglas Stanford, Herman Verlinde and Gabriel Wong for helpful discussions. 
We are grateful to the organizers of the workshop ``Random Geometry in Math and Physics" at the Simons Center for Geometry and Physics, Stony Brook for inviting us to attend and participate in the workshop, which helped initiate this collaboration.  The work of JC is supported in part by the Leinweber Institute for Theoretical Physics at UC Berkeley; and by the Department of Energy, Office of Science, Office of High Energy Physics through award DE-SC0025293 and QuantISED award DE-SC0019380.  BP is supported by the ERC Consolidator Grant GeoChaos No 101169611. DL is supported by the SNF starting grant “The Fundamental Description of the Expanding Universe”. DL is also supported by the Fonds National Suisse de la Recherche Scientifique (Schweizerischer Nationalfonds zur Förderung der wissenschaftlichen Forschung) through the Project Grant 200021\_215300 and the NCCR51NF40-141869 The Mathematics of Physics (SwissMAP). Part of this work is funded by the European Union. Views and opinions expressed are however those of the authors only and do not necessarily reflect those of the European Union or the European Research Council Executive Agency. Neither the European Union nor the granting authority can be held responsible for them. For the purpose of open access, the authors have applied a CC BY public copyright license to any Author Accepted Manuscript (AAM) version arising from this submission.

\paragraph{Statement of AI use:} 
The manuscript was entirely written by the authors. The conception, analysis and computations were all carried out by the authors who take full responsibility for the validity of their work. AI was used in an exploratory phase to help locate relevant mathematics literature on spherical cone manifolds. It was also used to write a code that numerically verifies a double sine integral identity. It helped determine the locations of the worldlines in the tetravalent theta graph from the holonomies of the figure-eight knot. Lastly, it helped in learning some standard facts about Picard-Lefschetz theory and in writing a Mathematica code for plotting the thimbles.

\appendix

\section{Special functions}\label{app:specialfunctions}
Here we collect some known facts about the double sine function and the (non-)compact quantum dilogarithm function which are needed in the main text. Many of the remarkable properties of the double sine/quantum dilogarithm have been collected in \cite{Eberhardt:2023mrq}.

The double sine function is defined as a ratio of Barnes double gamma functions,
\begin{equation}
    S_b(z) = \frac{\Gamma_b(z)}{\Gamma_b(Q-z)}.
\end{equation}
Hence clearly $S_b(Q-z) = 1/S_b(z)$.
The double sine function $S_b(z)$ can be expressed in terms of Faddeev's non-compact quantum dilogarithm function as
\begin{equation}\label{eq:quantumdilog}
\Phi_b(z)=S_b(\tfrac{Q}{2}+iz)e^{\frac{\pi i}{2}z^2+\frac{i\pi}{24}(Q^2-2)}  = \frac{e^{\frac{\pi i}{2}z^2+\frac{i\pi}{24}(Q^2-2)}}{S_b(\tfrac{Q}{2}-iz)}.
\end{equation}
For $b\in \mathbb{R}_{>0}$, the quantum dilogarithm admits the following integral representation
\cite{Faddeev:1993rs}:
\begin{equation}
   \log \Phi_b(z) = \int_{\mathbb{R}+i\epsilon} \frac{dw}{4w}\frac{e^{-2i zw}}{\sinh(bw)\sinh(b^{-1}w)}.
\end{equation}
This converges in the strip $|\mathrm{Im}(z)|<\frac{Q}{2}$, and analytically continues to a meromorphic function of $z$ in the complex plane. 

For $\mathrm{Im}(b^2)>0$, the non-compact quantum dilogarithm can be written as a ratio of compact quantum dilogarithms \cite{Kashaev_2011}, which are themselves $q$-Pochhammer symbols:
\begin{equation}\label{eq:quantdilog}
    \Phi_b(z) = \frac{\psi_{\tilde{q}}(e^{2\pi b^{-1}z})}{\psi_q(e^{2\pi  b z})}, \qquad \psi_q(x) = \frac{1}{(-q x;q^2)_\infty} = \prod_{n=0}^\infty (1+q^{2n+1}x)^{-1}.
\end{equation}
Here we defined the nome $q = e^{
\pi ib^2}$ and its modular transform $\tilde{q} =e^{-\pi i/b^2}$. Plugging this into \eqref{eq:quantumdilog}, we see that the double sine function with $\Im{(b^2)}>0$ can be written as
\begin{equation}\label{eq:doublesinepsi}
    S_b(\tfrac{Q}{2}+z) =\frac{\psi_q(e^{2\pi i b z})}{\psi_{\tilde{q}}(e^{2\pi i z/b})} e^{-\frac{\pi i}{2}z^2+\frac{i\pi}{24}(Q^2-2)} .
\end{equation}
This is equation \eqref{eq:doublesinepochhammer} in the main text. Substituting the definition of the compact quantum dilogs in terms of $q$-Pochhammer symbols, it also agrees with (B.46) in \cite{Eberhardt:2023mrq}.

Let us specialize to the case that $b^2$ is purely imaginary, and let $ b^2 = \frac{i}{\kappa}\in i\mathbb{R}_{>0}$ so that $q = e^{-\pi/ \kappa }\in (0,1)$.
The compact quantum dilogarithm has the following classical limit as $\kappa\to \infty$ \cite{Faddeev:1993rs}:
\begin{equation}\label{eq:compactpsilimit}
    \psi_q (x) \sim \exp(\frac{ \kappa}{2\pi}\mathrm{Li}_2(-x)).
\end{equation}
We adopt the standard choice for the branch cut of $\mathrm{Li}_2(x)$ to lie on $(1,\infty)$. As $q\to 1$, its S-dual approaches $\tilde{q}\to 0$, and hence $\lim_{\tilde q \to 0}\psi_{\tilde{q}} = 1$. Substituting these asymptotic expansions into \eqref{eq:quantdilog}, we obtain the following asymptotics of $\Phi_b$ as $\kappa\to \infty$:
\begin{equation}
    \log \Phi_b\left(\frac{u}{2\pi b}\right)\sim -\frac{\kappa}{2\pi}\text{Li}_2(-e^u), \qquad \kappa\to\infty. 
\end{equation}

Finally, with \eqref{eq:quantumdilog} we also obtain the semiclassical limit of the double sine function at real $q = e^{-\pi/\kappa}$, as $\kappa\to\infty$:
\begin{equation}\label{app:doublesineasymp}
    \log S_b(b^{-1}z) \sim -\frac{\kappa}{2\pi }\left( \Li_2(e^{-2\pi i z})-\pi^2(z^2-z)-\frac{\pi^2}{6}\right).
\end{equation}
This is equation \eqref{eq:doublesinekappa} in the main text.
We can use the dilogarithm identity
\begin{equation}
\mathrm{Li}_2(x)+\mathrm{Li}_2(1/x) = -\frac{\pi^2}{6}-\frac{1}{2}(\log(-x))^2
\end{equation}
to rewrite this in two equivalent forms
\begin{align}\label{eq:form2}
\log S_b(b^{-1}z) &\sim \frac{\kappa}{2\pi }\left( \Li_2(e^{2\pi i z})-\pi^2(z^2-z)-\frac{\pi^2}{6}\right)\\
    \log S_b(b^{-1}z) &\sim \frac{\kappa}{4\pi } (\text{Li}_2(e^{2\pi i z}) - \text{Li}_2(e^{-2\pi i z})).
\end{align}
Using $\kappa = \frac{i}{b^2}$, this is the same asymptotic form of the double sine as for real $b$ \cite{Eberhardt:2023mrq}.

\section{Saddle-point equation for $6j$ symbol}\label{app:saddlepoint6j}

The saddle-point equation of the $6j$ symbol asymptotics was written in \eqref{eq:saddle-equation}. It is a quadratic equation for $z = e^{-s/2}$. Its two roots can be written as
\begin{equation}\label{eq:root}
     z_0^\pm = \frac{-q_1 \pm \sqrt{q_1^2 - 4q_0q_2}}{2q_2} 
 \end{equation}
where $x_j = e^{-i\theta_j}$ and
 \begin{align}
     q_0 &= x_1^{-1}x_4^{-1} + x_2^{-1}x_5^{-1} + x_3^{-1}x_6^{-1}
     + x_1^{-1}x_2^{-1}x_6^{-1} + x_1^{-1}x_3^{-1}x_5^{-1} \\ \nonumber
    &\qquad + x_2^{-1}x_3^{-1}x_4^{-1} + x_4^{-1}x_5^{-1}x_6^{-1}
     + x_1^{-1}x_2^{-1}x_3^{-1}x_4^{-1}x_5^{-1}x_6^{-1},\\
q_1 &= -(x_1^{-1}-x_1)(x_4^{-1}-x_4)
      -(x_2^{-1}-x_2)(x_5^{-1}-x_5)
      -(x_3^{-1}-x_3)(x_6^{-1}-x_6),\\
q_2 &= x_1x_4 + x_2x_5 + x_3x_6
     + x_1x_2x_6 + x_1x_3x_5 + x_2x_3x_4  + x_4x_5x_6
     + x_1x_2x_3x_4x_5x_6.
 \end{align}
 Both $q_1$ and $q_0q_2$ are real and positive for $\theta_j\in (0,\pi)$, hence 
 the numerator of $z_0^\pm$ is real and negative. So in the spherical regime,
\begin{equation}
    \text{Im} \log z_0^\pm = -\text{arg}(-q_2).
\end{equation}
Moreover, the discriminant $q_1^2 - 4q_0q_2$ is positive in the spherical regime. Up to a numerical prefactor, it coincides with the determinant of the angle Gram matrix of the spherical tetrahedron:
\begin{equation}
    D \coloneqq q_1^2-4q_0q_2 = 16\det G
\end{equation}
where $G$ was defined in \eqref{eq:grammatrix}.

In the $6j$ symbol analysis of section~\ref{sec:observerOTOC}, which focused on the
continuation to $b_+$, only the upward thimble of the root $z_0^-$
intersects the integration contour (rotated with $b_+$).
This is the root with $|z_0^-|>1$.
Evaluating the real part of the action on this saddle gives plus
the volume function of Murakami's Theorem~1.1~\cite{murakami2012volume}, as in (\ref{eq:6jactionvol}).
For the $b_-$ continuation, the roles are reversed: using reflection
symmetry to keep the same $x_j=e^{-i\theta_j}$ convention, the contour
picks up the $z_0^+$ saddle, with $|z_0^+|<1$, where the real part of
the same potential gives minus Murakami's volume formula.
Its coefficient in the exponent also changes sign, so the two
continuations give the same growing real exponential.

\paragraph{Heavy-light limit.} For the OTOC analysis, we specialize to the heavy-light configuration of dihedral angles in \eqref{eq:otocx}, \eqref{eq:otocx2}. The individual terms in the saddle-point equation become
\begin{align}
  1-x_1x_2x_4x_5z
  &= 1-\exp(
        \delta
        -\frac{i\pi}{\kappa}\omega
)=
  -\delta
  +\frac{i\pi}{\kappa}\omega
  +O(\kappa^{-3}),
  \\
  1-x_2x_3x_5x_6z
  &=1-\exp(
        \delta
        +\frac{i\pi}{\kappa}\omega
)=
  -\delta
  -\frac{i\pi}{\kappa}\omega
  +O(\kappa^{-3}),\\
    1+x_1x_2x_3z
    =
    1+x_2x_4x_6z
    &=1- \exp(\delta-\frac{i\pi\nu}{\kappa})=
    \frac{i\pi\nu}{\kappa}
    +O(\kappa^{-2}),
    \\
    1+x_1x_5x_6z
    =
    1+x_3x_4x_5z
    &= 1- \exp(\delta-\frac{i\pi\nu'}{\kappa})= 
    \frac{i\pi\nu'}{\kappa}
    +O(\kappa^{-2}),
\end{align}
\begin{equation}
    (1-z)(1-x_1x_3x_4x_6z)
    =
    (1-x^{-2})(1-x^2)+O(\kappa^{-1})
    =
    4\sin^2\theta_H+O(\kappa^{-1}).
    \label{eq:regular-factor}
\end{equation}
Here we kept only the leading-in-$\kappa$ scaling: in the first two terms, the leading non-zero term is order $\kappa^{-2}$ (if we use that $\omega$ is order $\omega_c$), in the middle terms the first non-zero term is order $\kappa^{-1}$, whereas the last term is order 1 (if we keep $\theta_H$ fixed, so finite backreaction). Multiplying out the saddle-point equation \eqref{eq:saddle-equation} gives
\begin{equation}
    4\sin^2\theta_H
    \left(
       \delta^2
       +
       \left(
          \frac{\pi \omega}{\kappa}
       \right)^2
    \right)
    =
    \left(\frac{\pi}{\kappa}\right)^4
    \nu^2\nu'^2,
\end{equation}
which is equation \eqref{eq:saddlepointdelta} in the main text.

\section{The pure-recoil limit of the backreacted tetrahedron}

\label{sec:tetra-recoil}

In this appendix, we take the pure-recoil limit of the spherical tetrahedron in which all gravitational
deficit angles vanish while the proper masses of the observer segments and the probe masses are kept fixed. We will first show
that the classical volume of the doubled tetrahedron \eqref{eq:6jactionvol} reduces to the action
of a tetrahedral worldline graph on $S^3$. This is the additional input that
follows from the spherical $6j$ saddle. Once this limit has been taken, the
remaining relation between the tetrahedral graph and the recoil OTOC is the
pure-recoil calculation of \cite{Chen:2026boh}.

Throughout this appendix we use
\begin{equation}
\kappa=\frac{\ell_{\rm dS}}{4G_N},
\qquad
d_j=\frac{L_j}{\ell_{\rm dS}},
\end{equation}
and measure the particle masses in de Sitter units. Thus $m_j d_j$ is the
dimensionless action of a worldline segment. The recoil limit is
\begin{equation}
\alpha_j=2\pi\left(1-\frac{m_j}{\kappa}\right),
\qquad
\theta_j=\frac{\alpha_j}{2}
=\pi\left(1-\frac{m_j}{\kappa}\right),
\qquad
\kappa\rightarrow\infty,
\label{eq:recoil-scaling}
\end{equation}
with all six $m_j$ fixed. The physical cone deficits
$2\pi-\alpha_j=2\pi m_j/\kappa$ therefore vanish. We take the masses large
only after this limit, so that the resulting particles can be treated as
worldlines without retaining a finite gravitational backreaction.

The geometric saddle of the spherical $6j$ symbol obeys the Schl\"afli formula,
\begin{equation}
{\rm Re}\,S
=
\frac{{\rm Vol}(T_{\vec\alpha})}{\ell_{\rm dS}^3},
\qquad
d\,{\rm Re}\,S
=
\frac12\sum_{j=1}^6 d_j\,d\theta_j,
\label{eq:schlafli-recoil}
\end{equation}
where $T_{\vec\alpha}$ is a single spherical tetrahedron. To integrate this
relation we first determine the geometry at the endpoint $\kappa=\infty$.

Let $N_a$, $a=1,\ldots,4$, be the outward unit normals to the four faces of the tetrahedron.
If faces $a,b$ meet along the edge $j$, then
\begin{equation}
N_a\cdot N_b
=
-\cos\theta_j
=
\cos\left(\frac{\pi m_j}{\kappa}\right)
\longrightarrow 1.
\label{eq:normal-recoil}
\end{equation}
Thus all four face normals approach the same unit vector $N$. The tetrahedron
is the intersection of the four hemispherical half-spaces
\begin{equation}
T_{\vec\alpha}
=
\left\{
X\in S^3\,\big|\,N_a\cdot X\leq0,\quad a=1,\ldots,4
\right\}.
\end{equation}
In the limit these four inequalities become the single condition
$N\cdot X\leq0$. Hence the tetrahedron approaches a hemisphere in the recoil limit. Its boundary is the equatorial $S^2$ defined by
$N\cdot X=0$, and the four limiting vertices remain as four generally
distinct points on this $S^2$. It follows that
\begin{equation}
{\rm Re}\,S\longrightarrow \pi^2,
\qquad
d_j(\kappa)\longrightarrow d_j^{(0)},
\label{eq:hemisphere-endpoint}
\end{equation}
where the finite lengths $d_j^{(0)}$ depend on the fixed mass ratios.

We can now integrate the Schläfli relation along the trajectory
\eqref{eq:recoil-scaling}. Since $\frac{d\theta_j}{d\kappa}
=
\frac{\pi m_j}{\kappa^2},$
we have
\begin{equation} \label{eq:schlafli-kappa-derivative}
\frac{d}{d\kappa}{\rm Re}\,S
=
\frac{\pi}{2\kappa^2}
\sum_{j=1}^6 m_j d_j(\kappa).
\end{equation}
Integrating backwards from the hemisphere gives
\begin{equation} \label{eq:schlafli-integrated}
{\rm Re}\,S(\kappa)
=
\pi^2
-
\frac{\pi}{2}
\int_\kappa^\infty
\frac{dk}{k^2}
\sum_{j=1}^6 m_j d_j(k).
\end{equation}
Using $d_j(k)=d_j^{(0)}+\order{k^{-2}}$ at large $k$, 
\begin{equation} \label{eq:volume-worldline-expansion}
{\rm Re}\,S
=
\pi^2
-
\frac{\pi}{2\kappa}
\sum_{j=1}^6 m_j d_j^{(0)}
+
\order{\kappa^{-3}}.
\end{equation}
The first correction to the hemisphere volume is therefore the weighted
length of the limiting graph.

The gravitational saddle is obtained by doubling the tetrahedron. Since
${\rm Vol}(M_{\vec\alpha})=2{\rm Vol}(T_{\vec\alpha})$, its on-shell action
is
\begin{equation}
I_{\rm EH+WL}[M_{\vec\alpha}]
=
-\frac{2\kappa}{\pi}{\rm Re}\,S.
\label{eq:double-grav-action}
\end{equation}
Substituting \eqref{eq:volume-worldline-expansion},
\begin{align}
I_{\rm EH+WL}[M_{\vec\alpha}]
&=
-2\pi\kappa
+
\sum_{j=1}^6 m_j d_j^{(0)}
+
\order{\kappa^{-2}}
\nonumber\\
&=
-S_{\rm dS}
+
I_{\rm tet}
+
\order{\kappa^{-2}},
\label{eq:gravity-to-tetra-action}
\end{align}
where
\begin{equation}
S_{\rm dS}=2\pi\kappa,
\qquad
I_{\rm tet}
=
\sum_{j=1}^6 m_j d_j^{(0)}.
\label{eq:tetra-worldline-action}
\end{equation}
Thus, after dividing out the empty-sphere contribution $e^{S_{\rm dS}}$, the
classical exponential becomes
\begin{equation}
\frac{Z_{\rm geo}}{Z_{S^3}}
\sim
e^{-I_{\rm tet}}.
\label{eq:tetra-exponential}
\end{equation}
Notice that the factor of two from doubling is important: it cancels the
factor $1/2$ in the Schläfli formula and leaves precisely one copy of the
worldline action.

We should also check that the limiting lengths $d_j^{(0)}$ describe a
stationary worldline graph. Consider a vertex at which edges $i,j,k$ meet,
and denote their outward unit tangent vectors by $t_i,t_j,t_k$. The three
directions define the spherical link of the vertex. Applying the spherical
cosine rule to this link gives
\begin{equation}
t_i\cdot t_j
=
\frac{
\cos\theta_k+\cos\theta_i\cos\theta_j
}{
\sin\theta_i\sin\theta_j
}.
\label{eq:link-cosine-rule}
\end{equation}
Using
\begin{equation}
\cos\theta_i
=
-1+\frac{\pi^2m_i^2}{2\kappa^2}
+O(\kappa^{-4}),
\qquad
\sin\theta_i
=
\frac{\pi m_i}{\kappa}
+O(\kappa^{-3}),
\end{equation}
one finds the finite limit
\begin{equation}
t_i\cdot t_j
\longrightarrow
\frac{
m_k^2-m_i^2-m_j^2
}{
2m_im_j
},
\label{eq:tangent-mass-cosine}
\end{equation}
and cyclically for the other pairs. These relations imply $\left|
m_it_i+m_jt_j+m_kt_k
\right|^2
=
0.$
Since the tangent space is positive definite,
\begin{equation}
m_it_i+m_jt_j+m_kt_k=0.
\label{eq:force-balance-recoil}
\end{equation}
This is exactly the force-balance equation obtained by varying the
worldline action. Indeed, if the limiting vertices are
$X_a\in S^3\subset\mathbb R^4$, then
\begin{equation}
d_{ab}
=
\arccos(X_a\cdot X_b),
\qquad
t_{ab}
=
\frac{
X_b-\cos d_{ab}\,X_a
}{
\sin d_{ab}
},
\label{eq:graph-distance-tangent}
\end{equation}
and a tangent displacement $X_a\cdot\delta X_a=0$ gives
\begin{equation}
\delta I_{\rm tet}
=
-\sum_a
\delta X_a\cdot
\left(
\sum_{b\neq a}m_{ab}t_{ab}
\right).
\label{eq:worldline-variation-recoil}
\end{equation}
Thus the graph obtained from the spherical tetrahedron is not only weighted
by the correct action, but its limiting vertices are already at the
stationary locations of the tetrahedral worldline problem.

\section{Details on evaluating the trefoil Seifert partition function} \label{app:Trefoil}

In this appendix, we provide more details on evaluating the partition functions on the Seifert manifolds obtained by integer Dehn fillings on the trefoil knot complement in the semiclassical limit. In particular, we derive the expression (\ref{eq:trefoil-Seifert-bminus}) by evaluating the $P$-integral in (\ref{eq:trefoil-Seifert-VTQFT}) in the semiclassical limit. First, by a change of integration variable, $P=ib_-y$ note that the exact partition function in this sector can be expressed as
\begin{equation}
    Z_{\rm Vir}^{(b_-)}[Y_n]=\frac{1}{\pi}e^{\frac{i\pi}{12}\left(k-5\right)}\sqrt{\frac{2\kappa}{k}}\sin \left(\frac{\pi}{k}\right)e^{\frac{\pi n}{12k}(\kappa-\kappa^{-1})}J_k(\kappa)\,,
\end{equation}
where $J_k(\kappa)$ is given by the following integral,
\begin{equation}
    J_k(\kappa)=-\frac{i\pi}{\kappa}\int_{\Gamma}dy e^{-\frac{\pi n y^2}{3k \kappa}}\frac{A_3(y;\kappa)A_k(y;\kappa)}{\sin(\pi y)\sinh\left(\frac{\pi y}{\kappa}\right)}\,,
\end{equation}
with the functions in the numerator of the integrand defined by
\begin{equation}
    A_j(y;\kappa)=\cos \frac{2\pi y}{j} \cosh \frac{2\pi y}{j \kappa} - \cot \frac{\pi}{j} \sin \frac{2\pi y}{j} \sinh \frac{2\pi y}{j \kappa}\,.
\end{equation}
The zeroes of the function in the denominator give simple poles along the real and imaginary axes (some of which could be cancelled by the zeroes of the numerator) at
\begin{equation}
    y= m, \qquad y= i\kappa m, \qquad m \in \mathbb{Z}-\{0\}.
\end{equation}
Note that there is also a double pole at $y=0$ which can be safely cancelled by subtracting the integrand,
\begin{equation}
    J_k(\kappa) \to -i \int_\Gamma dy \left(\frac{\pi}{\kappa}e^{-\frac{\pi n y^2}{3k \kappa}}\frac{A_3(y;\kappa)A_k(y;\kappa)}{\sin(\pi y)\sinh\left(\frac{\pi y}{\kappa}\right)}-\frac{1}{\pi y^2}\right)\,.
\end{equation}
Notice that the subtraction does not affect the value of $J_k(\kappa)$ since the second term evaluates to zero, given that the contour,
\begin{equation}
    \Gamma = t e^{-i\delta}, \quad t \in \mathbb{R}, \quad 0<\delta<\frac{\pi}{4}\,,
\end{equation}
is an oblique ray through the origin at an angle to the real axis. It is transported by analytic continuation in such a way that it lies above the poles on the negative real axis and is below the poles on the positive real axis and does not cross any poles. Since the Gaussian factor in the integral decays at the asymptotics of $\Gamma$, the integral is convergent. We evaluate the integral by deforming the contour into small semicircular arcs around the poles separated by horizontal segments along the real axis. Thus, the integral evaluates to the following two terms coming from the two portions of the contour,
\begin{equation}
    J_k(\kappa) = 2\pi \sum_{m=1}^{\infty} r_m - i\, \text{Pr}\left[\int_\mathbb{R} dy \left(\frac{\pi}{\kappa}e^{-\frac{\pi n y^2}{3k \kappa}}\frac{A_3(y;\kappa)A_k(y;\kappa)}{\sin(\pi y)\sinh\left(\frac{\pi y}{\kappa}\right)}-\frac{1}{\pi y^2}\right)\right]\,.
\end{equation}
Since the integrand is even, we have combined the residues $r_m$ at the poles on the positive and negative real axis. The second term is the principal value of the integral on the real axis coming from the horizontal segments of the contour. The first term gives the real part of $J_k(\kappa)$ and the second term gives the imaginary part. Now, we evaluate the two terms in the semiclassical limit $\kappa \to \infty$ to get closed form expressions. Evaluating the residues at the poles in this limit, we find
\begin{equation}
    \text{Re}\,J_k(\kappa) \to 2\sum_{m=1}^{\infty}\frac{(-1)^m}{m}\cos \frac{2\pi m}{3}\cos \frac{2\pi m}{k}, \qquad \kappa \to \infty\,.
\end{equation}
Using the identity 
\begin{equation}
    \sum_{m=1}^{\infty}\frac{(-1)^m}{m}\cos(m\theta)=-\log \left |2 \cos \frac{\theta}{2}\right|\,,
\end{equation}
we see that the above sum evaluates to 
\begin{equation}
   \lim_{\kappa \to \infty} \text{Re}\,J_k(\kappa) = -\log \left(1-2\cos \frac{2\pi}{k}\right).
\end{equation}
To evaluate the imaginary part in this limit, note that
\begin{equation}
    \text{Im}\, J_k(\kappa) \to -\text{Pr}\left[\int_\mathbb{R} dy \left(e^{-\frac{\pi n y^2}{3k \kappa}}\frac{\cos \left(\frac{2\pi y}{3}\right)\cos \left(\frac{2\pi y}{k}\right)}{y\sin(\pi y)}-\frac{1}{\pi y^2}\right)\right]\,.
\end{equation}
 We have retained the Gaussian factor in the integrand to ensure convergence. To evaluate the above principal value, note the identity 
 \begin{equation}
     \text{Pr}\left[\int_{\mathbb{R}}dy \left (\frac{\cos (cy)}{y\sin (\pi y)}-\frac{1}{\pi y^2}\right)\right]=0, \qquad |c|<\pi \,.
 \end{equation}
This is derived by noting that the integrand is regular at the origin, so the principal value can be evaluated by averaging the integral over the contours slightly above and slightly below the real axis. These integrals can respectively be evaluated by closing the contours in the upper half and lower half planes. The contribution from the semicircular arcs vanish since the integrand decays for $|c|<\pi$. Since these closed contours do not enclose any poles, they evaluate to zero. Applying this identity, it is easy to see that
\begin{equation}
    \lim_{\kappa \to \infty} \text{Im}\, J_k(\kappa) =\pi.
\end{equation}
We have therefore derived (\ref{eq:trefoil-Seifert-bminus}).

\bibliographystyle{JHEP}
\bibliography{main}
\end{document}